\documentclass[twocolumn]{aastex62}

\usepackage{ulem,xcolor}
\usepackage{epsf}
\usepackage{graphicx}
\usepackage{natbib}
\usepackage{url}
\usepackage{mathrsfs}
\usepackage{amsmath}
\usepackage{threeparttable, booktabs}

\newcommand{\Msun}{\mbox{\,$M_{\odot}$}}
\newcommand{\Lsun}{\mbox{\,$L_{\odot}$}}
\newcommand{\kms}{\,km~s$^{-1}$}
\newcommand{\degree}{$^\circ$}
\newcommand{\nai}{\ion{Na}{1}}
\newcommand{\hi}{\ion{H}{1}}
\newcommand{\caii}{\ion{Ca}{2}}
\newcommand{\rproj}{$R_{\rm proj}$}
\newcommand{\feh}{$\rm[Fe/H]$}
\newcommand{\fehp}{$\rm[Fe/H]_{\rm phot}$}
\newcommand{\afe}{$\rm[\alpha/Fe]$}
\newcommand{\vsys}{$v_{\rm sys}$}
\newcommand{\rdisk}{R$_{\rm disk}$}
\newcommand{\sigmav}{$\sigma_{\rm v}$}

\newcommand{\sigmavoff}{$\sigma_{\rm v, offset}$}
\newcommand{\sighalo}{$\sigma_{\rm halo}$}
\newcommand{\sigdisk}{$\sigma_{\rm disk}$}

\newcommand{\vlos}{$v_{\rm los}$}

\newcommand{\vmodel}{$v_{\rm H\,\textsc{i}}$}
\newcommand{\voff}{$v_{\rm offset}$}
\newcommand{\fhalo}{$f_{\rm halo}$}
\newcommand{\frot}{$f_{\rm rot}$}
\newcommand{\frotdisk}{$f_{\rm rot, disk}$}
\newcommand{\frothalo}{$f_{\rm rot, halo}$}
\newcommand{\pdisk}{$P_{\rm disk}$}

\shorttitle{Stellar Kinematics in M33}
\shortauthors{Gilbert et al.}

\newcommand{\triangsystemic}{$=-180\pm3$}
\newcommand{\triangdmod}{$24.67$}
\newcommand{\rdiskcut}{15}
\newcommand{\rdiskcuttwo}{30}
\newcommand{\nslitmasks}{36}

\newcommand{\deltafehpmedinner}{0.09}
\newcommand{\deltafehpmedmiddle}{0.24}
\newcommand{\deltafehpmedouter}{0.28}
\newcommand{\twosidekstestmiddle}{0.01}
\newcommand{\twosidekstestouter}{0.0002}

\newcommand{\noRGBstars}{1667}

\newcommand{\medianRGBveluncertainty}{8.2}
\newcommand{\maxrprojRGBNarcmin}{45.7}
\newcommand{\maxrprojRGBSarcmin}{43.2}
\newcommand{\maxrdiskRGBNarcmin}{72.5}
\newcommand{\maxrdiskRGBSarcmin}{63.7}
\newcommand{\minrprojRGBNarcmin}{3.7}

\newcommand{\minrdiskRGBNarcmin}{3.7}

\newcommand{\noyoungstars}{286}

\newcommand{\medianyoungstarsveluncertainty}{6.3}

\newcommand{\sigvRGBdiskonly}{$43.9^{+0.8}_{-0.7}$}

\newcommand{\sigvyoungdiskonly}{$16.0^{+0.7}_{-0.6}$}
\newcommand{\sigvyoungdiskonlyintr}{$\sim 15$}

\newcommand{\plusthindisklagmodeldisksigv}{$21.1^{+0.7}_{-0.7}$}
\newcommand{\plusthindisklagmodeldisksigvintr}{$\sim 19$}
\newcommand{\plusthindisklagmodelfhalo}{$0.22^{+0.02}_{-0.02}$}
\newcommand{\plusthindisklagmodelfrotdisk}{$0.87^{+0.01}_{-0.01}$}
\newcommand{\plusthindisklagmodelfrothalo}{$0.11^{+0.08}_{-0.10}$}
\newcommand{\plusthindisklagmodelhalosigv}{$59.3^{+2.6}_{-2.5}$}
\newcommand{\plusthindisklagmodelhalosigvintr}{$\sim 59$}

\newcommand{\noRGBstarsintPHAT}{513}
\newcommand{\plusthindisklagmodeldisksigvintPHAT}{$26.3^{+1.7}_{-1.7}$}

\newcommand{\plusthindisklagmodelfhalointPHAT}{$0.28^{+0.05}_{-0.04}$}
\newcommand{\plusthindisklagmodelfrotdiskintPHAT}{$0.96^{+0.03}_{-0.04}$}
\newcommand{\plusthindisklagmodelfrothalointPHAT}{$-0.48^{+0.19}_{-0.16}$}
\newcommand{\plusthindisklagmodelhalosigvintPHAT}{$43.2^{+5.9}_{-5.0}$}
\newcommand{\rgbveluncintPHAT}{8.9}

\newcommand{\noRGBstarsintPHATS}{126}
\newcommand{\plusthindisklagmodeldisksigvintPHATS}{$18.4^{+5.3}_{-4.3}$}

\newcommand{\plusthindisklagmodelfhalointPHATS}{$0.56^{+0.11}_{-0.11}$}
\newcommand{\plusthindisklagmodelfrotdiskintPHATS}{$1.10^{+0.06}_{-0.07}$}
\newcommand{\plusthindisklagmodelfrothalointPHATS}{$0.09^{+0.19}_{-0.21}$}
\newcommand{\plusthindisklagmodelhalosigvintPHATS}{$51.5^{+5.4}_{-5.0}$}
\newcommand{\rgbveluncintPHATS}{12.3}

\newcommand{\noRGBstarsintPHATN}{387}
\newcommand{\plusthindisklagmodeldisksigvintPHATN}{$26.5^{+1.5}_{-1.4}$}

\newcommand{\plusthindisklagmodelfhalointPHATN}{$0.23^{+0.03}_{-0.03}$}
\newcommand{\plusthindisklagmodelfrotdiskintPHATN}{$0.93^{+0.03}_{-0.03}$}
\newcommand{\plusthindisklagmodelfrothalointPHATN}{$-0.71^{+0.12}_{-0.11}$}
\newcommand{\plusthindisklagmodelhalosigvintPHATN}{$35.5^{+4.8}_{-3.7}$}
\newcommand{\rgbveluncintPHATN}{7.8}

\newcommand{\noRGBstarsextPHAT}{631}
\newcommand{\plusthindisklagmodeldisksigvextPHAT}{$21.0^{+1.1}_{-1.0}$}

\newcommand{\plusthindisklagmodelfhaloextPHAT}{$0.17^{+0.04}_{-0.03}$}
\newcommand{\plusthindisklagmodelfrotdiskextPHAT}{$0.88^{+0.01}_{-0.01}$}
\newcommand{\plusthindisklagmodelfrothaloextPHAT}{$0.11^{+0.14}_{-0.19}$}
\newcommand{\plusthindisklagmodelhalosigvextPHAT}{$55.8^{+5.0}_{-5.1}$}
\newcommand{\rgbveluncextPHAT}{11.0}

\newcommand{\noRGBstarsextPHATS}{402}
\newcommand{\plusthindisklagmodeldisksigvextPHATS}{$22.3^{+1.4}_{-1.4}$}

\newcommand{\plusthindisklagmodelfhaloextPHATS}{$0.16^{+0.05}_{-0.04}$}
\newcommand{\plusthindisklagmodelfrotdiskextPHATS}{$0.86^{+0.02}_{-0.02}$}
\newcommand{\plusthindisklagmodelfrothaloextPHATS}{$0.05^{+0.20}_{-0.28}$}
\newcommand{\plusthindisklagmodelhalosigvextPHATS}{$55.3^{+6.8}_{-7.6}$}
\newcommand{\rgbveluncextPHATS}{14.0}

\newcommand{\noRGBstarsextPHATN}{229}
\newcommand{\plusthindisklagmodeldisksigvextPHATN}{$19.7^{+1.4}_{-1.3}$}

\newcommand{\plusthindisklagmodelfhaloextPHATN}{$0.16^{+0.05}_{-0.04}$}
\newcommand{\plusthindisklagmodelfrotdiskextPHATN}{$0.90^{+0.02}_{-0.02}$}
\newcommand{\plusthindisklagmodelfrothaloextPHATN}{$0.07^{+0.22}_{-0.26}$}
\newcommand{\plusthindisklagmodelhalosigvextPHATN}{$55.2^{+8.8}_{-8.6}$}
\newcommand{\rgbveluncextPHATN}{6.8}

\newcommand{\noRGBstarsintbreak}{1144}
\newcommand{\plusthindisklagmodeldisksigvintbreak}{$23.1^{+1.4}_{-1.3}$}

\newcommand{\plusthindisklagmodelfhalointbreak}{$0.24^{+0.05}_{-0.05}$}
\newcommand{\plusthindisklagmodelfrotdiskintbreak}{$0.90^{+0.01}_{-0.01}$}
\newcommand{\plusthindisklagmodelfrothalointbreak}{$-0.07^{+0.17}_{-0.28}$}
\newcommand{\plusthindisklagmodelhalosigvintbreak}{$54.4^{+3.8}_{-7.1}$}
\newcommand{\rgbveluncintbreak}{9.7}

\newcommand{\noRGBstarsextbreak}{523}
\newcommand{\plusthindisklagmodeldisksigvextbreak}{$18.8^{+0.8}_{-0.7}$}

\newcommand{\plusthindisklagmodelfhaloextbreak}{$0.10^{+0.02}_{-0.02}$}
\newcommand{\plusthindisklagmodelfrotdiskextbreak}{$0.84^{+0.01}_{-0.01}$}
\newcommand{\plusthindisklagmodelfrothaloextbreak}{$0.17^{+0.16}_{-0.18}$}
\newcommand{\plusthindisklagmodelhalosigvextbreak}{$67.2^{+8.3}_{-6.7}$}
\newcommand{\rgbveluncextbreak}{5.6}

\newcommand{\noRGBstarsextbreakS}{221}
\newcommand{\plusthindisklagmodeldisksigvextbreakS}{$23.0^{+1.3}_{-1.2}$}

\newcommand{\plusthindisklagmodelfhaloextbreakS}{$0.09^{+0.03}_{-0.02}$}
\newcommand{\plusthindisklagmodelfrotdiskextbreakS}{$0.91^{+0.02}_{-0.02}$}
\newcommand{\plusthindisklagmodelfrothaloextbreakS}{$-0.42^{+0.13}_{-0.12}$}
\newcommand{\plusthindisklagmodelhalosigvextbreakS}{$25.9^{+5.1}_{-2.4}$}
\newcommand{\rgbveluncextbreakS}{6.4}

\newcommand{\noRGBstarsextbreakN}{302}
\newcommand{\plusthindisklagmodeldisksigvextbreakN}{$16.7^{+0.8}_{-0.7}$}

\newcommand{\plusthindisklagmodelfhaloextbreakN}{$0.07^{+0.02}_{-0.02}$}
\newcommand{\plusthindisklagmodelfrotdiskextbreakN}{$0.81^{+0.01}_{-0.01}$}
\newcommand{\plusthindisklagmodelfrothaloextbreakN}{$0.01^{+0.24}_{-0.26}$}
\newcommand{\plusthindisklagmodelhalosigvextbreakN}{$79.3^{+16.0}_{-11.8}$}
\newcommand{\rgbveluncextbreakN}{5.2}

\begin{document}
\bibliographystyle{aasjournal}

\title{The TREX Survey: Kinematical Complexity Throughout M33's Stellar Disk and Evidence for a Stellar Halo \footnote{The data presented herein were obtained at the W.M. Keck Observatory,
which is operated as a scientific partnership among the California
 Institute of Technology, the University of California and the National
Aeronautics and Space Administration. The Observatory was made
possible by the generous financial support of the W.M. Keck
Foundation.}}

\correspondingauthor{Karoline M. Gilbert}
\email{kgilbert@stsci.edu}

\author[0000-0003-0394-8377]{Karoline M. Gilbert}
\affiliation{Space Telescope Science Institute, 3700 San Martin Dr., Baltimore, MD 21218, USA}
\affiliation{Department of Physics \& Astronomy, Bloomberg Center for Physics and Astronomy, Johns Hopkins University, 3400 N. Charles Street, Baltimore, MD 21218}

\author[0000-0001-8481-2660]{Amanda C. N. Quirk}
\affiliation{UCO/Lick Observatory, Department of Astronomy \& Astrophysics, University of California Santa Cruz, 
 1156 High Street, 
 Santa Cruz, California 95064, USA}

\author[0000-0001-8867-4234]{Puragra Guhathakurta}
\affiliation{UCO/Lick Observatory, Department of Astronomy \& Astrophysics, University of California Santa Cruz, 
 1156 High Street, 
 Santa Cruz, California 95064, USA}

\author[0000-0002-9599-310X]{Erik Tollerud}
\affiliation{Space Telescope Science Institute, 3700 San Martin Dr., Baltimore, MD 21218, USA}

\author[0000-0002-3233-3032]{Jennifer Wojno}
\affiliation{Department of Physics \& Astronomy, Bloomberg Center for Physics and Astronomy, Johns Hopkins University, 3400 N. Charles Street, Baltimore, MD 21218}

\author[0000-0002-1264-2006]{Julianne J. Dalcanton}
\affiliation{Department of Astronomy, University of Washington, Box 351580, U.W., Seattle, WA 98195-1580, USA}
\affiliation{Center for Computational Astrophysics, Flatiron Institute, 162 Fifth Avenue, New York, NY 10010, USA}

\author[0000-0002-0786-7307]{Meredith J. Durbin}
\affiliation{Department of Astronomy, University of Washington, Box 351580, U.W., Seattle, WA 98195-1580, USA}

\author[0000-0003-0248-5470]{Anil Seth}
\affiliation{Dept. of Physics Astronomy, 
University of Utah, 
115 South 1400 East, 
Salt Lake City, UT 84112, USA}
 
\author[0000-0002-7502-0597]{Benjamin F. Williams}
\affiliation{Department of Astronomy, University of Washington, Box 351580, U.W., Seattle, WA 98195-1580, USA}

\author{Justin T. Fung}
\affiliation{The Harker School, 500 Saratoga Ave., San Jose, CA 95129, USA}
\author{Pujita Tangirala}
 \affiliation{Saint Francis High School, 1885 Miramonte Avenue
Mountain View, California 94040, USA}
 \author{Ibrahim Yusufali}
 \affiliation{Duke University, 2080 Duke University Road, Durham, North Carolina 27708, USA}

\begin{abstract}
We present initial results from a large spectroscopic survey of stars throughout M33's stellar disk.  We analyze a sample of \noRGBstars\ red giant branch (RGB) stars extending to projected distances of $\sim 11$~kpc from M33's center ($\sim 18$~kpc, or $\sim 10$ scale lengths, in the plane of the disk). The line-of-sight velocities of RGB stars show the presence of two kinematical components. One component is consistent with rotation in the plane of M33's \hi\ disk and has a velocity dispersion (\plusthindisklagmodeldisksigvintr~\kms) consistent with that observed in a comparison sample of younger stars, while the second component has a significantly higher velocity dispersion. A two-component fit to the RGB velocity distribution finds that the high dispersion component has a velocity dispersion of
\plusthindisklagmodelhalosigv~\kms\ and rotates very slowly in the plane of the disk (consistent  with  no  rotation at the $<1.5\sigma$ level), which favors interpreting it as a stellar halo rather than a thick disk population.  A spatial analysis indicates that the fraction of RGB stars in the high-velocity-dispersion component decreases with increasing radius over the range covered by the spectroscopic sample. Our spectroscopic sample establishes that a significant high-velocity-dispersion component is present in M33's RGB population from near M33's center to at least the radius where M33's \hi\ disk begins to warp at 30\arcmin\ ($\sim 7.5$~kpc) in the plane of the disk. This is the first detection and spatial characterization of a kinematically hot stellar component throughout M33's inner regions. 

\end{abstract}


\section{Introduction}\label{sec:intro} 
Extended, old stellar distributions are found nearly ubiquitously in star forming dwarf galaxies \citep[e.g.,][and references therein]{stinson2009}. Multiple physical mechanisms have been identified as potential contributors to stellar halos in low-mass galaxies, including merging of lower-mass subhalos \citep{helmi2012,starkenburg2015,starkenburg2016} and purely internal heating mechanisms \citep{stinson2009,maxwell2012, el-badry2016,kado-fong2021}.   
However, a definitive kinematical detection of a stellar halo component around a relatively isolated
low-mass galaxy has remained elusive. 

The Triangulum galaxy (M33) provides a unique opportunity to study the resolved stellar dynamics of a low-mass \citep[$M_*\sim 4.8\times10^9$\Msun;][]{corbelli2014} disk galaxy, and to identify and measure the kinematical properties of a low-mass galaxy's stellar halo. It is located $\sim 230$~kpc from M31, and proper motion measurements suggest M33 is 
on first infall into the M31 system \citep{patel2017b, patel2017a, vandermarel18}.
M33 therefore should be relatively undisturbed, although its extended \hi\ disk suggests a minor kinematical warp at $r\gtrsim 7$~kpc \citep[e.g.,][]{corbelli1997,putman2009,corbelli2014,kam2017}, equivalent to $\gtrsim4$ disk scale lengths \citep[as measured at 3.6\micron;][]{verley2009,kam2015}. 

Studies of M33 have led to mixed conclusions regarding the presence for a stellar halo.
The existence of metal-poor RR Lyrae stars in M33's disk, including at large deprojected disk radii, has been interpreted as evidence of a stellar halo \citep{sarajedini2006,pritzl2011,tanakul2017}. 
M33 also hosts a population of disk stellar clusters with a large spread in line-of-sight velocities, which has been interpreted as a halo cluster population \citep{schommer1991,chandar2002}. 
An outer break in the exponential surface density profile of M33's disk at $\sim 36'$ \citep[8~kpc; $\sim 4.5$ disk scale lengths][]{ferguson2007} 
has been interpreted 
as a transition to either 
a stellar halo component
\citep{barker2011, cockcroft2013} or 
a large
extended disk \citep{grossi2011}.  

In contrast to the above evidence in favor of a halo, an analysis of the final Pan-Andromeda Archaeological Survey (PAndAS) dataset \citep{mcconnachie2018} did not detect a stellar halo component around M33.   Employing a spatially varying contamination model for the PAndAS dataset \citep{martin2013pandas} and a statistical spatial analysis of the resolved stellar density distribution, \citet{mcmonigal2016} determined an upper 
limit on the maximum surface brightness of a smooth stellar halo of $\mu_{\rm V} = 35.5$ mag arcsec$^{-2}$, and
limited any halo luminosity to $< 10^6$~\Lsun\ over a range of 0.8\degree\ to 3.75\degree\ in radius; they extrapolated this to a total halo luminosity upper limit of $\sim2\times10^6$\Lsun.  A recent deep, narrow-band imaging survey characterizing M33's planetary nebulae also found no evidence of an extended stellar halo population \citep{galera-rosillo2018}. 

The most definitive evidence for the presence of a stellar halo would be identifying a population of old stars whose kinematics are more consistent with an extended, kinematically hot spheroidal distribution than with a rotating disk. To date, the only measurements of individual velocities of red giant branch stars in the literature were presented by \citet{mcconnachie2006m33} and \cite{hood2012}.   
The velocity distribution measured by \citeauthor{mcconnachie2006m33} 
includes a tail of 12 stars with large negative line-of-sight velocities, consistent with the presence of a component in excess of the stellar disk, 
providing the first tentative evidence of a halo in M33 from stellar velocities. \citeauthor{hood2012} 
identified 11 potential halo stars in the tail of the M33 stellar velocity distribution. 
Both estimated a dispersion of \sigmav$\sim 50$\kms\ for their respective halo star samples.

We have expanded greatly on this earlier work, and present here first results from a spectroscopic survey of stars throughout M33's disk. Section~\ref{sec:data} briefly describes the TRiangulum EXtended Survey (TREX), the spectroscopic data reduction, and the sample selection.  Section~\ref{sec:models} presents the velocity distribution of RGB stars and young, red stars 
in M33's disk and the results of fitting models to the velocity distribution of the stellar populations.  Section~\ref{sec:fehphot} compares photometric metallicity estimates of stars in each of the two components in the RGB stellar sample. 
Section~\ref{sec:disc} summarizes our results and places them within the context of recent observations of M33 and simulations of low-mass galaxies. 

We assume a systemic velocity for M33 of \vsys\triangsystemic\kms\ \citep{kam2017}, and a distance modulus of $m-M=$\,\triangdmod\ \citep{sarajedini2006,degrijs2014}, corresponding to a distance of 859~kpc (resulting in a scale of $1' = 250$~pc). 
For the purposes of this work, we define `inner disk' as the region interior to both M33's stellar disk break and the beginning of the \hi\ warp at 30\arcmin\ in the plane of the disk \citep{kam2017}.

\section{Spectroscopic Dataset}\label{sec:data}

\begin{figure*}[tbh]
\includegraphics[width=1.0\textwidth]{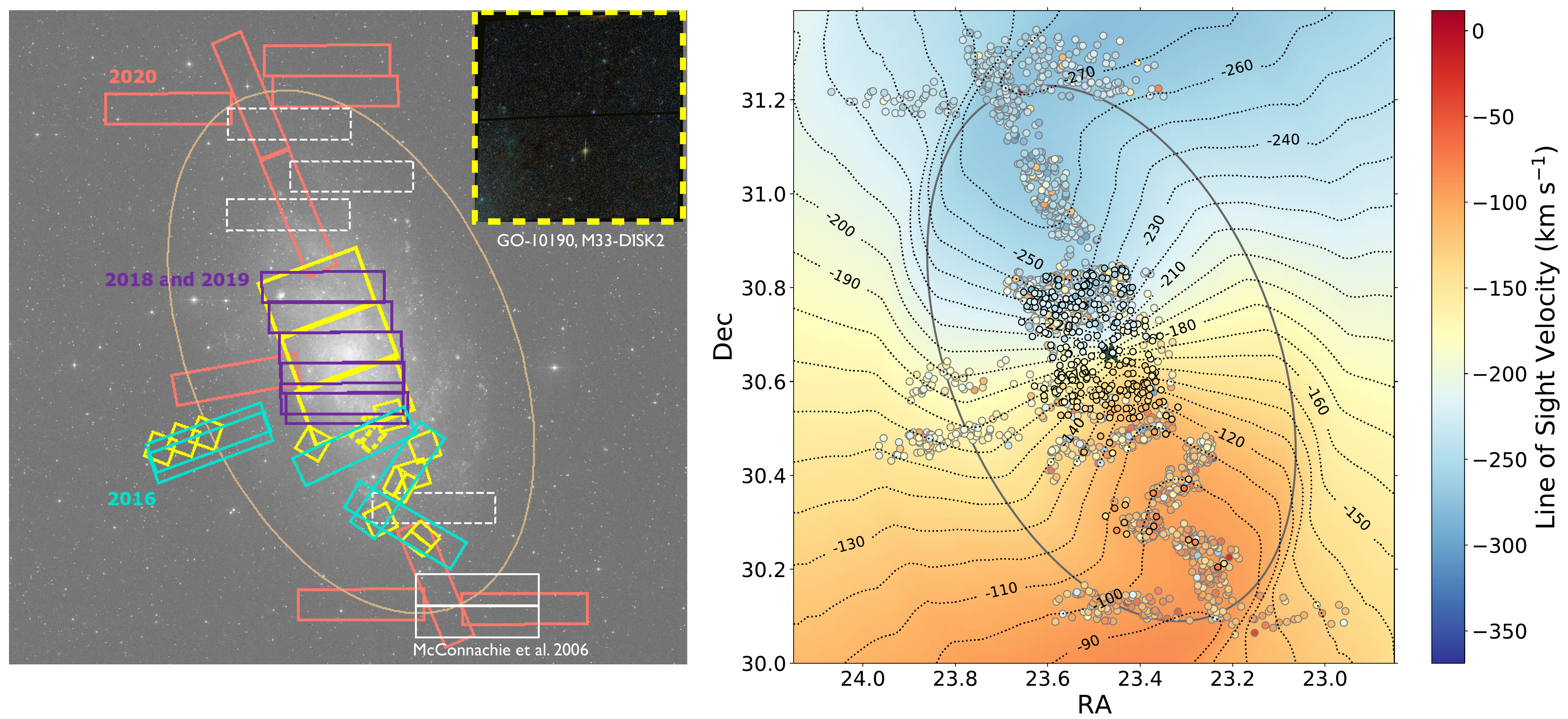}
\caption{(\textit{Left:}) Footprints of the \nslitmasks\ spectroscopic slitmasks 
obtained in our M33 survey (Section~\ref{sec:data}) in 2016 (turquoise), 2018 and 2019 (purple), and 2020 (coral); we observed between one and three masks at each location. 
Also shown is the extent of 
the contiguous, 6-band UV to NIR HST imaging obtained by the PHATTER survey \citep[large yellow rectangles]{williams2021}, 
other archival HST optical broadband imaging used for target selection (small yellow squares), and spectroscopic masks with potential halo stars 
observed by \citet[solid white]{mcconnachie2006m33} and \citet[dashed white]{hood2012}.
The inset shows the stellar density at the edge of the PHATTER survey: 
M33-DISK2 is the HST/ACS field shown with a dashed yellow outline.  Fields interior to this have higher stellar density, necessitating HST-based target selection.
(\textit{Right:}) Location on the sky of the RGB star sample (grey outlined points) and young star comparison sample (black outlined points; Section~\ref{sec:data}).  The RGB star sample extends from \minrprojRGBNarcmin\arcmin\ to \maxrprojRGBNarcmin\arcmin\ ($\sim 0.9$ to 11.4~kpc) in projected distance from M31's center. The stellar points are color-coded by their measured line-of-sight velocity.  The contours and underlying color map shows the M33 \hi\ disk velocity as a function of position assuming the \hi\ disk model of \citet{kam2017}.  The large star denotes M33's center.
 In both panels, the ellipse denotes the approximate location of the break in the surface density profile of M33's disk, observed at 36\arcmin\ \citep{ferguson2007}.
}
\label{fig:roadmap}
\end{figure*}

Full details of the TREX spectroscopic survey will be presented in Quirk et al.\ (in preparation).  We summarize here the aspects most relevant for the present work.  

Targets for spectroscopy were selected using a collection of heterogeneous imaging data.  Due to the high stellar density, 
HST imaging was used 
in all regions where it was available (Figure~\ref{fig:roadmap})
to identify stars which would remain isolated in typical ground-based observing conditions. 
The ten spectroscopic masks observed in 2016 were designed to connect archival HST fields with moderately deep imaging in two broadband optical filters \citep[using photometric reductions by][]{williams2009}. 
We used contiguous HST imaging from the Panchromatic Hubble Andromeda Treasury: Triangulum Extended Region \citep[PHATTER;][]{williams2021} as the basis of target selection for the majority of the area covered by the 15 masks observed in 2018 and 2019. 
In areas where HST imaging was not available 
targets were selected from stellar catalogs derived from CHFT/MegaCam imaging using the MegaPipe reduction pipeline \citep{gwyn2008}.  In the 11 masks observed in 2020, the public release of the PAndAS catalog \citep{mcconnachie2018} was used for all target selection.

Spectra of the target stars were obtained with the DEIMOS Spectrograph \citep{faber2003} on the Keck~II 10-m telescope.  The \nslitmasks\ multi-object spectroscopic slitmasks analyzed here were observed with either the 
the 600~line~mm$^{-1}$ grating ($R\sim 2000$) 
and a central wavelength setting of 7200~\AA\@, 
resulting in a wavelength range of 
$\lambda\lambda\sim$~4600\,--\,9800~\AA\@  (22 masks), or the   
1200~line~mm$^{-1}$ grating ($R\sim 6000$) and a central wavelength setting of 7800~\AA\@, resulting in a wavelength range of 
$\lambda\lambda\sim$~6300\,--\,9800~\AA\@ (14 masks).  The choice of grating depended on whether the mask targeted young, blue stars in addition to older red stars (600~line~mm$^{-1}$ grating, 2016 and 2018 masks), or heavily prioritized red giant branch stars (1200~line~mm$^{-1}$ grating, 2019 and 2020 masks). 

The locations of the masks are shown in Figure~\ref{fig:roadmap}. The density of stars in M33's disk allowed for multiple slitmasks with non-overlapping target sets to be designed with the same mask center. The spectroscopic targets extend 
to projected distances from M33's center of \rproj$=$\maxrprojRGBNarcmin\arcmin\ (11.4~kpc) in the northeastern half of the disk, and \rproj$=$\maxrprojRGBSarcmin\arcmin\ (10.8~kpc) in the southwestern half.  This translates to deprojected distances from M33's center in the plane of the disk (computed using the \citet{kam2017} disk model, Section~\ref{sec:models}) of \rdisk$=$\maxrdiskRGBNarcmin\arcmin\ (18.1~kpc) in the northeastern half, and \rdisk$=$\maxrdiskRGBSarcmin\arcmin\ (15.9~kpc) in the southwestern half.  

Spectra were reduced using the {\tt spec2d} and {\tt spec1d} software \citep{cooper2012,newman2013}.  These routines perform flat-fielding, night-sky emission line removal, extraction
of one-dimensional spectra, and redshift measurement.  Line-of-sight velocities were measured by cross-correlating the spectra against stellar templates \citep{simon2007}, and transformed to the heliocentric frame. A correction for imperfect centering of the star within the slit was applied, using the observed position of the atmospheric A-band absorption feature relative to night-sky emission lines  \citep{simon2007,sohn2007}. Since the individual A-band correction measurements for this dataset were found in some cases to have unreasonably high values, we calculated and applied a median A-band correction as a function of the position of the target's slit on the mask.  This method corrects for any errors in alignment of the slitmask, as well as any differential relative astrometry errors between guide and alignment stars and target stars in the cases where the mask included a mix of targets from ground- and HST-based stellar catalogs.  The mean of the absolute magnitudes of the A-band corrections for the RGB sample defined below was $7.3$~\kms, with a standard deviation of $6.5$\kms.
Measurement uncertainties for each star were estimated by combining the random uncertainty estimated by the velocity measurement cross-correlation routine with systematic uncertainties of $5.6$~\kms\ and $2.2$~\kms\ for stars observed with the 600~line~mm$^{-1}$ and 1200~line~mm$^{-1}$ gratings respectively \citep{collins2011, simon2007}.  

\begin{figure*}[tbh]
\includegraphics[width=0.49\textwidth]{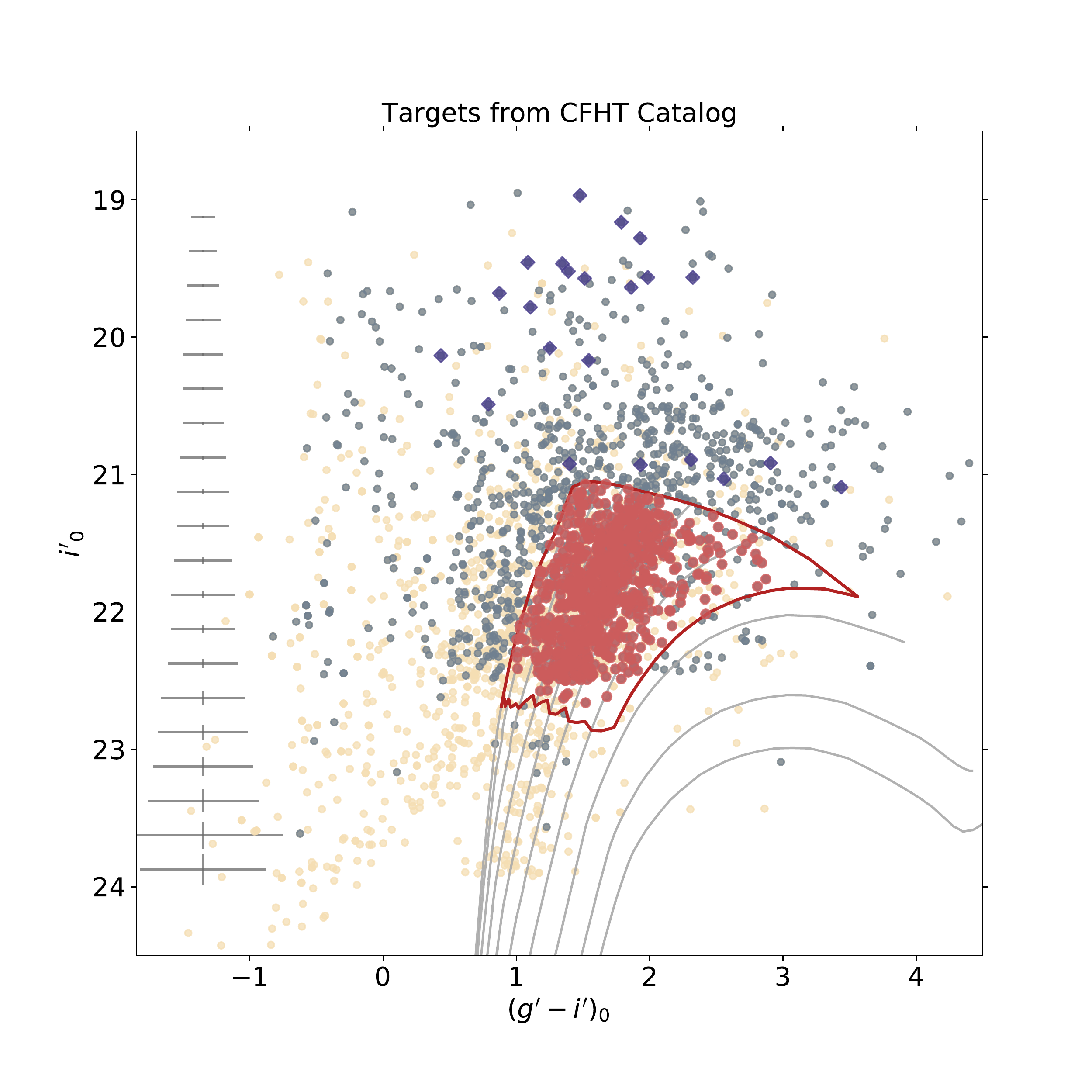}
\includegraphics[width=0.49\textwidth]{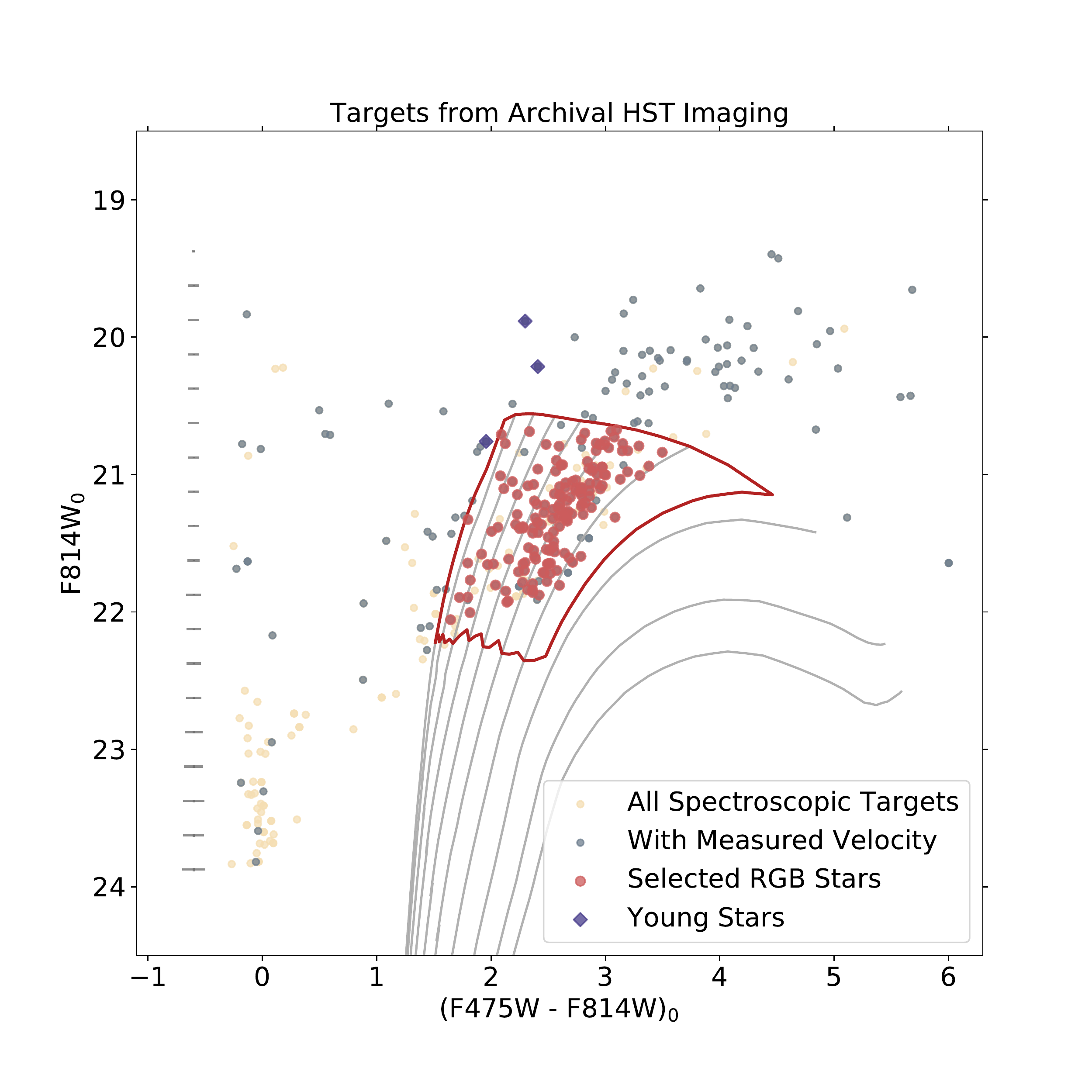}
\includegraphics[width=0.49\textwidth]{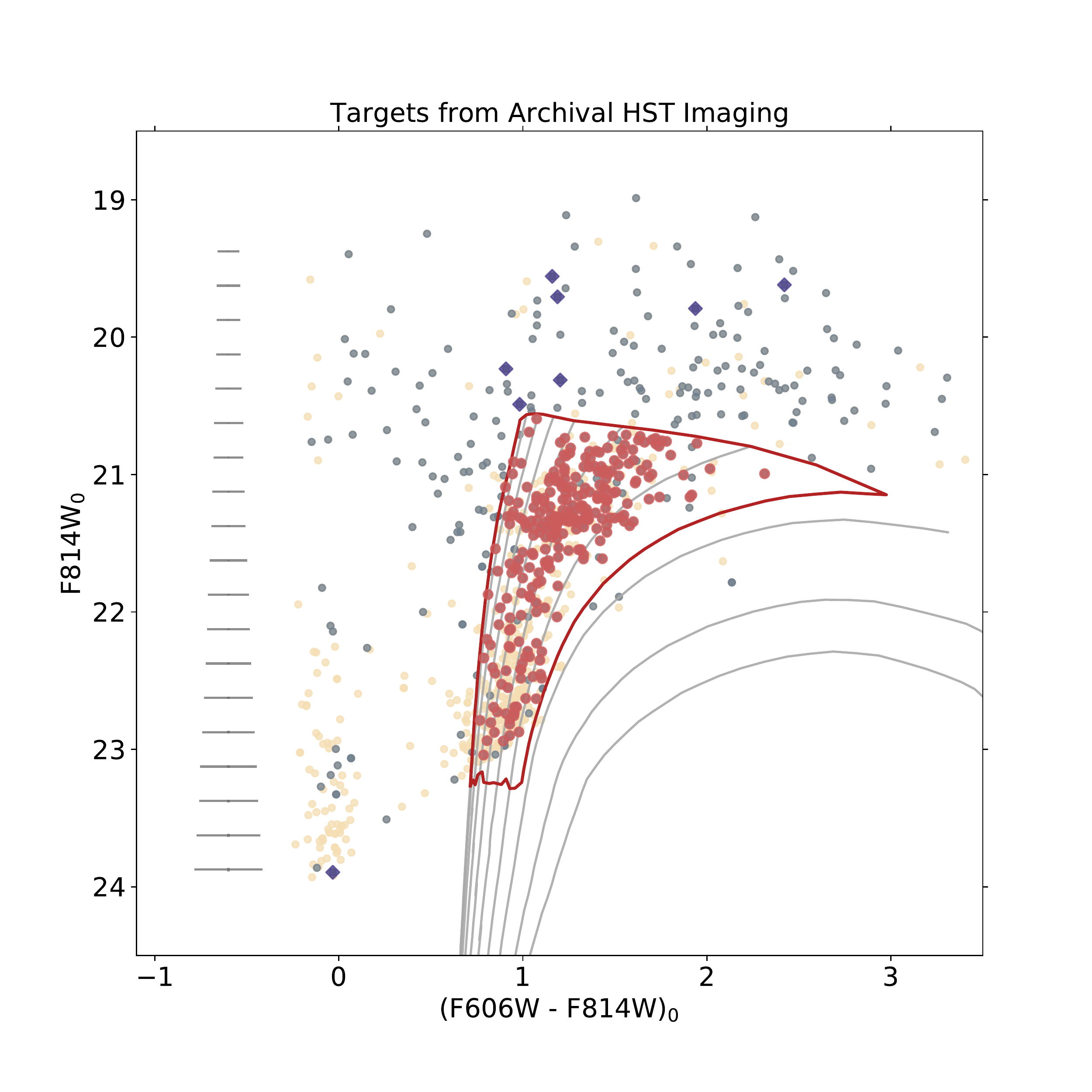}
\includegraphics[width=0.49\textwidth]{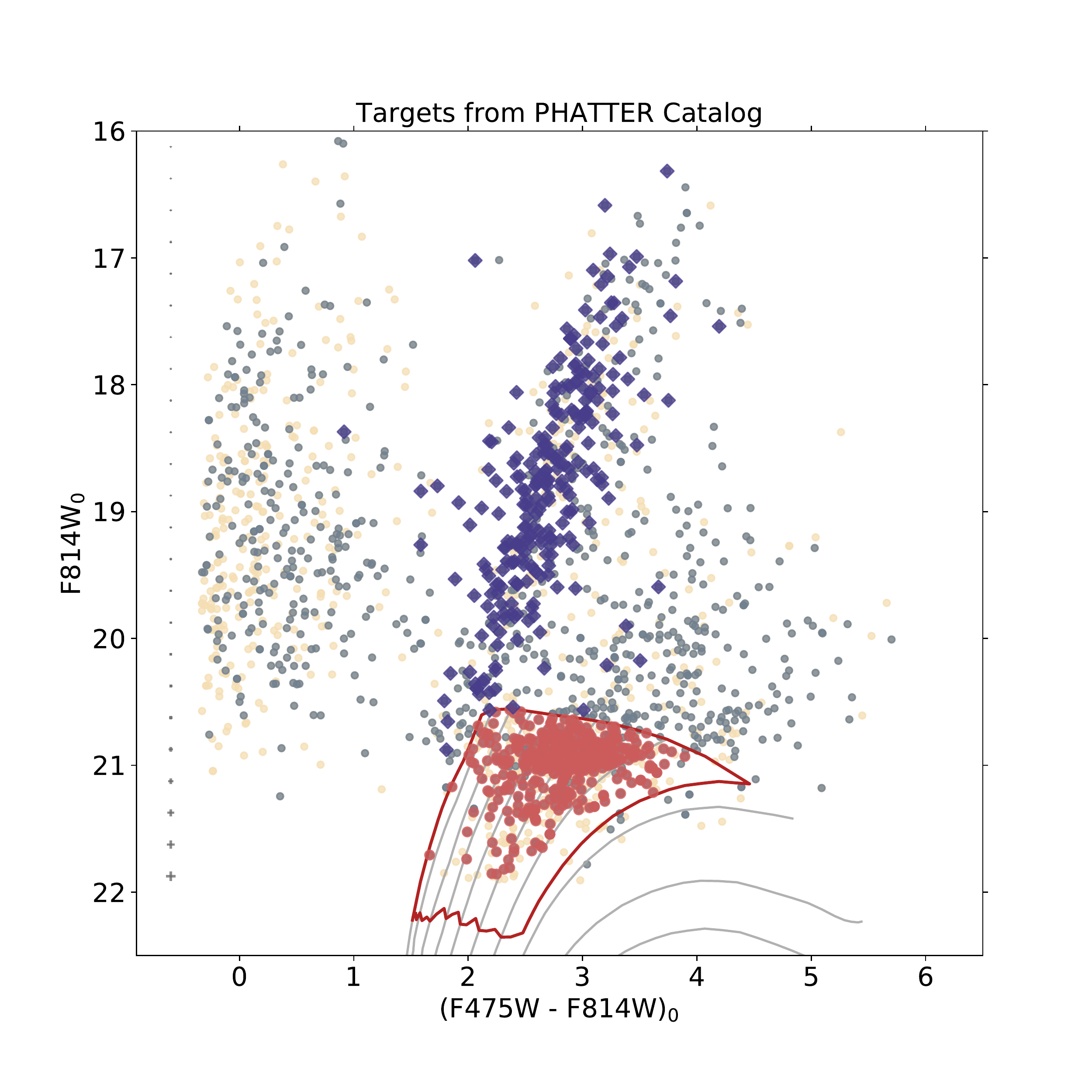}
\caption{
Color-magnitude diagrams for all spectroscopic targets (beige circles), as well as those with measured velocities (grey circles), and those selected for the RGB sample (larger red circles) and young red star comparison sample (blue diamonds; Section~\ref{sec:data}).  Average photometric uncertainties as a function of magnitude are shown on the left hand side of each panel.
A consistent RGB sample selection region was identified across the various filter combinations using 10 Gyr, \afe\,$=0$ Padova isochrones \citep{marigo2017}; the RGB selection box as well as select isochrone tracks 
 are shown in each panel.  The metal-rich extent of the RGB selection box shown here was selected to fully encompass the final RGB sample, and was used in estimating the MW contamination within the RGB sample region (Appendix~\ref{sec:app_mwcont}). 
}
\label{fig:cmds}
\end{figure*}

We used selection boxes in color-magnitude space, defined using Padova isochrones\footnote{\tt http://stev.oapd.inaf.it/cgi-bin/cmd\_3.2} \citep[10~Gyr, \afe\,$=0$;][]{marigo2017}, to ensure consistent selection of an RGB star sample across the multiple filter sets from which the spectroscopic targets were drawn (Figure~\ref{fig:cmds}).  Stars falling below the tip of the RGB, and redder than the \feh\,$=-2.32$ isochrone, were included in the RGB sample. We remove foreground MW stars by excluding stars with signs of the surface-gravity sensitive \nai\ doublet absorption feature in their spectrum; we estimate that MW contaminants comprise $<1$\% of the final RGB sample (Appendix~\ref{sec:app_mwcont}).  We also removed stars with possible spectral characteristics of weak CN or carbon stars, and a small number of spectra whose S/N indicates the automated reduction misidentified another, significantly brighter star in the slit for the target star.  

We isolated a comparison sample of younger red stars, 
selected on the basis of spectroscopic characteristics. 
The young star sample consists of stars whose spectra are consistent with those of ``weak CN'' stars.\footnote{These stars have a much weaker CN spectral absorption feature at $\sim7900$\AA\ than is typically seen in carbon stars.  They also have spectral absorption features such as TiO and the Ca triplet that are generally not detectable in carbon star spectra but are typically seen in stars with oxygen rich atmospheres.  Comparisons to stellar models indicate weak CN stars are massive (5-10\Msun) stars in a core helium burning phase \citep{guhathakurta2017}.}   
As shown in Figure~\ref{fig:cmds}, the weak CN 
stars are found in the red helium burning sequence. 
Unlike the blue massive (and also young) stars in the dataset, the weak CN 
stars have an abundance of absorption features in the red region of the spectrum, where the majority of the features that drive the velocity determination for RGB stars reside. 
These young, red stars sample the kinematics of M33's disk with similar velocity measurement systematics as the RGB sample. However, since they are on average brighter than the RGB stars, they have higher average spectral S/N and correspondingly lower average velocity uncertainties than the RGB sample. 

In total, \noRGBstars\ and \noyoungstars\ stars pass the sample selection criteria for RGB and young stars, respectively.  The median velocity measurement uncertainty for the RGB and young star samples are \medianRGBveluncertainty\kms\ and  \medianyoungstarsveluncertainty\kms, respectively. 
The spatial distributions of both samples are shown in the right panel of Figure~\ref{fig:roadmap}. Due to a combination of target selection strategies and intrinsic spatial variation in the age of the stellar populations, the young star sample is more centrally concentrated than the RGB sample.  

\section{The Velocity Distribution of M33's Stellar Disk}\label{sec:models}

We performed an initial analysis and characterization of the kinematical distribution of stars in M33's stellar disk using the simplifying assumption that the stars are rotating 
in the same thin plane as the \hi\ 
disk. 
In this section, we present the observed kinematics of the young star and RGB star spectroscopic samples with the results of a single component fit to both samples (Section~\ref{sec:one_comp_fits}), and the results of a two-component fit to the full RGB sample (Section~\ref{sec:two_comp_fits}).  We then divide the RGB sample into radial and spatial subsamples and compare the results of fitting our two-component model to these subsamples (Section~\ref{sec:models_bins}). Finally, we discuss the limitations of the current model and prospects for improvement (Section~\ref{sec:model_limitations}).   

\subsection{Velocity Distributions of the Young and RGB Star Samples Compared to a One Component Model}\label{sec:one_comp_fits}

\begin{figure*}[tbh]
\includegraphics[width=1.0\textwidth]{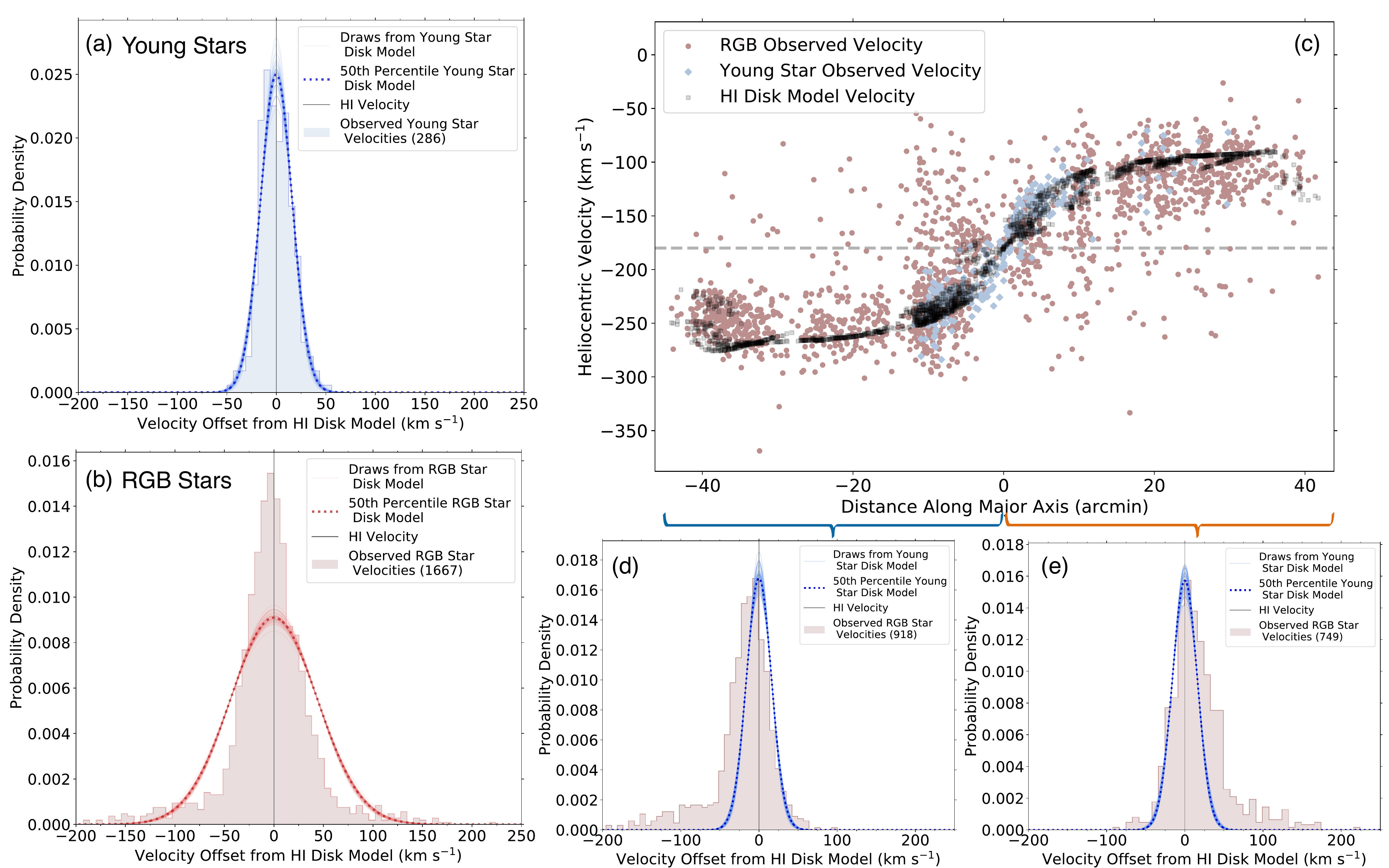}
\caption{
Distributions of the offset of the line-of-sight stellar velocities from the model \hi\ disk velocities (Section~\ref{sec:models}, \voff) for the (\textit{a}) young star comparison sample and (\textit{b}) RGB stars.  Significant tails to large absolute values of \voff\ are seen in the RGB sample, but not in the young star comparison sample.  The best-fit stellar disk model to each population 
is shown as a thick curve, with fifty random draws from the MCMC chains shown as thin curves.  A population rotating with the \hi\ disk will be well represented by a single Gaussian in \voff\ centered at $\sim 0$\kms.   
A one-component disk model provides a poor fit to the RGB sample, but a reasonable fit to the young star comparison sample. (\textit{c}) The line-of-sight velocities of the RGB and young stars as a function of distance along the semi-major axis, compared to the line-of-sight velocities of the \hi\ model computed at the location of every star. M33's systemic velocity is shown as a dashed line. 
The tail of the RGB velocity distribution extends asymmetrically to less negative velocities than the disk on the approaching (northeast) side of M33 (\textit{d}), and to more negative velocities on the receding (southwest) side (\textit{e}).  This is a natural consequence of a component that rotates more slowly than the \hi\ disk, or does not rotate with the \hi\ disk. A comparison of the RGB velocity distribution in the northeast and southwest halves of M33's disk with the model fit to the \voff\ distribution of the young stars, scaled to match the peak of the RGB star probability distribution, shows that the RGB sample contains a population consistent with rotating with the young stars and \hi\ disk in addition to the asymmetric tails.  
}
\label{fig:velhist_wmodel_diskonly}
\end{figure*}

Figure~\ref{fig:velhist_wmodel_diskonly} shows the distributions of the velocities of the RGB and young star samples, relative to the velocity of M33's \hi\ disk.  For each star $j$, we computed the difference between the line-of-sight velocity of the star and the line-of-sight velocity of the \hi, computed at the sky location of star $j$ using the tilted ring model by \citet{kam2017}:   
\begin{equation}
    v_{{\rm offset},j} = v_{{\rm H\,\textsc{i}}, j} - v_{{\rm los},j}.
\end{equation}

The line-of-sight velocity of the \hi\ disk model is computed via 
\begin{equation}\label{eqn:hidiskvel}
v_{{\rm H\,\textsc{i}}, j} =  v_{\rm sys} + v_{{\rm H\,\textsc{i}\ rot},j}*\cos(\theta_{j})\sin(i_{j})
\end{equation}
where \vsys\ is the systemic velocity of M33 ($-180$~\kms), $v_{{\rm H\,\textsc{i}\ rot},j}$ is the rotation speed of the \hi\ disk at 
the deprojected angular distance of star $j$ from M33's center in the plane of the disk (\rdisk),
$\theta_j$ is the azimuthal angle of star $j$ computed in the plane of the \hi\ disk, and $i_j$ is the inclination of the \hi\ disk at 
the \rdisk\ of star $j$.

We interpolated within the \citeauthor{kam2017} tilted ring model (their Table~4), assuming a series of infinitesimally thin rings, each with its own inclination, position angle, and rotation speed.  
We calculated $\theta_j$ and assigned an \hi\ disk inclination $i_j$ and \hi\ rotation speed $v_{\rm H\,\textsc{i}\ rot,j}$  based on star $j$'s 
\rdisk. The azimuthal angle $\theta_j$ is computed from the position angle (PA$_j$) and \hi\ disk inclination ($i_j$) at star $j$'s location: 
\begin{equation}
\theta_{j} = \frac{\beta}{\alpha\cos(i_{j})}
\end{equation}
\begin{equation}
\alpha = \eta \cos(\rm PA_{\it j}) + \xi \sin(\rm PA_{\it j})
\end{equation}
\begin{equation}
\beta = \xi \cos(\rm PA_{\it j}) - \eta \sin(\rm PA_{\it j})
\end{equation}
where $\xi$ and $\eta$ are the M33-centered sky coordinates of star $j$.

We computed an initial estimate of \rdisk\ for each star based on global values for M33's disk inclination and PA, and then revised the \rdisk\ estimate using the inclination and PA values from \citeauthor{kam2017}'s Table 4 for the star's initial \rdisk\ estimate. For a small fraction of stars, further iteration was required: we iterated until the change in inclination and PA from the previous and current estimates of \rdisk\ converged, with convergence defined as a change in $i$ of less than 0.01 degrees, and a change in PA of less than 0.35 degrees (each corresponding to 0.1 times the typical resolution of changes in $i$ and PA in \citeauthor{kam2017}'s Table 4).  The converged \rdisk\ estimate, and corresponding \hi\ model values for $i_j$ and PA$_j$, were used to compute \vmodel, and subsequently \voff, for each star $j$.  The above formalism assumes circular orbits.

A stellar population that rotates with the gas disk 
will have a \voff\ distribution centered at the velocity of the gas disk (\voff\,$=0$~\kms). 
Both the young stars and the RGB stars show a relatively narrow distribution centered near \voff\,$=0$~\kms\ (Figure~\ref{fig:velhist_wmodel_diskonly}).  In contrast with the \voff\ distribution of young stars, the RGB distribution also shows significant tails to large positive and negative values of \voff\@. 

We first evaluated the hypothesis that the stellar velocities can be modelled with a single rotating disk.  We fit a simple single component disk model to each sample, assuming a Gaussian distribution in \voff\@ centered at \voff\,$=0$~\kms, with a velocity dispersion of \sigmavoff. 
In other words, the stars are assumed to rotate in the plane of the \hi\ disk with the same rotation speed as the \hi\ disk, with
no mean offset from the \hi\ line-of-sight disk velocity at any given location within the disk ($\langle$\voff$\rangle=0$~\kms). 
The likelihood of the one-component model for an individual star is thus given by
\begin{equation}
\mathscr{L}_{j} = \mathscr{N}(v_{{\rm offset},j} | 0, \sigma_{\rm v,offset}) 
\label{eq:onecomp}
\end{equation}
where $\mathscr{N}(v_j | \mu, \sigma)$ denotes a normalized Gaussian with mean $\mu$ and dispersion $\sigma$, evaluated at the velocity of star $j$ ($v_j$) in the specified frame of reference (e.g., \voff\ in Equation~\ref{eq:onecomp}).

We used a Markov-Chain Monte Carlo (MCMC) implementation \citep[{\tt emcee} version 3.0.2;][]{foreman-mackey_emcee, emcee-sw} to converge on the best-fit velocity dispersion, assuming a scale free prior ($1/\sigma$) for \sigmavoff.
All model parameter values quoted in this work are the 50th percentile of the marginalized 1-dimensional posterior probability distribution function, derived from the MCMC chains, and the quoted uncertainties are the 16th and 84th percentiles.  
The details of the MCMC implementation are included in Appendix~\ref{app:mcmc_implementation}.

The \voff\, distribution of young stars in M33's disk appears well characterized by a single disk component, with a 
velocity dispersion of  \sigmav$=$\sigvyoungdiskonly~\kms, 
resulting in an estimated intrinsic velocity dispersion of \sigvyoungdiskonlyintr\kms, as shown in Figure~\ref{fig:velhist_wmodel_diskonly}(a). 

In contrast, the distribution of RGB stars is not well characterized by a single component.  A single Gaussian fit to the RGB population results in an estimated dispersion of \sigdisk$=$\sigvRGBdiskonly, which simultaneously considerably overestimates the observed spread in velocities about the \hi\ disk velocity for the majority of the stars, while underestimating the spread in \voff\ of the tails of the distribution (Figure~\ref{fig:velhist_wmodel_diskonly}(b)).  When split into the receding (southwestern) and approaching (northeastern) halves of M33's disk, the RGB sample shows strong asymmetry in the \voff\ distributions of the high dispersion component (Figure~\ref{fig:velhist_wmodel_diskonly}, panels c\,--\,e).\footnote{The sign convention used for \voff\ 
has been adopted for consistency with the typical convention used when measuring the asymmetric drift of stars as their lag in rotation speed relative to $v_{\rm circ}$, the circular velocity of the dark matter halo, or $v_{\rm gas}$, where the rotation speed of the gas is being used as a proxy for $v_{\rm circ}$. We note that the impact of using this convention in M33 is that a lag in the rotation velocities of the stars relative to the \hi\ model will result in a negative line-of-sight velocity offset (\voff) in the northeastern, approaching side of M33's disk, and a positive \voff\ in the southwestern, receding side of M33's disk.}  This asymmetry demonstrates that the high dispersion component is likely due to a population that rotates either significantly more slowly than, or does not rotate with, the \hi\ disk, since a component rotating with the \hi\ disk would have a velocity distribution symmetric about the \hi\ model velocity regardless of whether it was observed on the receding or approaching side of the disk.  Figure~\ref{fig:velhist_wmodel_diskonly} also shows that the majority of the RGB stars are well described by a low-dispersion component rotating with a speed comparable to that of the \hi\ disk, although a slight asymmetry in the peak of the \voff\ distribution is also apparent in the lower dispersion component (Figure~\ref{fig:velhist_wmodel_diskonly}, panels d and e).

\subsection{Results of Fitting a Two Component Model to the RGB Sample}\label{sec:two_comp_fits}

\begin{figure*}[tbh]
\includegraphics[width=0.49\textwidth]{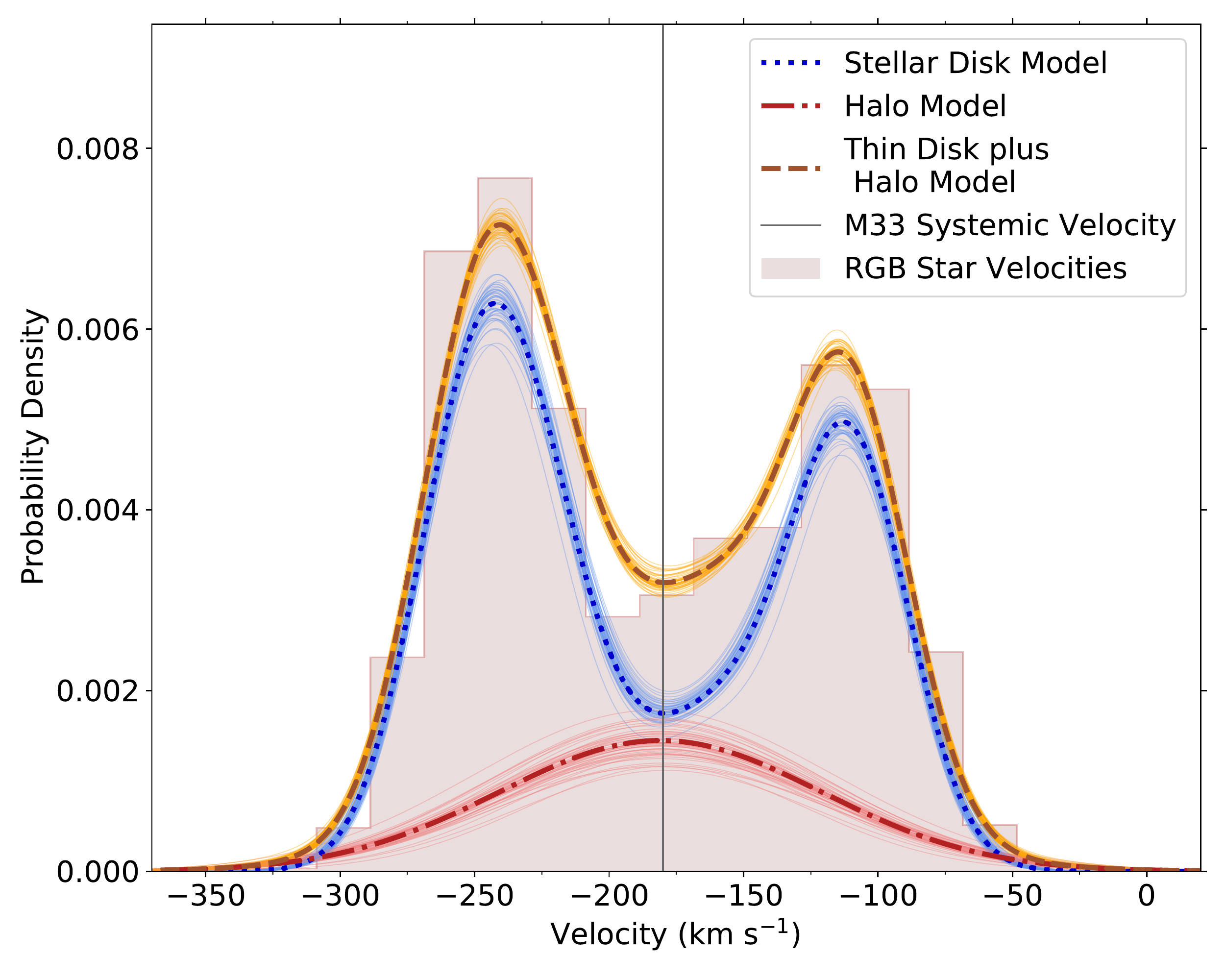}
\includegraphics[width=0.49\textwidth]{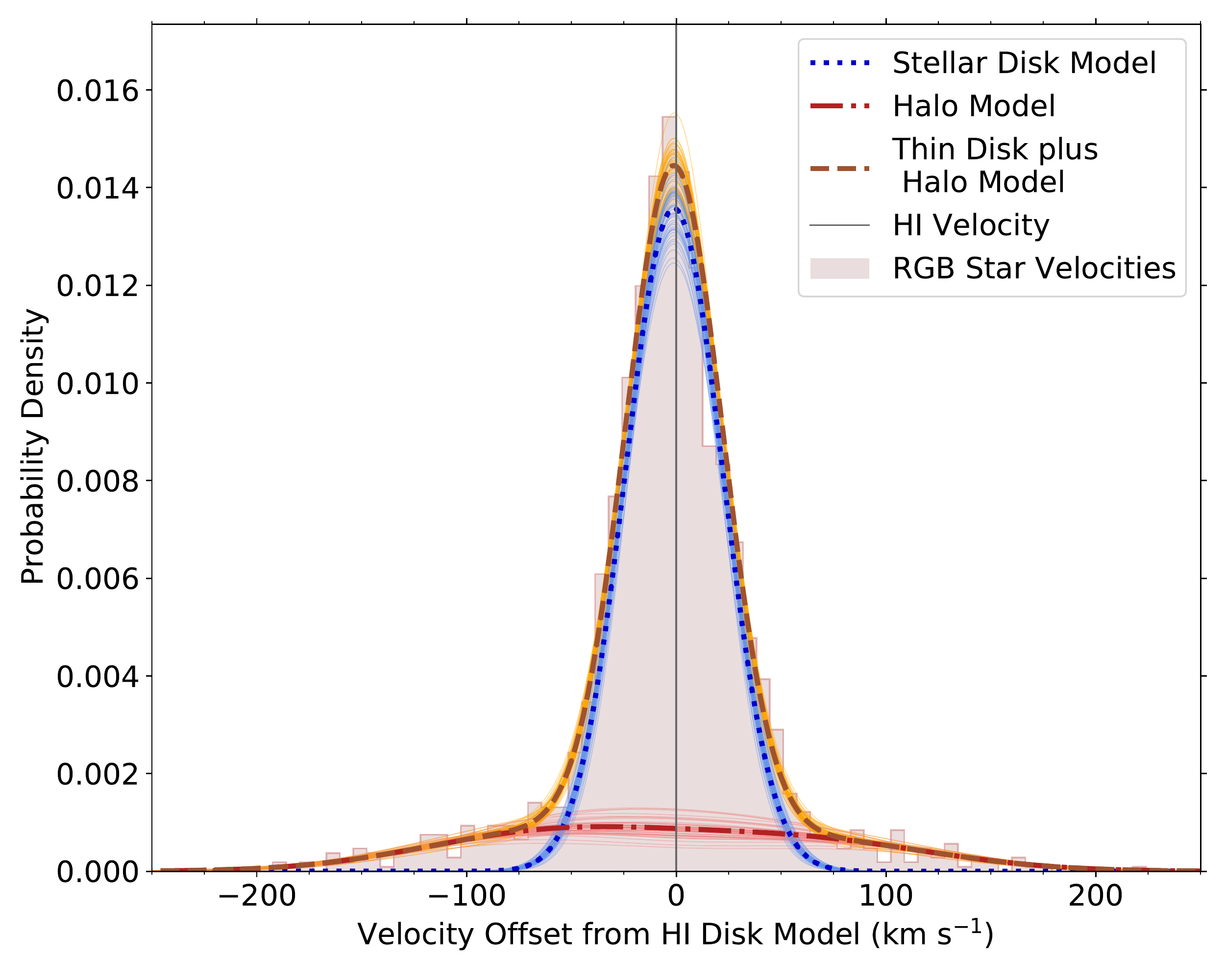}
\caption{Statistical visualization of the results of the two-component model fit to the full RGB sample (Section~\ref{sec:models}, Appendix~\ref{sec:app_velmodel}), compared to the observed line-of-sight velocities (left panel) and the velocity offset of each RGB star from the \hi\ disk model (right panel).
Results from the MCMC chains are shown as in Figure~\ref{fig:velhist_wmodel_diskonly} (details of the computation of the model visualization are provided in Appendix~\ref{sec:app_velmodel_viz}). 
The second model component has a large velocity dispersion (\sigmav$=$\plusthindisklagmodelhalosigv~\kms) and rotates very slowly in the plane of the disk (\frothalo$=$\plusthindisklagmodelfrothalo), which favors interpreting it as a stellar halo rather than a classical thick disk population.  The high velocity dispersion component comprises a significant fraction of the RGB sample (\fhalo$=$ \plusthindisklagmodelfhalo) over deprojected disk radii of \minrdiskRGBNarcmin\arcmin\ to \maxrdiskRGBNarcmin\arcmin\ ($\sim 0.9$ to 18.1~kpc).
}
\label{fig:velhist_wmodel_halo}
\end{figure*}

Given the obvious failure of a single disk model, we explored the nature of the RGB distribution using a two-component model.
We retained the simplifying assumption that any stellar components of M33 remain tightly coupled to the \hi\ disk, assuming the same kinematic center, as well as the same position angle and inclination of the stellar components with deprojected disk radius, as derived from the observed \hi\ velocity measurements \citep{kam2017}.
However, we relaxed the assumption that the stellar components had to be rotating at the same speed as the \hi\ disk ($\langle$\voff$\rangle=0$~\kms), which we enforced for the one-component model. This choice was motivated by the 
proposed internal disk heating mechanisms that may contribute to a kinematically hot component in low mass galaxies \citep{stinson2009,maxwell2012,el-badry2016}, combined with the observed asymmetry in the \voff\ distributions between the two halves of M33's disk (Figure~\ref{fig:velhist_wmodel_diskonly}, panels c\,--\,e).
We thus modelled both stellar components as rotating in the plane of the \hi\ disk, each with a rotation speed characterized as a fraction of the rotation speed of the \hi\ disk model (\frot)\footnote{We approximated the rotation speed of the two RGB components as a fractional speed of the \hi\ disk, rather than employing a single asymmetric drift value for each component, to account for the fact that our RGB sample spans a large range of radii, from \rdisk$=$\minrdiskRGBNarcmin\arcmin\ to \rdisk$=$\maxrdiskRGBNarcmin\arcmin. Over this radial range the rotation curve of M33's \hi\ disk increases from $\sim 58$ to $125$~\kms\ \citep{kam2017}.}. 

For a star $j$, we compute a modified line-of-sight model disk velocity for each of the two model components, by adding the parameter \frot\ to Equation \ref{eqn:hidiskvel}:

\begin{equation}
v_{{\rm model\ los,} j} =  v_{\rm sys} + f_{\rm rot}*v_{{\rm H\,\textsc{i}\ rot},j}*\cos(\theta_{j})\sin(i_{j})
\label{eq:vmodel_los}
\end{equation}

From this, we define a modified \voff\ as
\begin{equation}
    v^\prime_{{\rm offset}, j} = v_{{\rm model\ los,} j} - v_{{\rm los,} j}
\label{eq:vprime_offset}
\end{equation} 
for each of the two model components. 

We assume that at any given location in M33's disk, the stellar line-of-sight velocities of each model component are distributed as a Gaussian centered at the computed model component's line-of-sight velocity ($v^\prime_{{\rm offset}, j}$ = 0) with width $\sigma$.  
This is analogous to the one-component model, which assumed a Gaussian distribution centered at the velocity of the \hi\ disk model, \voff$=0$~\kms. 
We fit for the model parameters \frot\ and $\sigma$ for each of the two components, as well as the fraction of stars in the second, asymmetric, high dispersion component.

Thus, the individual likelihoods for the two-component model, with each component rotating at some fraction of the speed of the \hi\ disk, 
with velocity dispersion $\sigma$, are given by

\begin{equation}
\begin{split}
\mathscr{L}_{j} = (1.-f_{\rm halo})\mathscr{N}(v^\prime_{{\rm disk\ offset}, j}(f_{\rm rot, disk}) | 0, \sigma_{\rm disk}) \\
 + f_{\rm halo}\mathscr{N}(v^{\prime\prime}_{{\rm halo\ offset}, j}(f_{\rm rot, halo}) | 0, \sigma_{\rm halo})
\end{split}
\label{eq:twocomps_rotating}
\end{equation}

where \fhalo\ denotes the fraction of the population present in the second, high dispersion component, and the computed velocity offsets for star $j$ are functions of the model parameters \frotdisk\ and \frothalo, the sky location of star $j$, and the \citet{kam2017} \hi\ disk model, per Equations~\ref{eq:vmodel_los} and \ref{eq:vprime_offset}. While this likelihood is formulated in a manner very similar to that of a Gaussian Mixture Model (GMM), where \fhalo\ would be the mixing fraction \citep{mclachlan2000,f-s2019,bouveyron2019}, it is critical to note the difference between Equation~\ref{eq:twocomps_rotating} and a standard GMM.  Because each component has its own \frot\ model parameter, this means that the two Gaussian components in the likelihood function for the two-component model are each evaluated in \textit{different} velocity reference frames. For a given star $j$, unless the model parameters \frotdisk\ and \frothalo\ happen to be identical, $v^\prime_{{\rm disk\ offset}, j}$ is \textit{not} equal to $v^{\prime\prime}_{{\rm halo\ offset}, j}$.  Equivalently, at a given star $j$'s sky position, the line-of-sight velocity that corresponds to an offset from the model velocity of $v^\prime_{{\rm offset}, j} = v_{{\rm model\ los,} j} - v_{{\rm los,} j} = 0$~\kms\ will be different for the disk and halo components. Thus, we do not implement standard GMM techniques, but instead implement an MCMC algorithm to determine the model parameters and their uncertainties (Appendix~\ref{app:mcmc_implementation}).

Initial fits were run with priors on \frot\ that only allowed values from 0 to 1.  These fits resulted in marginalized 1d posterior probability distributions that were highly skewed to 1 for the first component (\frotdisk) and to 0 for the second component (\frothalo).  For the fits presented here, we relaxed the priors on \frot, allowing the model to explore whether, in the plane of the \hi\ disk, the kinematically hot component is favored to rotate at speeds comparable to that of the \hi\ disk, to rotate slowly or not at all, or even to counter-rotate (by allowing negative values of \frot). The model parameter \frothalo\ therefore has the potential to differentiate whether the high dispersion component looks more like a classical stellar halo (non-rotating or slowly rotating in the disk plane) or thick disk (rotating slower than the thin stellar disk, but still with significant rotation in the disk plane). Full details of the priors implemented in the MCMC algorithm are provided in Appendix~\ref{app:mcmc_implementation}. 

The results of fitting the two-component model to the full RGB spectroscopic sample are shown in Figure~\ref{fig:velhist_wmodel_halo} (details of how we produce visualizations of the model in \vlos\ and \voff\ space are provided in Appendix~\ref{sec:app_velmodel_viz}).  The model appears to reasonably describe the aggregate properties of M33's RGB population.  The primary disk component has a velocity dispersion similar to that of the young star sample (\sigdisk$=$\plusthindisklagmodeldisksigv~\kms; with an estimated intrinsic dispersion of \plusthindisklagmodeldisksigvintr~\kms), and rotates at a high fraction of the speed of the \hi\ disk (\frotdisk$=$\plusthindisklagmodelfrotdisk).\footnote{For stars with the smallest values of \rdisk\ in our spectroscopic sample, this \frotdisk\ translates to a rotation speed only $\sim 5$~\kms\ slower than the \hi\ disk, increasing to a difference in rotation speed of $\sim 15$~\kms\ for stars with \rdisk\,$\sim 40'$.  These differences in rotation speed translate to a smaller difference in the line-of-sight velocity (or equivalently, \voff) due to geometric effects: we observe only a fraction of the rotation velocity vector for a given star, set by the inclination of M33's disk at the location of the star and the position angle of the star in the plane of the disk (Appendix~B).  Therefore, the location of the resulting apparent mean peak in the data and the model representation shown in Figure~4 (as well as later figures depicting subsamples of the data) depends strongly on the distribution on the sky of the stellar sample.}  In contrast, the second component has a significantly larger velocity dispersion (\sighalo$=$\plusthindisklagmodelhalosigv~\kms; with an estimated intrinsic dispersion of \plusthindisklagmodelhalosigvintr~\kms) and rotates very slowly compared to the \hi\ disk (\frothalo$=$\plusthindisklagmodelfrothalo; consistent with no rotation at the $<1.5\sigma$ level).  The high velocity dispersion component comprises \plusthindisklagmodelfhalo\ of the spectroscopic RGB sample.             

Using simple scaling relations and assuming a stellar mass ratio of 10:1 between M31 and M33 and a halo dispersion of 145\kms\ for M31, \citet{mcconnachie2006m33} argued that $\sim 50$~\kms\ would be the expected stellar halo velocity dispersion for a galaxy of M33's mass.  Using the scaling relations for dispersion supported systems found by \citet{zahid2018} in the Illustris cosmological hydrodynamical simulations, and assuming recent literature estimates for the virial mass of M31 \citep[$\sim 1.4\times10^{12}$;][]{watkins2010,patel2017b} and M33 \citep[$1.3$\,--\,$2.1\times10^{11}$\Msun;][]{patel2017a}, the predicted line-of-sight stellar velocity dispersion within the half light radius of the M31 and M33 halos are $\sim 100$~\kms\ \citep[consistent with the results of][]{gilbert2018} and $\sim 50$~\kms, respectively.  
The lack of significant rotation of the high velocity dispersion component in the plane of the disk 
makes it unlikely to be reasonably classified as a classical thick disk. Furthermore, this second component is seen only in the older RGB sample and not in the young red star comparison sample.  We therefore refer to this second component as M33's halo throughout the remainder of the paper, following both observational and theoretical precedent (Section~\ref{sec:intro}).  However, as discussed below, the origin of the high velocity dispersion component remains undertermined.  If the origin of this component is (primarily) internal heating, rather than satellite accretion, any empirical distinction between a `halo' and `thick disk' is likely to be ill-defined \citep[e.g.,][]{dorman2013}.  

Finally, we note that we performed a Bayesian model comparison analysis on the full RGB sample in order to assess the evidence for more than one component in the RGB velocity distribution.  We found that while a one component model is strongly disfavored by the data, our data and present model do not provide strong evidence for favoring three components over two components.  Details of these statistical tests are provided in Appendix~\ref{app:models_number_comps}.

\subsection{Radial and Spatial Trends}\label{sec:models_bins}

\begin{figure*}[tbh]
\includegraphics[width=0.98\textwidth]{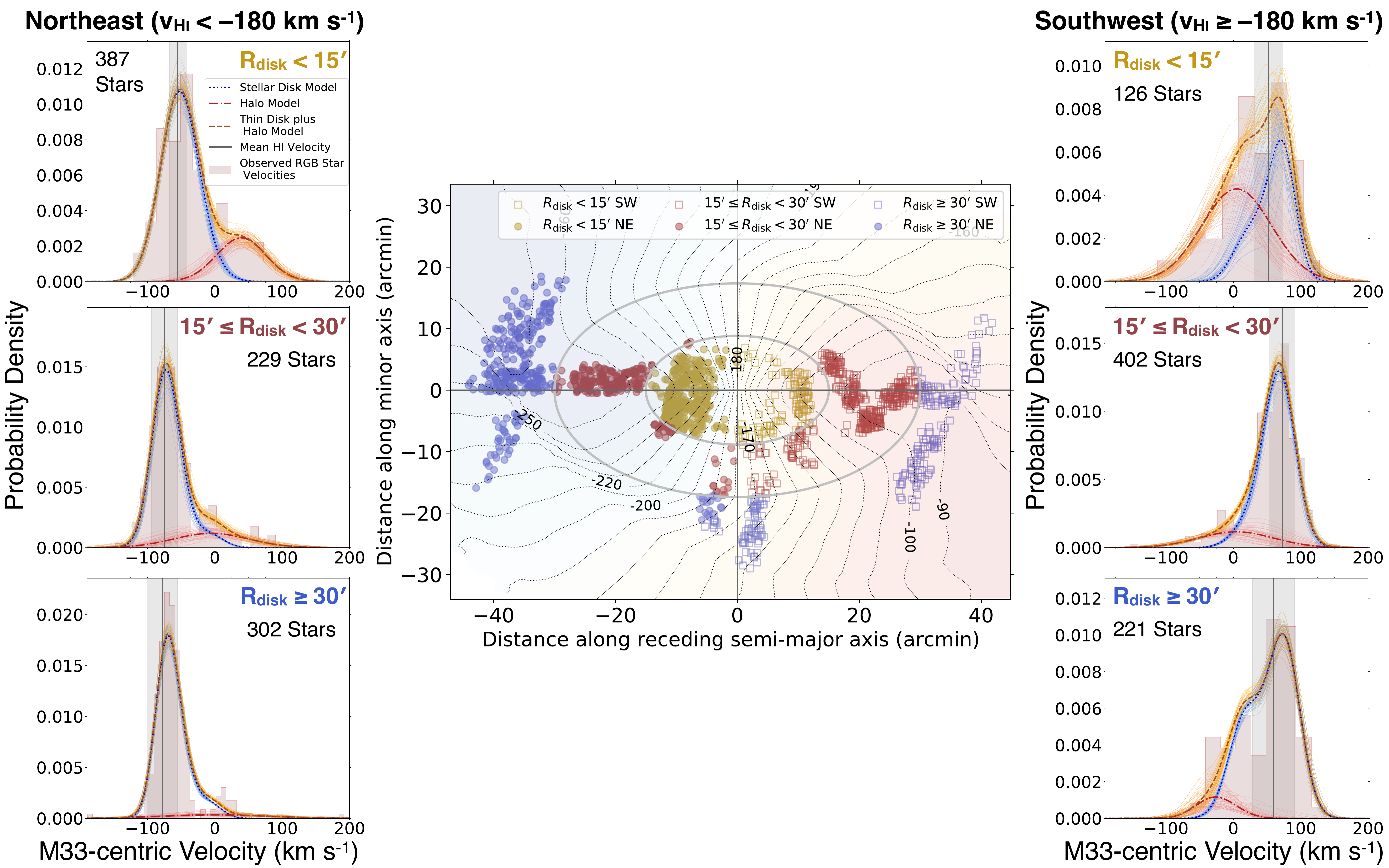}
\caption{ 
Line-of-sight velocity histograms centered at the systemic velocity of M33 ($v=$\vlos$-$\vsys), with fits to the observed velocities overlaid (left and right panels; fits displayed as in Figure~\ref{fig:velhist_wmodel_halo}), for subsets of the RGB sample (shown in center). The sample has been divided into three bins of projected radius in the plane of the disk, each of which is also split into two bins according to whether the star is in the approaching (northeastern) or receding (southwestern) halves of M33's disk, defined using the \hi\ model as described in Section~\ref{sec:models_bins}.  The \hi\ model is shown in the center panel, as in Figure~\ref{fig:roadmap}, with contour lines shown every 10~\kms. 
The mean and standard deviation of model \hi\ velocities, computed at the locations of the stars included in each subsample (e.g., Figure~\ref{fig:velhist_wmodel_diskonly}, panel c), are denoted in each velocity distribution panel by a solid grey line and shaded grey region, respectively.
The analysis of these six spatial bins reveals that the relative fraction of disk and halo changes with radius, with the halo component fraction greatest in the innermost bins and declining with radius.  A halo component is independently detected at significant fractions in all spatial bins interior to \rdisk$=$\rdiskcuttwo\arcmin. We discuss additional complexities in the dataset made evident by this spatial analysis in Section~\ref{sec:models_bins}.  
}
\label{fig:fits_by_spatialposition}
\end{figure*}

\begin{figure}[tbh!]
\includegraphics[width=0.385\textwidth]{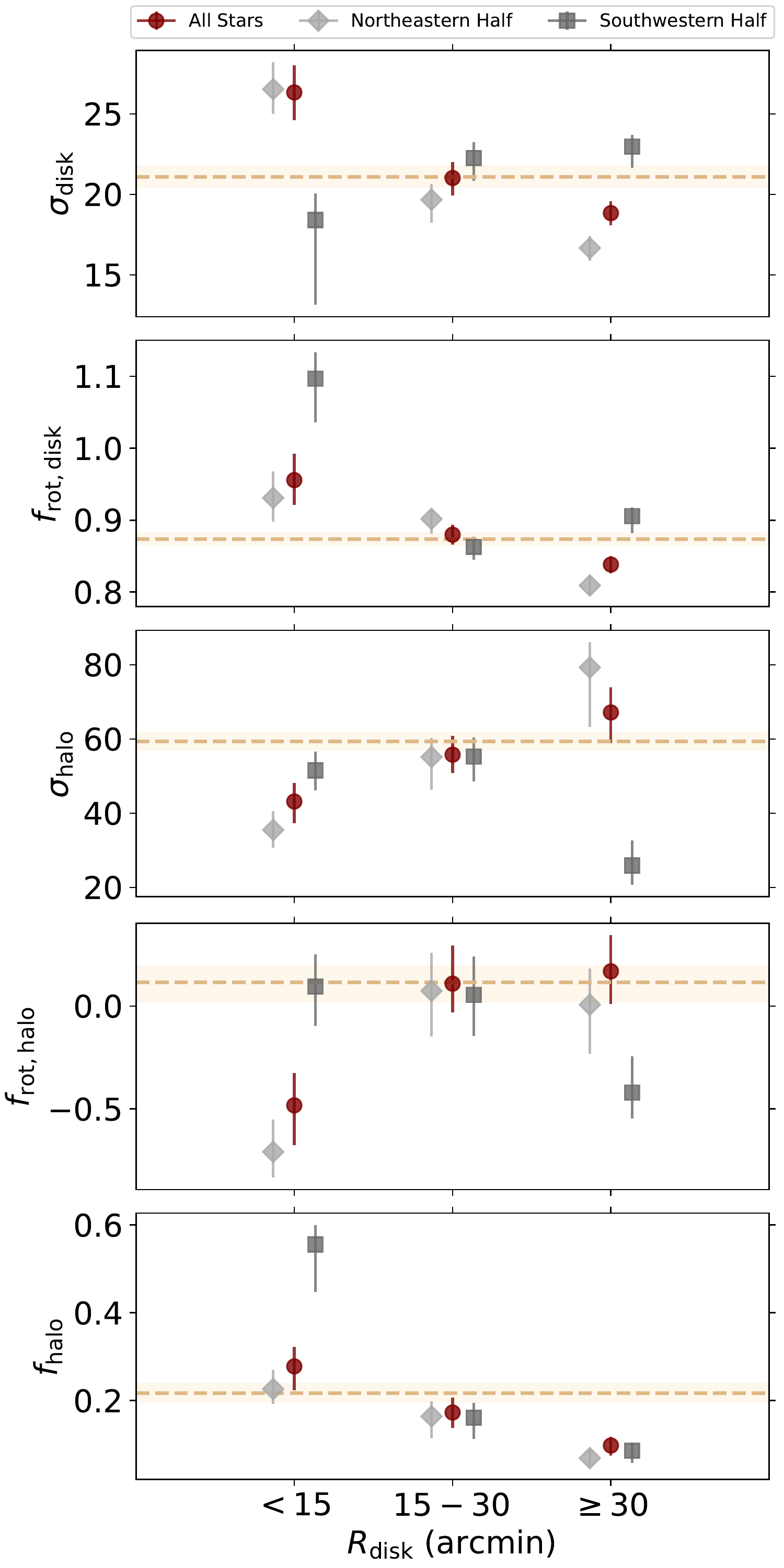}
\caption{Results of the independent two-component model fits to the subsamples shown in Figure~\ref{fig:fits_by_spatialposition}. 
Points denote the 50th percentile of the marginalized 1-dimensional posterior probability distributions, with the 16th and 84th percentiles denoted by the error bars.  The dashed line (shading) in each panel shows the 50th percentile (16th/84th percentile) of each parameter for the fit to the full RGB star sample (Figure~\ref{fig:velhist_wmodel_halo}). In each radial bin, points have been arbitrarily positioned in the abscissa.  The fits to all stars in each radial bin (red circles) indicate trends with \rdisk\ in many of the model parameters, most notably a decreasing halo fraction with increasing radius (bottom panel).  Statistically significant differences in model parameters derived from the separate fits to the northeastern (approaching) and southwestern (receding) halves of M33's disk indicate the presence of additional complexities within the M33 stellar populations in the innermost and outermost radial bins (Section~\ref{sec:models_bins}).  While this may indicate true differences in the stellar populations between the northeast and southwest halves, several limitations of the current model will need addressed (Section~\ref{sec:model_limitations}) before the physical nature of these differences can be meaningfully explored.
}
\label{fig:params_by_spatialposition}
\end{figure}

We performed an initial investigation into the presence of trends with radius in the M33 RGB populations by splitting the RGB sample into three radial bins:  \rdisk\,$<$\,\rdiskcut\arcmin, \rdiskcut\arcmin\,$\le$\,\rdisk\,$<$\,\rdiskcuttwo\arcmin, and \rdisk\,$\ge$\,\rdiskcuttwo\arcmin\ (Figure~\ref{fig:fits_by_spatialposition}).  
The choice of these bins was motivated in part by the properties of M33's disk, and in part by the spatial sampling of our spectroscopic survey.  The warp in the \hi\ model (see contours in Figure~\ref{fig:roadmap}) begins at \rdisk$=$\rdiskcuttwo\arcmin. This is slightly interior to the radius where the break in the disk surface brightness profile is observed \citep[$\sim 36$\arcmin;][]{ferguson2007}. Thus, \rdisk$=$\rdiskcuttwo\arcmin\ provides a fairly natural break-point for discussions of the outer versus inner disk of M33, while also denoting the radius at which one might reasonably expect larger deviations between the \hi\ and the stellar disk.  Dividing the inner disk sample at \rdisk$=$\rdiskcut\arcmin\ divides this region into half (by radius), and approximately identifies stars interior and exterior to the maximum deprojected disk radius covered by the PHATTER survey, enabling the kinematical analysis presented here to provide future context for interpreting results from the PHATTER survey.  These divisions result in comparable numbers of RGB stars in each radial bin (\noRGBstarsintPHAT, \noRGBstarsextPHAT\ and \noRGBstarsextbreak\ in the inner, middle, and outer radial bins, respectively).  

In addition, to test for consistency of the data and robustness of the model, we further subdivided the radial bins, fitting the approaching (northeastern) and receding (southwestern) halves of M33's disk separately.  Given that our data extend beyond the disk warp (Figure~\ref{fig:roadmap}), we do not simply use the location of M33's minor axis to subdivide the radial bin samples. The sub-samples are instead defined using the \hi\ disk model, with M33's systemic velocity as the boundary: if \vmodel\,$<$\vsys\ at the star's location, the star is placed in the approaching (northeastern) sub-sample, and if \vmodel\,$\ge$\vsys\ it is placed in the receding (southwestern) sub-sample.  In practice, this definition impacts the spatial bin of only 9 stars located at large minor axis distances (Figure~\ref{fig:fits_by_spatialposition}). 

We performed independent fits to the sub-samples of stars in each spatial bin (all stars within a radial bin, as well as stars in the approaching and receding halves of that radial bin separately) using the two-parameter model described above.  The results indicate that there is variation in the properties of the halo component over the range of radii covered by the spectroscopic data (Figures~\ref{fig:fits_by_spatialposition} and ~\ref{fig:params_by_spatialposition}). 

Some general overall radial trends are apparent in the fits to the samples of all stars within each of the three radial bins (red points, Figure~\ref{fig:params_by_spatialposition}).  The disk component dispersion (\sigdisk), fractional disk rotation (\frotdisk), and halo component fraction (\fhalo) all decrease with radius, while the halo dispersion (\sighalo) increases with radius.  These trends are also reflected in the majority of the fits to the approaching and receding halves of M33's disk.  
There is no apparent trend in the fractional rotation of the halo component (\frothalo), and the parameter values for the majority (two-thirds) of the sub-sample fits are consistent with no rotation of the halo component in the plane of the disk.

The model fits to the three sub-samples in the middle radial bin (\rdiskcut\arcmin\,$\le$\,\rdisk\,$<$\,\rdiskcuttwo\arcmin) return fully consistent sets of best-fit parameter values (Figure~\ref{fig:params_by_spatialposition}) and produce model probability distribution functions that reproduce well the general shape of the velocity distribution of each sub-sample (Figure~\ref{fig:fits_by_spatialposition}, also see Figure~\ref{fig:app_fits_by_spatialposition} in Appendix~\ref{sec:app_velmodel}). 

The model fits to the innermost radial bin (\rdisk\,$<$\,\rdiskcut\arcmin) display significant divergence in several of the model parameters amongst the three sub-sample fits (Figure~\ref{fig:params_by_spatialposition}).\footnote{We note that in the innermost radial bin there are only \noRGBstarsintPHATS\ RGB stars in our spectroscopic sample in the southwestern (receding) half of the disk (compared to \noRGBstarsintPHATN\ RGB stars in the northeastern (approaching) half), so the model fit to the full inner radial bin is dominated by stars in the northeast.}  Most notably, the fit to the northeastern half of the inner radial bin favors strong counter-rotation of the halo component in the plane of the disk, while the fit to the southwestern half is consistent with no rotation.  The counter-rotation favored by the model in the fit to the northeastern inner subsample appears consistent with the velocity distribution of the data in this sub-sample, which shows a broad tail to less negative line-of-sight velocities, consistent with a large velocity dispersion component peaked at a value significantly greater than M33's systemic velocity of \vsys\triangsystemic~\kms\ (upper left panel, Figure~\ref{fig:fits_by_spatialposition}).  In contrast, the model returns a best-fit \frothalo\ fully consistent with no rotation for the southwestern half of the inner radial bin.  However, we note that there are significant uncertainties on several of the model parameters for the southwestern, inner subsample: future work will increase the size and spatial distribution of the RGB sample in this region, improving constraints on the kinematics and enabling a determination as to whether the discrepancies noted above are due to the current limited sample size in the southwest, or whether the distribution of stars in the southwest is not well described by the  model. 
While the possibility of strong counter-rotation of a high dispersion component in M33's innermost regions is intriguing, given the limitations of the current model (Section~\ref{sec:model_limitations}), as well as the current discrepancies between the model parameters returned by the separate fits to the northeastern and southwestern halves, further analysis will be required to determine the robustness of this result.    

The model fits to the three sub-samples in the outermost radial bin (\rdisk\,$\ge$\,\rdiskcuttwo\arcmin) also show significant divergence in several of the model parameters, including in \sigdisk\ and \frotdisk, as well as the halo parameters. While statistically significant, the difference in the halo component parameters between the receding and approaching sides should be interpreted with caution. The halo component fractions are quite low in the fits to the sub-samples in the outermost bin (\fhalo$\lesssim 0.1$), meaning that the halo component parameters are being determined based on (statistically) $\sim 20$ stars each in the northeast and southwestern halves of the disk. 
There is clearly a small population of stars in both the northeastern and southwestern halves of the disk that are well removed from the bulk of the stellar velocity distribution as well as the expected \hi\ disk velocity, with a significant velocity dispersion (Figure~\ref{fig:velhist_wmodel_diskonly}).  However, in the velocity distributions, there are also apparent clusters, each containing a small number of stars, which are well removed from the disk velocity and with similar line-of-sight velocities or \voff  (Figures~\ref{fig:fits_by_spatialposition} and \ref{fig:app_fits_by_spatialposition}).  A robust determination of whether the high dispersion component identified by the model in the outer radial bin actually represents a true continuation of the halo population detected at \rdisk$<$\rdiskcuttwo\arcmin, or is attempting to fit other substructures, will require additional data. Given these considerations, it is reasonable to consider the \fhalo\ determined for the outer bin subsamples as an approximate upper limit on the potential halo component at \rdisk$>$\rdiskcuttwo\arcmin\@.

Given the small halo fractions found in the fits to the outermost radial bin and its northeastern and southwestern sub-samples, we performed a final additional fit to the full RGB spectroscopic sample interior to \rdisk$<$\rdiskcuttwo\arcmin\ (\noRGBstarsintbreak\ stars).  We find results comparable to the fit to the entire RGB sample: a similar halo fraction (\fhalo$=$\plusthindisklagmodelfhalointbreak), a similarly large halo dispersion (\sighalo$=$\plusthindisklagmodelhalosigvintbreak~\kms) and a fractional halo rotation fully consistent with no rotation (\frothalo$=$\plusthindisklagmodelfrothalointbreak). 

The best-fit parameter results for all two-component model fits discussed above are summarized in Table~\ref{tab:fitparams}. 
Additional velocity distributions, with model results overlaid, for the radial and spatial subsamples, as well as 1- and 2-dimensional marginalized posterior probability distribution functions for the model parameters, are shown in Appendix~\ref{sec:app_velmodel} in Figures~\ref{fig:app_corner_all} to \ref{fig:app_corner_spatialbins}.  The 2-dimensional marginalized posterior probability distribution functions demonstrate that while the model parameters are generally well constrained by the fits, there are non-trivial correlations between some model parameters, as well as significant non-Gaussianities in some of the posterior distribution functions. 

\begin{deluxetable*}{lcrrrrrr}
\tablecaption{Two Component Model Fits to M33 RGB Stars\label{tab:fitparams}}
\tablehead{
%    \hline \hline
    & & & \multicolumn{2}{c}{Disk Parameters} & \multicolumn{3}{c}{Halo Parameters}\\
     \cmidrule(lr){4-5}\cmidrule(lr){6-8}
    & \multicolumn{1}{c}{$N_{\rm stars}$} & 
    \multicolumn{1}{c}{$\delta_v$\tablenotemark{a}} &
    \multicolumn{1}{c}{\sigdisk} & \multicolumn{1}{c}{\frotdisk} &  \multicolumn{1}{c}{\sighalo} & \multicolumn{1}{c}{\frothalo} & \multicolumn{1}{c}{\fhalo}\\ %\hline
    & & 
    \multicolumn{1}{c}{(\kms)} &
    \multicolumn{1}{c}{(\kms)} &  &  \multicolumn{1}{c}{(\kms)} 
    &  &
}
    %previously all were: & \multicolumn{6}{c}{Full Radial Range}\\ \hline
    % but I think left justified is easier to read
\startdata
%\\[1mm]
\sidehead{Full Radial Range} \hline
    All Stars    & \noRGBstars &
    \medianRGBveluncertainty &
    \plusthindisklagmodeldisksigv &  
    \plusthindisklagmodelfrotdisk &  \plusthindisklagmodelhalosigv & 
    \plusthindisklagmodelfrothalo &
    \plusthindisklagmodelfhalo\\ \tableline
\sidehead{\rdisk$<30'$} \hline
    All Stars & \noRGBstarsintbreak & 
    \rgbveluncintbreak &
    \plusthindisklagmodeldisksigvintbreak & 
    \plusthindisklagmodelfrotdiskintbreak &  \plusthindisklagmodelhalosigvintbreak & 
    \plusthindisklagmodelfrothalointbreak &
    \plusthindisklagmodelfhalointbreak \\ 
%    \\[0mm]
    \hline
\sidehead{\rdisk$<15'$} \hline
    All Stars & \noRGBstarsintPHAT & 
    \rgbveluncintPHAT &
    \plusthindisklagmodeldisksigvintPHAT & 
    \plusthindisklagmodelfrotdiskintPHAT &  \plusthindisklagmodelhalosigvintPHAT & 
    \plusthindisklagmodelfrothalointPHAT &
    \plusthindisklagmodelfhalointPHAT \\
     Northeast & \noRGBstarsintPHATN &
     \rgbveluncintPHATN &
     \plusthindisklagmodeldisksigvintPHATN & 
    \plusthindisklagmodelfrotdiskintPHATN &  \plusthindisklagmodelhalosigvintPHATN & 
    \plusthindisklagmodelfrothalointPHATN &
    \plusthindisklagmodelfhalointPHATN \\
    Southwest & \noRGBstarsintPHATS &
    \rgbveluncintPHATS &
    \plusthindisklagmodeldisksigvintPHATS & 
    \plusthindisklagmodelfrotdiskintPHATS &  \plusthindisklagmodelhalosigvintPHATS & 
    \plusthindisklagmodelfrothalointPHATS &
    \plusthindisklagmodelfhalointPHATS \\ \hline
\sidehead{$15'\leq$\rdisk$<30'$} \hline
    All Stars & \noRGBstarsextPHAT & 
    \rgbveluncextPHAT &
    \plusthindisklagmodeldisksigvextPHAT & 
    \plusthindisklagmodelfrotdiskextPHAT &  \plusthindisklagmodelhalosigvextPHAT & 
    \plusthindisklagmodelfrothaloextPHAT &
    \plusthindisklagmodelfhaloextPHAT \\
     Northeast & \noRGBstarsextPHATN & 
     \rgbveluncextPHATN &
    \plusthindisklagmodeldisksigvextPHATN & 
    \plusthindisklagmodelfrotdiskextPHATN &  \plusthindisklagmodelhalosigvextPHATN & 
    \plusthindisklagmodelfrothaloextPHATN &
    \plusthindisklagmodelfhaloextPHATN \\
    Southwest & \noRGBstarsextPHATS & 
    \rgbveluncextPHATS &
    \plusthindisklagmodeldisksigvextPHATS & 
    \plusthindisklagmodelfrotdiskextPHATS &  \plusthindisklagmodelhalosigvextPHATS & 
    \plusthindisklagmodelfrothaloextPHATS &
    \plusthindisklagmodelfhaloextPHATS \\ \hline
\sidehead{\rdisk$\geq 30'$} \hline
    All Stars & \noRGBstarsextbreak &
    \rgbveluncextbreak &
    \plusthindisklagmodeldisksigvextbreak & 
    \plusthindisklagmodelfrotdiskextbreak &  \plusthindisklagmodelhalosigvextbreak & 
    \plusthindisklagmodelfrothaloextbreak &
    \plusthindisklagmodelfhaloextbreak \\
     Northeast & \noRGBstarsextbreakN & 
     \rgbveluncextbreakN &
    \plusthindisklagmodeldisksigvextbreakN & 
    \plusthindisklagmodelfrotdiskextbreakN &  \plusthindisklagmodelhalosigvextbreakN & 
    \plusthindisklagmodelfrothaloextbreakN &
    \plusthindisklagmodelfhaloextbreakN \\
    Southwest & \noRGBstarsextbreakS &
    \rgbveluncextbreakS &
    \plusthindisklagmodeldisksigvextbreakS & 
    \plusthindisklagmodelfrotdiskextbreakS &  \plusthindisklagmodelhalosigvextbreakS & 
    \plusthindisklagmodelfrothaloextbreakS &
    \plusthindisklagmodelfhaloextbreakS \\
%    \hline
\enddata
\tablenotetext{a}{The median velocity uncertainty of the stars included in the fit (Section~\ref{sec:data}).}
\tablecomments{
%\item Note. \textemdash\ 
The 50th-percentile values (uncertainties are the 16th- and 84th-percentile values) from the marginalized one-dimensional posterior probability distribution functions for each parameter in the two component model (Section~\ref{sec:models}, Appendix~\ref{sec:app_velmodel}): disk and halo velocity dispersions (\sigdisk, \sighalo), the fraction of the \hi\ rotation velocity at which the disk and halo are rotating (\frotdisk, \frothalo; negative values of \frot\ correspond to counter-rotation in the plane of the \hi\ disk), and the fraction of the population in the halo component (\fhalo). 
%Uncertainties are the 16th and 84th-percentile values of the marginalized one-dimensional posterior probability distribution functions.  
Spatial divisions of the sample are made as shown in Figure~\ref{fig:fits_by_spatialposition} and described in Section~\ref{sec:models}, and are based on deprojected radius in the plane of M33's disk (\rdisk).  The median velocity uncertainty, $\delta_v$, of the different subsamples of stars differs primarily due to the varying proportions of stars observed with the 600 and 1200 l/mm gratings in each subsample, as well as variations in the observing conditions and target selection. %(some masks observed stars farther down the RGB luminosity function than others). 
}
\end{deluxetable*}

\subsection{Limitations of the Current Model}\label{sec:model_limitations}
Our model makes a number of simplifying assumptions, all of which may affect the best-fit values of the halo component parameters.  The tight coupling to the \hi\ disk assumed in the model\,---\, requiring that the RGB disk component has a velocity that is tied to the \hi\ model velocity\,---\,means we have assumed the \hi\ and RGB stars have the same position angle and inclination with radius (including the same disk warp at large radii). By tying our stellar models to the \hi\ model, we have also assumed the \hi\ and RGB stars have the same kinematical center and that the \hi\ and RGB rotation curves simultaneously rise and flatten with each other: e.g., that they have the same break radius, and the same behavior at large radii.\footnote{The spatially resolved fits provide some sensitivity to potential failures of this assumption by fitting \frot\ independently in each spatial bin.}  In addition, M33's disk population may be more complex than the simple one-component model assumed here.  Furthermore, while the model allows for a halo component that does not rotate, counter-rotates, or rotates with the disk at a fraction of the disk rotation speed, the stellar halo could rotate completely independently from the rotation of the disk.  

While our simplified model provides a reasonable first approximation of the overall stellar distribution in M33's disk (Figure~\ref{fig:velhist_wmodel_halo}), as discussed above there are indications in the spatially resolved fits (Figures~\ref{fig:fits_by_spatialposition} and \ref{fig:params_by_spatialposition}; Section~\ref{sec:models_bins}) of complexities in the RGB population that may not be fully captured by the current model.  Quirk et al.\ (in preparation) will measure and analyze the asymmetric drift of the M33 disk as a function of deprojected disk radius and of stellar age, including for the RGB stars analyzed here. Future work will explore relaxing the simplifying assumptions we have made here in studying the disk and halo populations of M33, including modelling the RGB population (disk and halo) independently of the gas disk, implementing more complex disk models (e.g., allowing multiple disk components), and testing the halo population for signs of rotation unconnected to the rotation of the disk.  

Finally, we note that the models considered here are formulated to constrain the major components of the velocity distribution in M33's disk, and are not sensitive to or designed for identifying or characterizing any minor substructures that may be present.  Future work, incorporating additional observations, will explore the evidence for or against the presence of substructure within the TREX spectroscopic survey.

\section{Photometric Metallicity Distribution of M33's Disk and Halo}\label{sec:fehphot}

\begin{figure}
\includegraphics[width=0.45\textwidth]{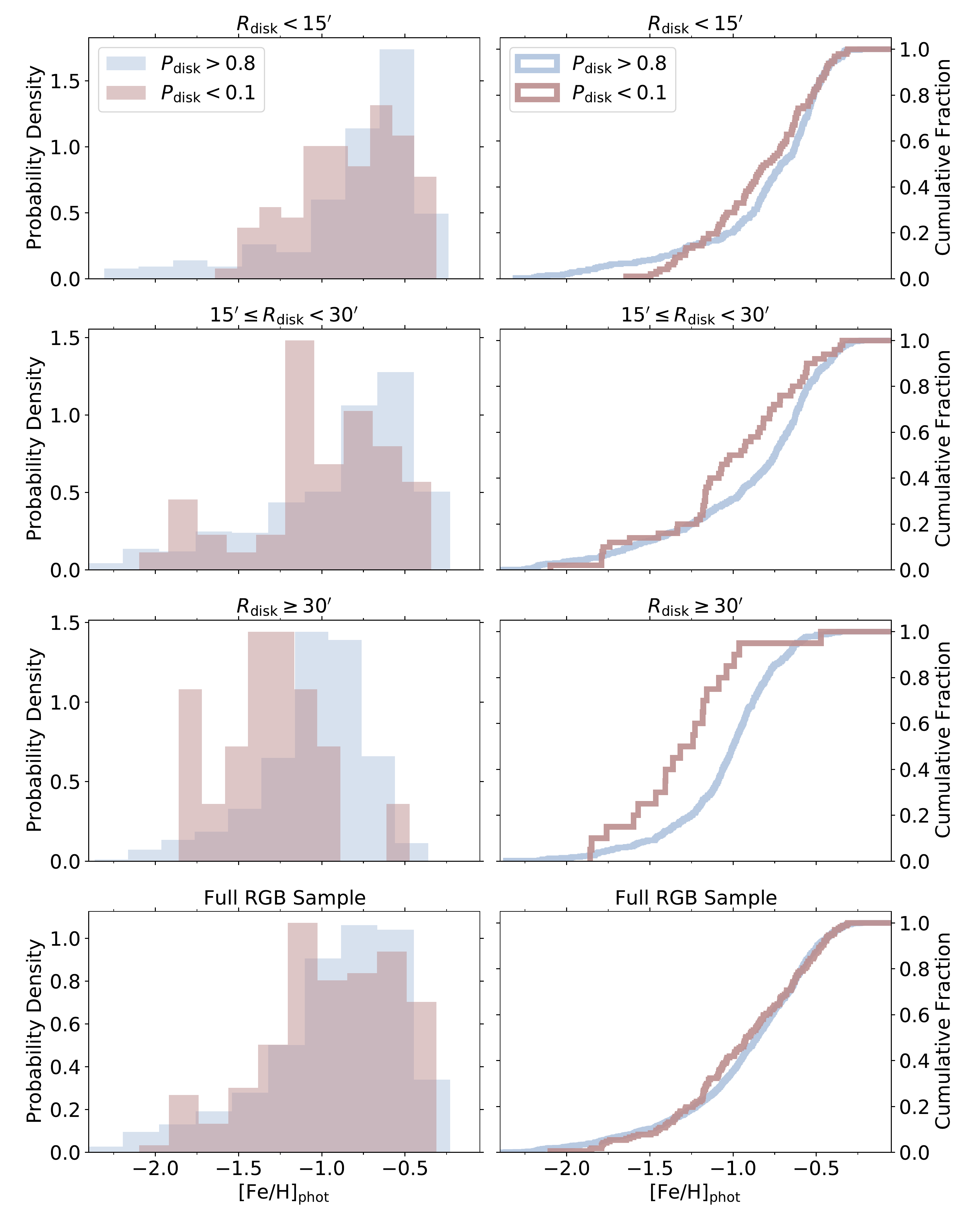}
\caption{Photometric [Fe/H] distributions in normal (left panels) and cumulative (right panels) form, for RGB stars with high probabilities of belonging to the disk ($P_{\rm disk}>0.8$) and halo ($P_{\rm disk}<0.1$) components, for stars in each of the three radial bins 
as well as the full sample. 
\fehp\ was computed by comparing a star's position in the CMD with theoretical stellar isochrones, using the Padova isochrone models \citep{marigo2017} as shown in Figure~\ref{fig:cmds}, assuming an age of 10~Gyr and \afe\,$=0$.  As a whole the \fehp\ distributions of the halo and disk are remarkably similar.  However, the data indicate that the difference between the distributions increases with increasing radius. 
}
\label{fig:fehphot}
\end{figure}

We investigated differences between the CMD distributions of RGB stars likely to belong to the disk and halo components by translating their CMD positions to photometric estimates of \feh\ using interpolation within a grid of theoretical stellar isochrones. The resulting \fehp\ distributions are shown in Figure~\ref{fig:fehphot}. We used the Padova isochrones \citep{marigo2017} with an age of 10~Gyr and solar $\alpha$-enhancement (\afe$=0.0$) to calculate \fehp. Stars were assigned as members of the disk (halo) component if their probability of membership in the disk component was \pdisk$>0.8$ (\pdisk$<0.1$).  The probability a given star belongs to the disk component was computed from the MCMC chains for the fit to the spatial subsample in which the star falls (Figure~\ref{fig:fits_by_spatialposition}, Section~\ref{sec:models_bins}).  The distribution of disk probabilities for the RGB spectroscopic sample is highly peaked at values \pdisk$>0.8$ and $<0.1$, so this conservative assignment includes 90\% of the RGB sample. 

The full RGB sample, as well as the inner radial bin, shows a remarkable similarity between the \fehp\ distributions of stars likely to belong to the disk and halo components. The difference between the \fehp\ distributions of stars likely to belong the disk and halo components increases with radius.  This appears due to a slightly greater variation with radius of the halo distribution than the disk distribution.  The differences between the halo and disk distributions are relatively small: the differences in the median \fehp\ of the disk and halo component are \deltafehpmedinner, \deltafehpmedmiddle, and \deltafehpmedouter\ in the inner, middle, and outer radial bins, respectively.  A 2-sided Kolmogorov-Smirnov test returns an increasing probability that the two distributions are drawn from a different parent distribution with radius: while the null hypothesis cannot be rejected for the inner radial bin, the p-values for the middle and outer radial bins are \twosidekstestmiddle\ and \twosidekstestouter, respectively.  

There are several limitations to this approach which should be kept in mind when interpreting the \fehp\ distributions.  First, only the average foreground (MW) extinction is applied to the observed photometry \citep{schlegel1998,schlafly2011}.  The amount of internal M33 extinction is likely to vary over the distances covered by our spectroscopic survey. Internal extinction along the line of sight not accounted for will result in an overestimate of a star's \feh.  Second, this does not account for changes in the mean age of the RGB population with distance from M33's center \citep{williams2009, barker2011}, or any potential differences between the mean ages (or $\alpha$-enhancements) of the disk and halo component populations; differences in age and \afe\ can impact the estimated \fehp\ by up to several tenths of a dex \citep[e.g.,][]{gilbert2014}. Finally, different radial bins contain targets selected from a differing mix of photometric sources (Section~\ref{sec:data}), drawn from different observatories and instruments, utilizing different measurement methodologies, and hence with different photometric random uncertainties and systematic biases. Since the impact of many of these limitations are expected to be a function of radius, comparisons of the disk (or halo) component's \fehp\ distributions with radius must be interpreted with caution.  However, these effects should be minimized for the comparisons between the disk and halo components within any given radial bin, as shown in Figure~\ref{fig:fehphot}, since it is reasonable to expect that stars highly probable to be in the disk or halo components within a given spatial region will be affected in approximately the same way for many of the above limitations.  

Given the above limitations, the comparison of the \fehp\ distributions of M33's disk and halo components within a given radial region is the most likely to be robust. Namely, while in the innermost regions there is no statistical difference between the \fehp\ distributions of stars with disk-like kinematics and halo-like kinematics, the difference in the \fehp\ distributions of these two components becomes larger with increasing radius, with statistically distinct \fehp\ distributions between the halo and disk components in the outermost bin.  Future work will use spectral synthesis techniques to alleviate the impact of many of the limitations noted above.

\section{Discussion and Conclusions}\label{sec:disc}

We have established the presence throughout M33's inner disk (\rdisk$<$\rdiskcuttwo\arcmin) of at least two distinct RGB populations, including a significant, kinematically hot component. The asymmetric tails of the velocity distributions of RGB stars in the northeastern and southwestern halves of M33's disk indicate it is unlikely the high velocity dispersion component is rotating with the \hi\ disk at any appreciable fraction of the rotation speed of the \hi. 

The RGB disk component, 
which is rotating at a significant fraction of the speed of the \hi\ disk (\frotdisk$=$\plusthindisklagmodelfrotdisk), 
has an intrinsic velocity dispersion of \sigdisk\plusthindisklagmodeldisksigvintr~\kms.  The second RGB population, which we interpret as a stellar halo, has a significantly higher velocity dispersion (\sighalo$=$\plusthindisklagmodelhalosigv~\kms; estimated intrinsic dispersion of \plusthindisklagmodelhalosigvintr~\kms) and rotates very slowly in the plane of the \hi\ disk (\frothalo$=$\plusthindisklagmodelfrothalo~\kms; consistent with no rotation at the $<1.5\sigma$ level). 
This second component has a dispersion consistent with that expected for a stellar halo in a galaxy of M33's mass.  

The stellar halo component comprises a significant fraction (\fhalo$=$\plusthindisklagmodelfhalo) of the full RGB spectroscopic sample, and a greater fraction (\fhalo$=$\plusthindisklagmodelfhalointPHAT) when only stars within \rdisk$<$\rdiskcut\arcmin\ are modelled.  The percentage of stars in the dynamically hot component within \rdisk$<$\rdiskcut\arcmin\ is of order $\sim 2-3$ times higher than theoretical expectations from the EAGLE simulations for the fraction of stellar mass formed ex-situ and residing within the stellar half-mass radius for a galaxy of M33's stellar mass \citep{davison2020}. However, we note that this should be viewed as an upper limit on the fraction of ex-situ stars in M33's central regions, as we do not know whether stars in the halo component were primarily accreted or formed in-situ, and furthermore we have measured only the fraction of stars in our RGB sample which belong to this component.  The \fhalo\ estimates resulting from dividing the sample into three radial bins indicate that the high velocity dispersion component comprises a \textit{decreasing} fraction of M33's RGB star population as radius increases, over the radial range covered by our spectroscopic sample.  This is also at odds with expectations for an accreted stellar halo from cosmological simulations, which show the fraction of stellar mass that is formed ex-situ (accreted) is expected to increase with radius \citep[e.g.,][for recent Illustris and EAGLE results, respectively]{rodriguez-gomez2016, davison2020}. 

The similarity in \fehp\ distributions between stars likely to belong to the disk and halo components, especially in the inner regions of M33's disk, may favor an in-situ origin for these stars, as stars from a significantly less massive satellite would be expected to have a significantly lower metallicity \citep{kirby2013}.  This is consistent with the interpretation by \citet{beasley2015} that the GCs they observe in M33's inner regions are most likely to have a disk origin: their GC sample has metallicities ranging from [M/H]$\sim -1.75$ to $\sim -0.5$ and a large velocity dispersion ($98.2\pm25.0 $~\kms) about the \hi\ disk velocity. Interestingly, recent simulations of M33's stellar and gaseous disk imply that strong stellar feedback is needed to match the observed disk properties \citep{dobbs2018}, and stellar feedback is one avenue by which in-situ stellar halos in low-mass galaxies are thought to form \citep{stinson2009, el-badry2016}.  Of course, the high velocity dispersion RGB population may represent a mix of stars formed in-situ and ex-situ, the relative fractions of which may change with radius.  Future abundance measurements using our spectroscopic RGB sample may shed further light on this question. 

Our kinematical analysis of M33's RGB population as a function of radius provides further insights into several unresolved discussions in the literature. The stellar age gradient in M33's disk is observed to reverse at large radii, changing from decreasing mean stellar age with radius interior to \rdisk$\sim 9$~kpc, to increasing mean stellar age with radius exterior to $\sim 9$~ kpc \citep{williams2009,barker2011}.  Using a simulated M33 analogue, \citet{mostoghiu2018} hypothesized that the age gradient reversal at large radii is due primarily to stellar accretion of smaller, ancient satellite galaxies, rather than radial migration of stars formed in-situ. \citeauthor{mostoghiu2018} predicted that this will be observable as a population of old stars with a large velocity dispersion, which should comprise a rapidly increasing fraction of the total stellar population beyond the radius of the disk break and age reversal; their simulations predict this fraction would start at $\gtrsim15$\% and increase to $\sim 30$\% over an equivalent radial range which encompasses the 16th to 84th percentile of stars (by radius) in our outermost radial bin (\rdisk\,$>$\rdiskcuttwo\arcmin; which has stars ranging from \rdisk$=$7.5 to 18~kpc).  The simulated M33 analogue has the accreted stellar fraction increasing from $\sim 30$\% to $\sim 75$\% in the equivalent radial range of the most distant stars in our sample.
In stark contrast, in the outermost radial bin the best-fit halo fraction is only \fhalo$=$\plusthindisklagmodelfhaloextbreak\ of the RGB stars, with the vast majority of the stars appearing to be consistent with a kinematically cold, rotating, disk-like component (Figures~\ref{fig:fits_by_spatialposition}, \ref{fig:app_fits_full_radial_bins}, and \ref{fig:app_fits_by_spatialposition}). Moreover, 
the evidence for a stellar halo component beyond the warp in the \hi\ disk should currently be viewed as inconclusive (Section~\ref{sec:models_bins}).
The current dataset thus favors the interpretation that the outer disk break marks a transition to a large extended disk \citep{grossi2011}, rather than to an accreted stellar halo component.  
The observed decrease in \fhalo\ with radius also suggests a possible explanation for the lack of detection of an extended stellar halo in M33 \citep{mcmonigal2016,galera-rosillo2018}: if the halo density profile is sufficiently steep, then it may be essentially undetectable beyond the stellar disk with current datasets. 

A weak, short bar has been observed in the innermost regions of M33's disk.  Previous measurements based on unresolved stellar light measured a bar with semi-major axis of $\sim 200$~pc \citep{ regan1994,corbelli2007}.  In 
the PHATTER survey, the bar is clearly visible in the resolved old (RGB star) and intermediate-age (AGB star) populations \citep{williams2021}, with an effective radius $\lesssim 500$~pc (Lazzarini et al. in prep).  Future papers on M33's recent SFH (Lazzarini et al., in prep) and M33's structure in resolved stellar populations (Smercina et al., in prep) will further quantify its properties. 
The innermost stars in our RGB sample are at a disk radius approximately twice that, so we do not directly observe the region of M33's bar in the current dataset.  However, we note that the development of a strong, more extensive bar, with semi-major axis of $\sim 2$\,--\,3~kpc, has proven difficult to avoid in simulations which try to recreate the properties of M33's disk \citep{sellwood2019}. \citeauthor{sellwood2019}\ note that while increasing the random motion of stars in the simulated disk or having a population of counter-rotating stars are two ways to increase stability against the formation of a bar, the existing observational kinematical studies \citep[based on measurements of the width of the \caii\ Triplet in unresolved spectra of M33's inner regions;][]{kormendy1993,corbelli2007}  did not support either.  The significant fraction of the RGB sample in the inner regions of M33's disk found to be in the high velocity dispersion component, as well as the potential counter-rotation of this population, may help to resolve the question of why M33's disk appears to be stable against the formation of a significant bar.

Our identification of a high velocity dispersion component throughout M33's inner stellar disk provides the first unambiguous kinematical detection and spatial characterization of a stellar halo in a relatively isolated galaxy of significantly lower mass than the Milky Way or M31. 
As discussed in Section~\ref{sec:intro}, there are multiple proposed mechanisms for forming a stellar halo in galaxies of M33's mass, ranging from accretion of smaller stellar systems to various internal and external heating mechanisms, each of which is expected to imprint clues of the halo's origin into the kinematical properties of the halo.  The M33 dataset presented here  provides the first opportunity to make concrete comparisons between the theoretical predictions and observations of a low mass, relatively isolated disk galaxy.

The presence or absence of rotation is likely to be a key observational constraint for determining the origins of M33's halo.  While rotation of the halo component in the plane of the disk appears to be minimal for stars within the range \rdiskcut\arcmin\,$\le$\,\rdisk\,$<$\,\rdiskcuttwo\arcmin, the model favors counter-rotation in the halo component for stars with \rdisk$<$\rdiskcut\arcmin.  
While this may indicate true physical complexity in the kinematics of the stars in M33's innermost regions, it could also be driven by limitations in the current model. 
Moreover, a stellar halo component could rotate in a different plane than the \hi\ or stellar disk.

In addition to future model improvements (Section~\ref{sec:model_limitations}), we also plan to increase our spectroscopic coverage in both the innermost and outermost regions of the current survey.  These combined model and data improvements will enable a detailed investigation into both the robustness, and the origin, of the intriguing kinematics observed in M33's innermost disk. 
Additional data in the outskirts of M33's stellar disk will enable a rigorous evaluation of whether a stellar halo component is present at distances that are beyond the warp in the \hi\ disk and the break in the stellar density profile.
These measurements will provide vital observational constraints on models of disk evolution and stellar halo formation in low-mass disk galaxies.

\acknowledgments

The authors recognize and acknowledge the very significant cultural role and reverence that the summit of Mauna Kea has always had within the indigenous Hawaiian community. We are most fortunate to have the opportunity to conduct observations from this mountain.  

This research made use of Astropy, a community-developed core Python package for Astronomy  \citep{astropy2013,astropy2018}\footnote{http://www.astropy.org}. 

We thank Stephen Gwyn for reducing the MegaCam photometry used in target and sample selection.

Support for this work was provided by NSF grants AST-1909066 (K.M.G.), AST-1909759 (P.G.), and DGE-1842400 (A.C.N.Q.).  
J.T.F., P.T., and I.Y. carried out this research under the auspices of the Science Internship Program at the University of California Santa Cruz.
The analysis pipeline used to reduce the DEIMOS data was developed at UC Berkeley with support from NSF grant AST-0071048.

\facilities{Keck:II (DEIMOS), HST(ACS), \newline CFHT(MegaCam)}
\software{emcee \citep{foreman-mackey_emcee, emcee-sw}, Astropy \citep{astropy2013},
 Matplotlib \citep{matplotlib}, numpy \citep{numpy},
corner \citep{corner-sw_v1, corner}}

\bibliography{m31}

\appendix

\section{Milky Way Contamination of the RGB sample}\label{sec:app_mwcont}

We removed clear Milky Way contaminants from the RGB sample by 
removing stars with visually identified \nai\ absorption, which is a surface-gravity sensitive feature observed in MW dwarf stars along the line of sight to the M31 system.  We expect minimal MW contaminants based on the both the high density of M33 stars in these fields, and on our visual 
inspection of the data, which 
identified only a small fraction ($\sim 4$\%) of stars in the total spectroscopic sample with strong \nai\ absorption.  

We quantified this expectation using the Besancon MW model \citep{robin2003,robin2014,robin2017,amores2017}; the version utilized for this analysis was the August 12, 2019 update.  Within our RGB selection region, the MW line-of-sight velocity distribution is expected to peak at \vlos\,$\sim -30$\kms, with $\sim 47$\% of the MW model distribution at \vlos$>-50$\kms\ (the fraction of MW model stars with \vlos$>-50$\kms\ increases slightly for masks with a brighter limiting magnitude).  
In our RGB selection region, 50 stars (3\%) were removed due to identified \nai\ absorption; 31 of these had \vlos$>-50$\kms.  
These numbers are dominated by the masks at large projected radii observed in 2020: in masks observed before 2020, 16 stars (1\%) were removed due to identified \nai\ absorption. 
In our final RGB sample, only four stars have \vlos$>-50$\kms.  
Conservatively assuming these are all MW stars, 
we estimate an upper limit of residual MW contamination in our RGB sample of $\sim 8$~stars, equivalent to $\sim 0.5$\% of the sample.  

While selecting on the presence of \nai\ absorption is effective in removing dwarf stars in the MW disk, it does not remove distant main sequence turn-off stars in the MW halo along the line of sight to M33.  MW halo stars are expected to span a large range of velocities (hundreds of \kms), centered relatively near the systemic velocity of M33, although most of these stars are expected to be bluer than the M33 RGB (see \citet{gilbert2012} for a discussion of the expected velocity distribution of MW halo stars along the line of sight to M31).  Again using the Besancon model, we have estimated that only $\sim 10$\% ($\sim 17$\%) of the stars in the MW model which fall within our RGB selection box will have distances greater than 8~kpc (6~kpc).  If we assume the 50 stars in our RGB selection region with \nai\ absorption represent 90\% (83\%) of the MW contaminants along the line of sight, then we would expect on order $\sim 6$ ($\sim 10$) MW halo stars in total in our RGB sample.    

\section{Velocity Models}\label{sec:app_velmodel}
This appendix provides details on several technical aspects related to the velocity models discussed in Section~\ref{sec:models}, including our MCMC implementation and choice of priors (\ref{app:mcmc_implementation}), how we calculate the visualizations of the two-component model fits in \vlos\ and \voff\ space (\ref{sec:app_velmodel_viz}), and a Bayesian evidence-based analysis of the number of components favored by the RGB spectroscopic sample within the current model framework (\ref{app:models_number_comps}).

\subsection{Details of the MCMC Implementation}\label{app:mcmc_implementation}

As discussed in Section~\ref{sec:one_comp_fits}, we used the software package {\tt emcee} \citep[version 3.0.2;][]{foreman-mackey_emcee, emcee-sw} to determine the posterior probability distributions for the one and two component model parameters, given the individual likelihoods for each model (Equations~\ref{eq:onecomp} and \ref{eq:twocomps_rotating})
and our assumed priors.

The total likelihood of the observed dataset given a set of model parameters $\theta$ (\sigdisk, \frotdisk, \sighalo, \frothalo, and \fhalo\ for the two-component model; \sigmavoff\ for the one-component model) is the product of the individual likelihoods:

\begin{equation}
\mathscr{L}_{\theta} = \prod_{j=1}^{N_{\rm stars}} \mathscr{L}_{j,\theta},
\label{eqn:likelihood}
\end{equation}

where the individual likelihoods for the one and two component models are given by Equations~\ref{eq:onecomp} and \ref{eq:twocomps_rotating}, respectively.
For a set of model parameters $\Theta$, we compute
\begin{equation}
p(\Theta | {v_j}_{j=1}^N, I) \propto p({v_j}_{j=1}^N | \Theta, I)\,p(\Theta | I)
\label{eq:bayes}
\end{equation}

where ${v_j}_{j=1}^N$ is the set of observed line-of-sight velocities, $I$
represents our prior knowledge, and $p({v_j}_{j=1}^N | \Theta, I)$ is the likelihood
term (Equation~\ref{eqn:likelihood}). 

We used uniform priors for 
\frotdisk, \frothalo, and \fhalo,
and a scale free prior ($1/\sigma$) for \sigdisk\ and \sighalo.  The minimum and maximum bounds on the range of \sigdisk\ and \sighalo\ were generous: 1 to 100 for \sigdisk, and 5 to 200 for \sighalo.  We also included a prior enforcing that the second component has a larger velocity dispersion than the first component: \sighalo\,$>$\,\sigdisk.  
The fraction of stars in the halo component, \fhalo, was allowed to vary from 0 to 1.  
The parameters governing the rotation speed of the disk and halo components, \frotdisk\ and \frothalo, were allowed to vary from 0 to 1.5 for \frotdisk, allowing the model to explore rotation speeds equal to or slightly faster than that of the \hi\ disk model and ensure no bias was introduced around \frotdisk$=1$ by the priors; and from $-1.0$ to 1.0 for \frothalo, allowing the model to explore counter-rotation of the halo component with respect to the \hi\ disk, and to ensure no bias was introduced around \frothalo$=0$ by the priors.  

\begin{figure*}[tb!]
\plotone{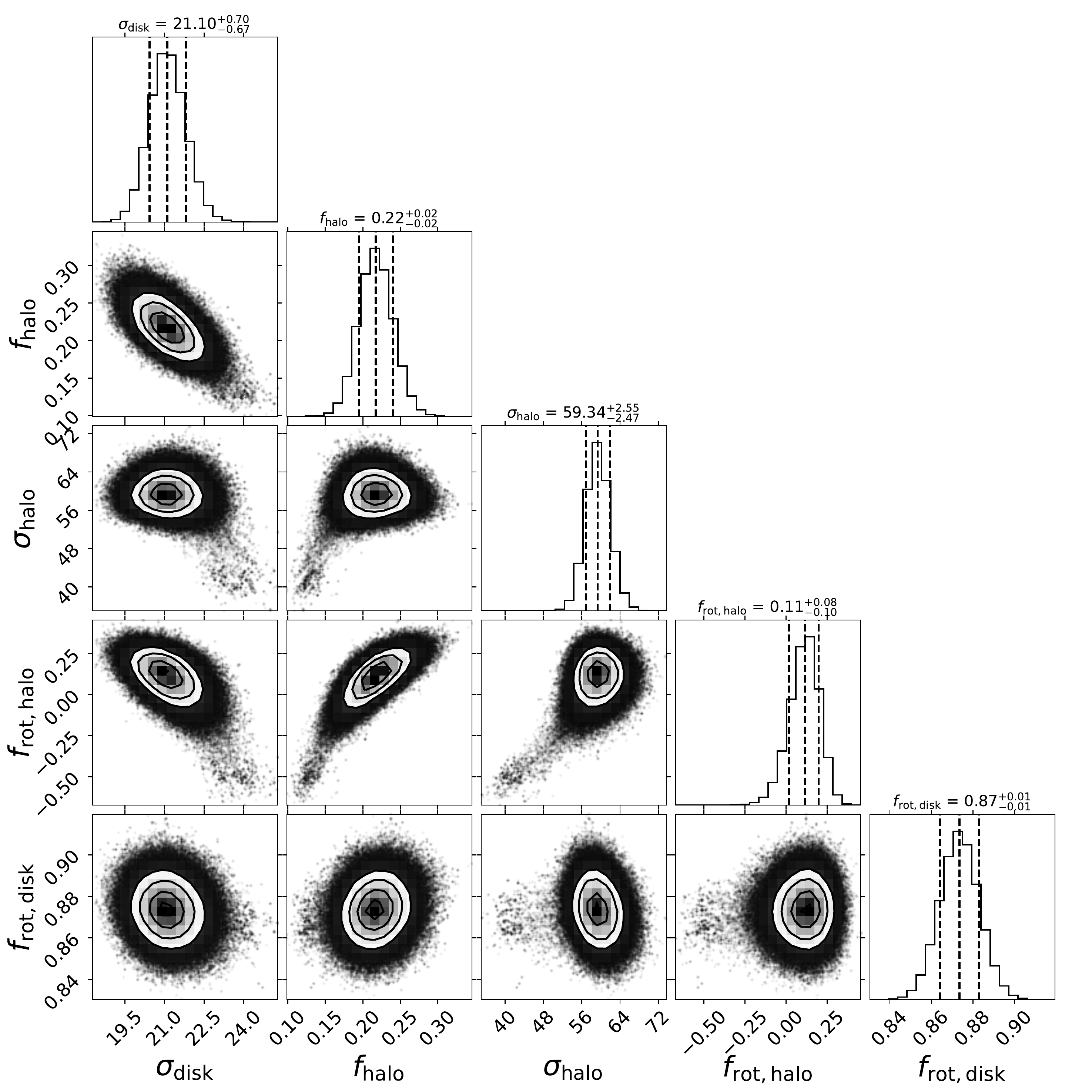}
\caption{
Marginalized one- and two-dimensional posterior probability distribution functions for each of the fit parameters in the two-component model (Sections~\ref{sec:models} and \ref{sec:app_velmodel}), fit to the full spectroscopic RGB sample.  The fit parameters include the disk and halo velocity dispersions (\sigdisk, \sighalo), the fraction of the \hi\ rotation velocity at which the disk and halo are rotating (\frotdisk, \frothalo; negative values of \frot\ correspond to counter-rotation in the plane of the \hi\ disk), and the fraction of the population in the halo component (\fhalo). Dashed lines and column headings show the 16th, 50th, and 84th percentiles of the marginalized 1-dimensional posterior probability distribution functions for each model parameter. While the parameter values are well constrained by the model, the two-dimensional posterior probability distribution functions show clear correlations between several of the model parameters.  
}
\label{fig:app_corner_all}
\end{figure*}

The 2-d and 1-d marginalized posterior probability distributions for the two-component model fit to the full RGB dataset is shown in Figure~\ref{fig:app_corner_all}. The length of the MCMC chains were set by the number of steps required to obtain an estimate of the auto-correlation time; they were run for a length of at least 50 times the longest parameter's estimated auto-correlation time for each fit, as estimated by {\tt emcee}, or for 2000 steps, whichever was longer. All MCMC runs used 300 walkers.  
We discarded the first 500 steps of the chains, which results in using a burn-in length which is a factor of 5 greater than the single longest autocorrelation time estimate in all of the fits presented here. 
We confirmed that even for the longest autocorrelation times encountered in our fits, our choices for burn-in, chain length, and number of walkers results in a total number of MCMC steps which exceeds the minimum number of steps needed to obtain the values of the (e.g.) 18th/64th percentiles of the distribution to a precision of 1\%.    

\subsection{Visualization of the Velocity Models}\label{sec:app_velmodel_viz}
In the two-component model there is no single velocity reference frame in which the data can be visualized against analytical descriptions of the model. Therefore, we provide visualizations of the data against the two-component model in the two velocity frames that we believe are the most intuitive and useful: line-of-sight velocity, and the offset of the observed stellar velocities from the line-of-sight velocities of the \hi\ disk model, \voff\@.  For visualizations of the model components in the line-of-sight velocity frame (\vlos), we computed a sum of Gaussians, one for each star $j$ in the sample, with means $\mu_ j = v_{{\rm model\ los,} j}$ for each component of the two-component model (computed using \frot\ for that component) and dispersion equal to the $\sigma$ of the component, weighted according to the fraction of the population in that component in the model. 
A similar procedure was used to provide visualizations of the two-component model in \voff: we computed a sum of Gaussians with means $\mu_j = v_{{\rm H\,\textsc{i}}, j} - v_{{\rm model\ los,} j}$, (i.e., the offset of the model component's mean line-of-sight velocity from the \hi\ disk model line-of-sight velocity at the location of star $j$), 
and dispersion equal to the $\sigma$ of the component, again weighted according to the fraction of the population in that component in the model.  

\begin{figure*}[tb!]
\plottwo{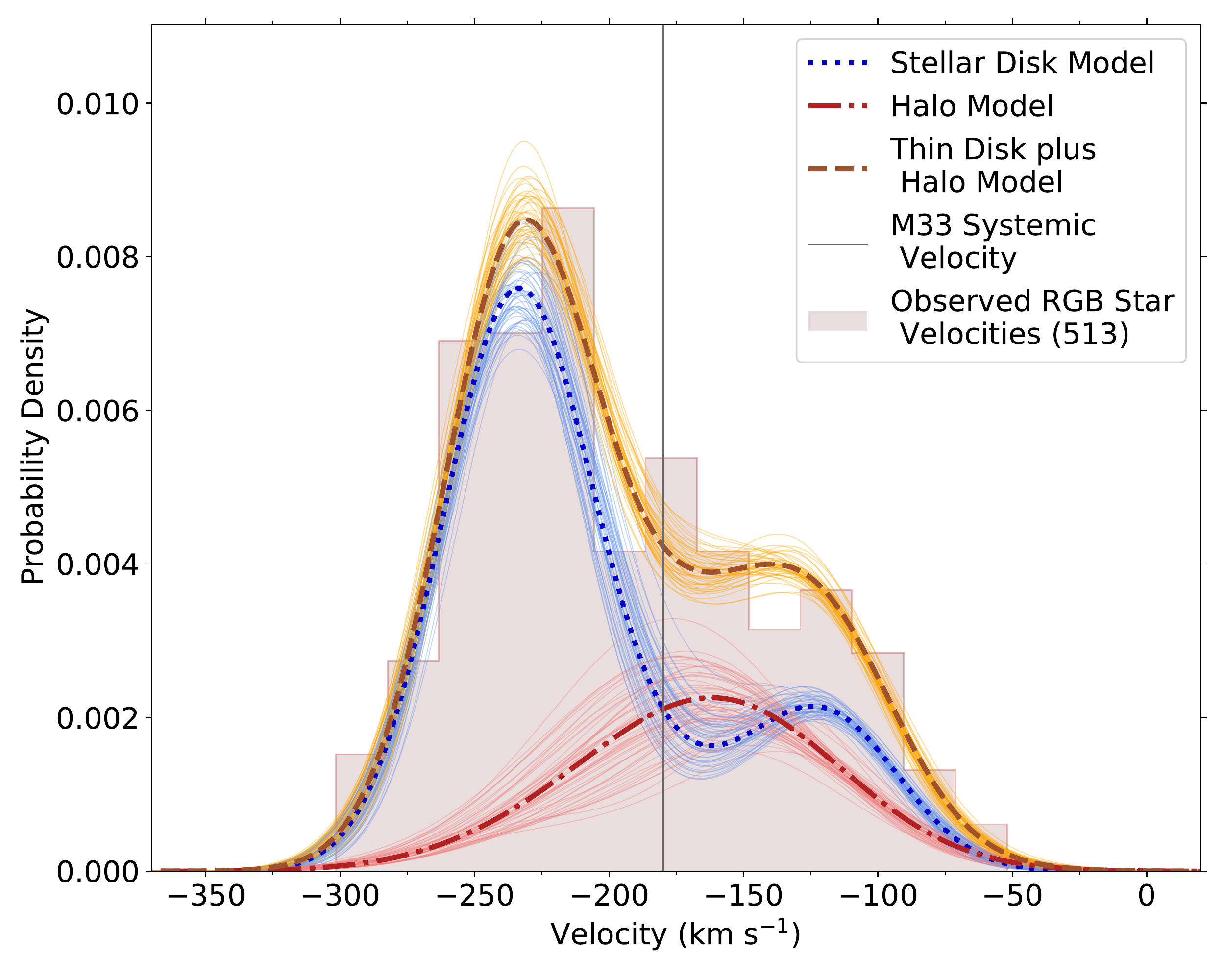}{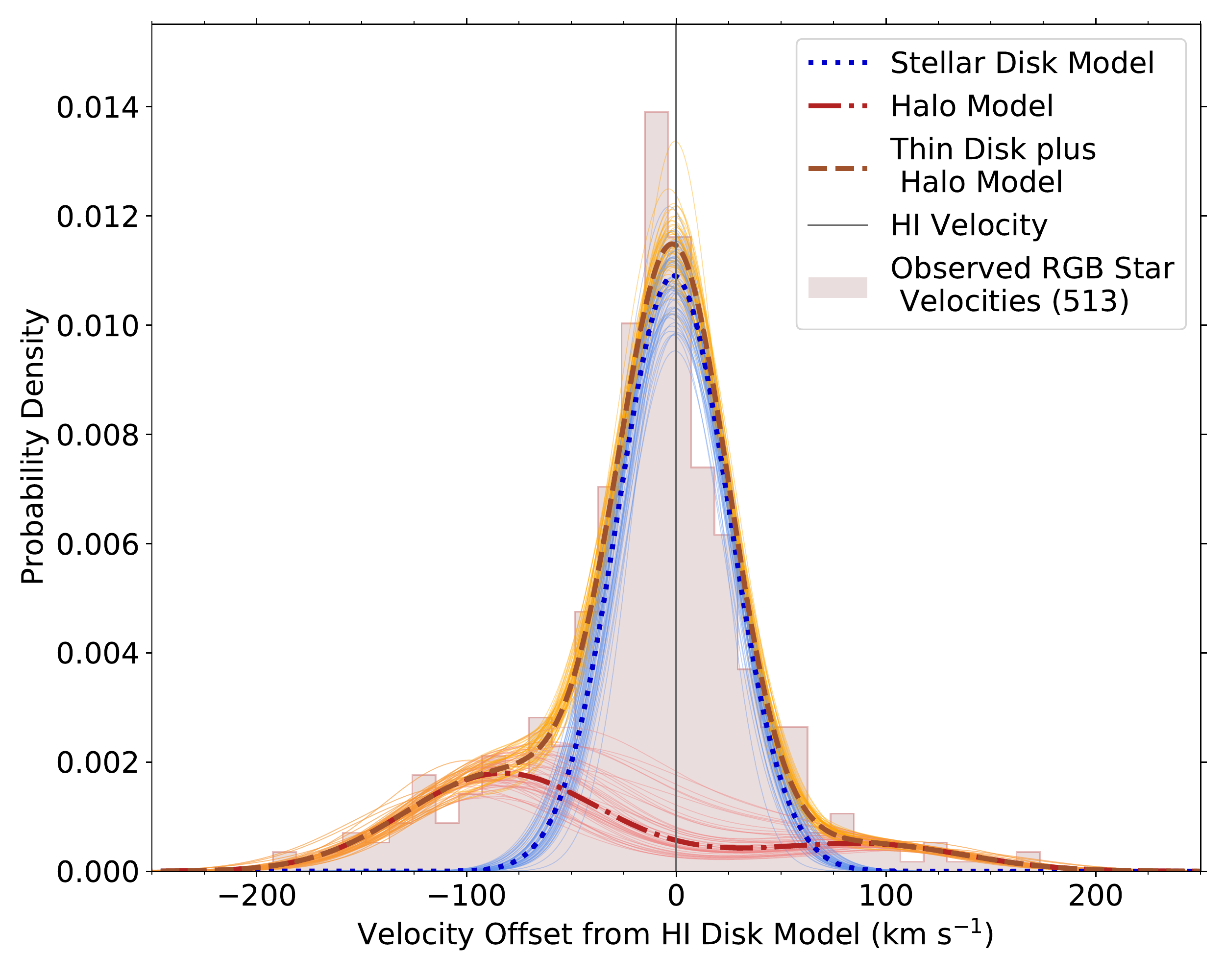}
\plottwo{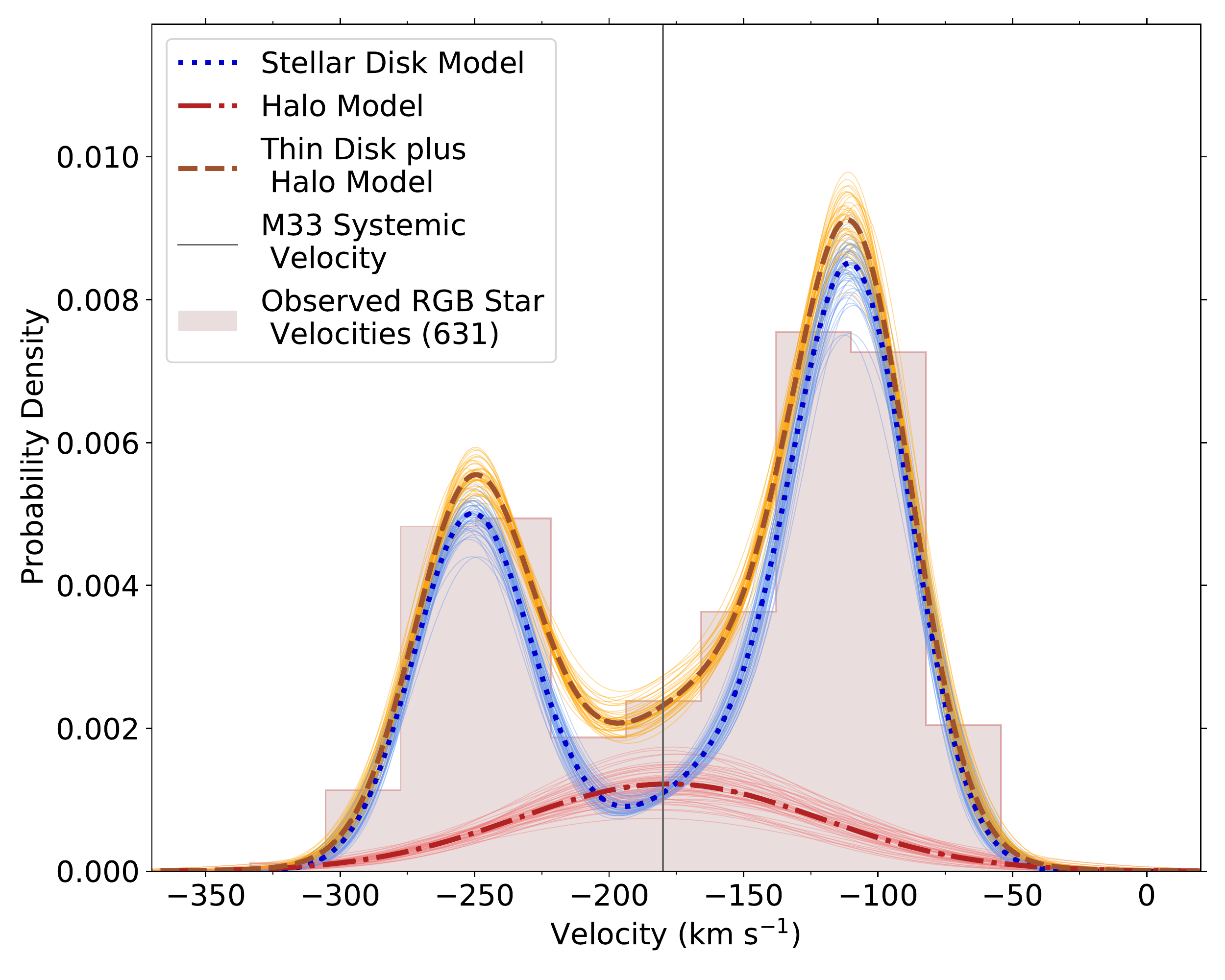}{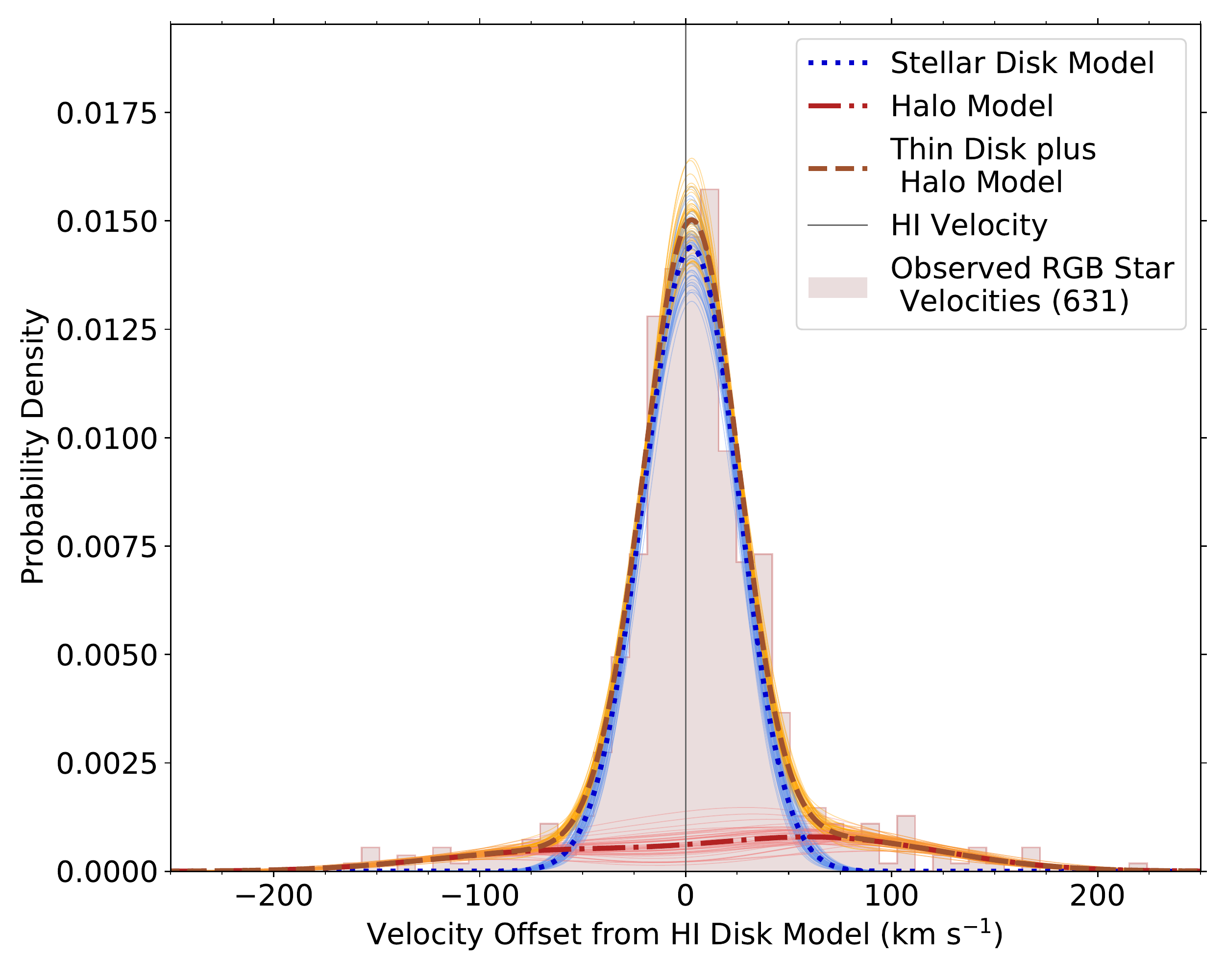}
\plottwo{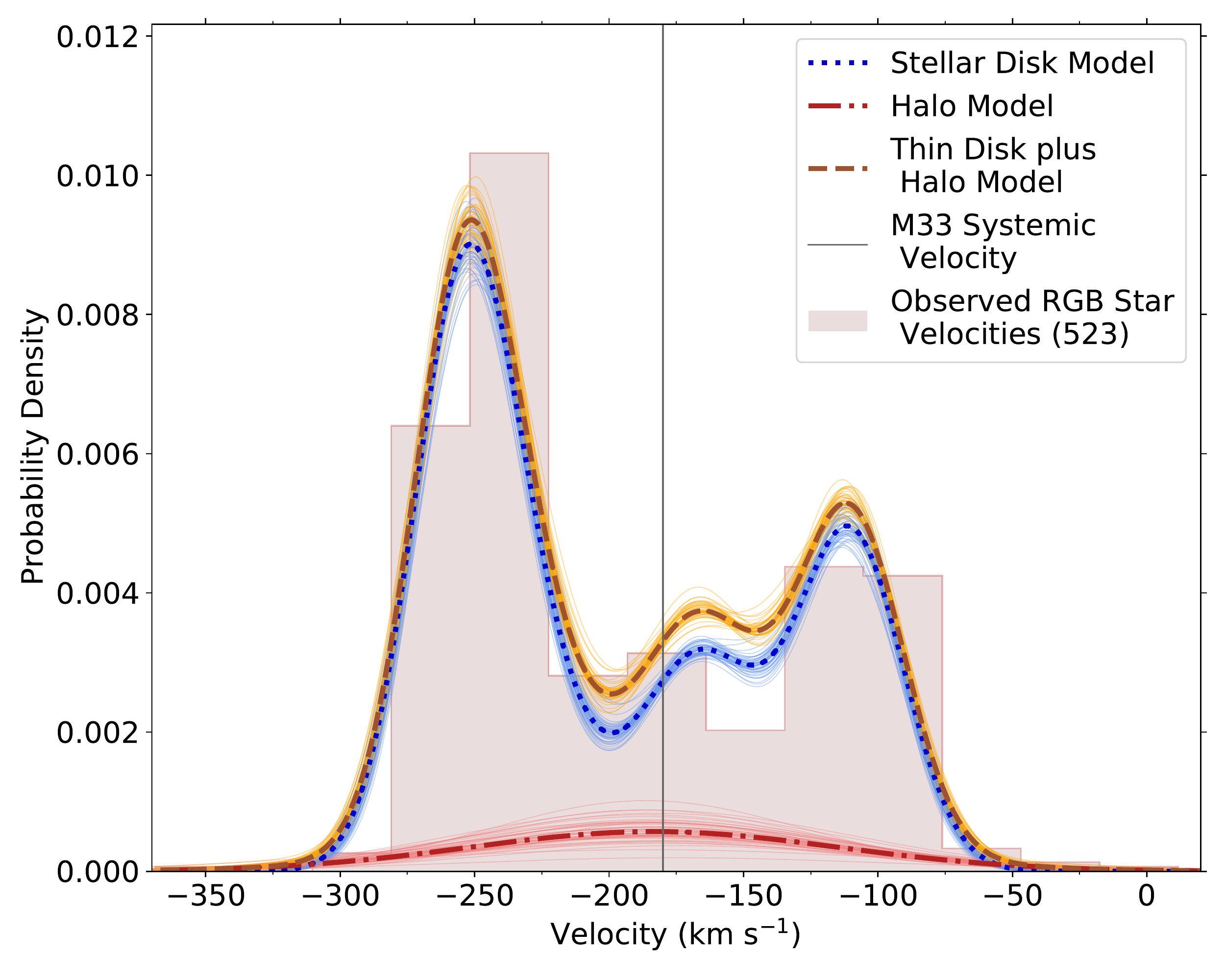}{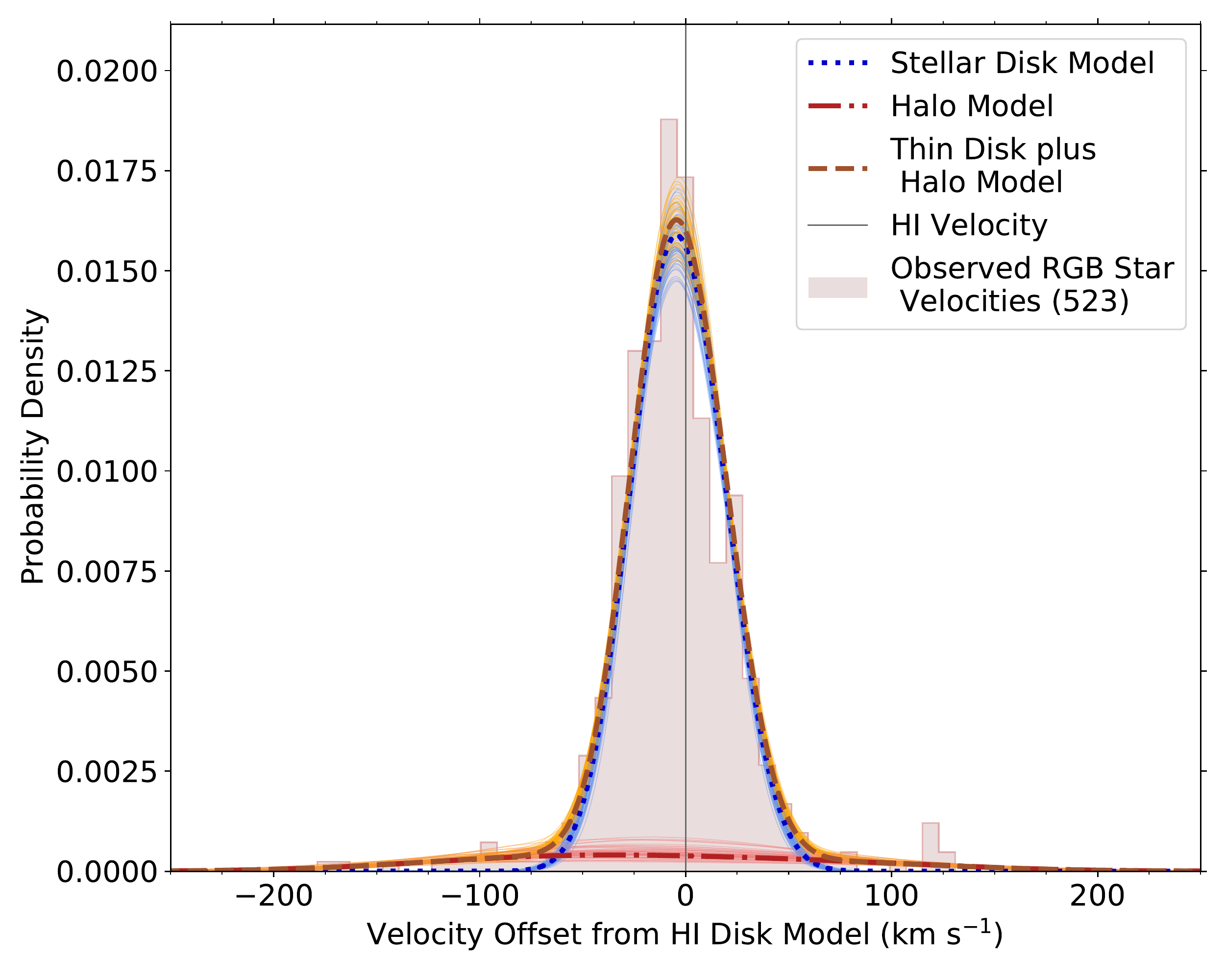}
\caption{
Distributions in \vlos\ (left panels) and \voff\ (right panels) of all RGB stars in each radial bin shown in Figure~\ref{fig:fits_by_spatialposition} (Section~\ref{sec:models_bins}), along with the results of the model fits to each sub-sample (shown as described in Figure~\ref{fig:velhist_wmodel_halo}).  The top row shows fits to the innermost radial bin (\rdisk$<$\rdiskcut\arcmin); the middle row to the middle radial bin (\rdiskcut\arcmin$\leq$\rdisk$<$\rdiskcuttwo\arcmin); and the bottom row to the outer radial bin (\rdisk$\geq$\rdiskcuttwo\arcmin).  The asymmetry in the strength of the peaks in \vlos\ (driven by the disk component), and in the tails of the distribution in \voff\ (driven by the halo component) is due to the unequal sample distribution between the northeastern and southwestern halves of the disk: we have obtained spectra for greater numbers of RGB stars in the northeast in the inner and outer radial bins, and in the southwest in the middle radial bin (Table~\ref{tab:fitparams}).
}
\label{fig:app_fits_full_radial_bins}
\end{figure*}

\begin{figure*}[tb!]
\plotone{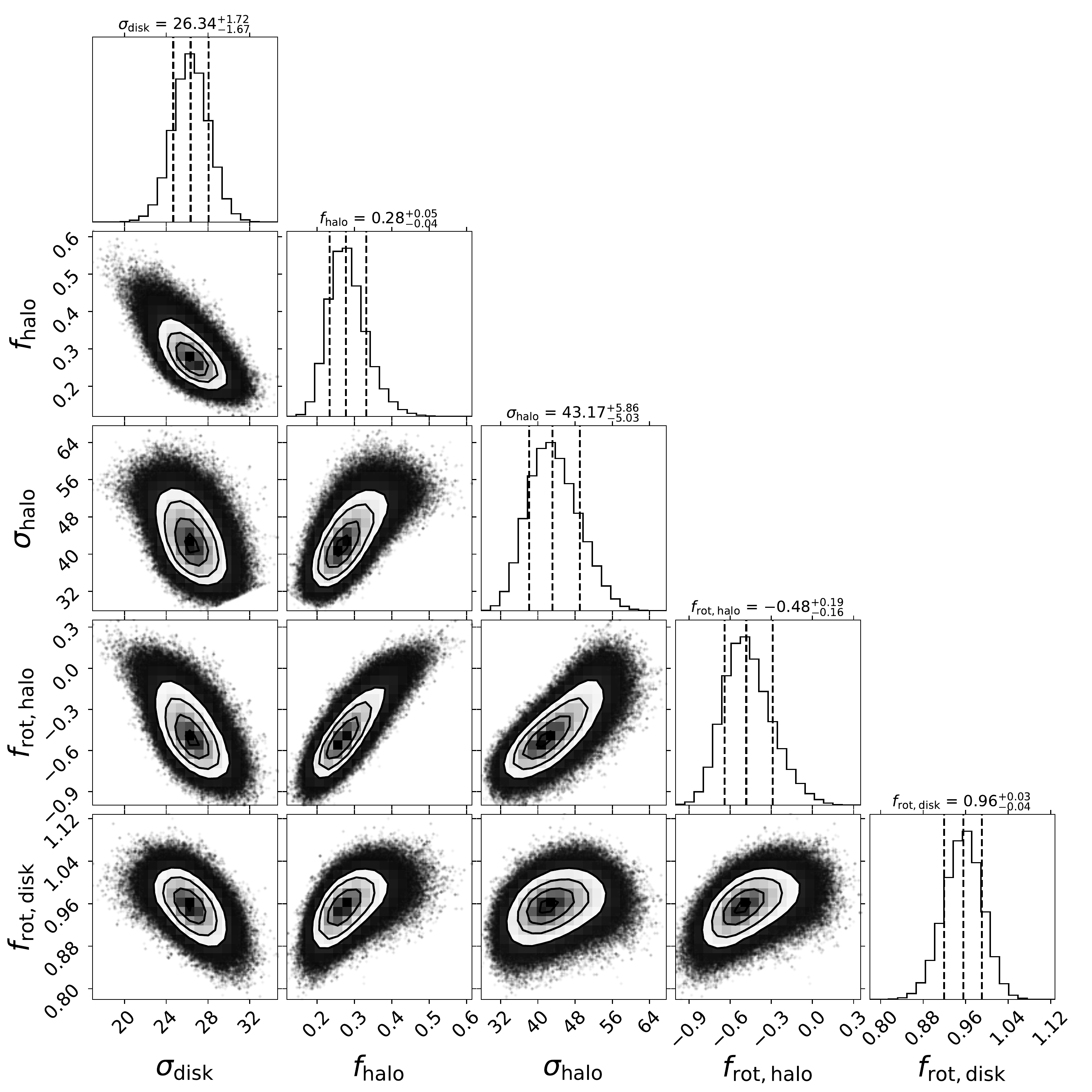}
\caption{
Same as Figure~\ref{fig:app_corner_all} for the independent two-component fit to the innermost (\rdisk$<15$\arcmin) of the three radial bins shown in Figure~\ref{fig:fits_by_spatialposition} and discussed in Section~\ref{sec:models_bins}. 
}
\label{fig:app_corner_radial_bins}
\end{figure*}

\begin{figure*}[tb!]
\plotone{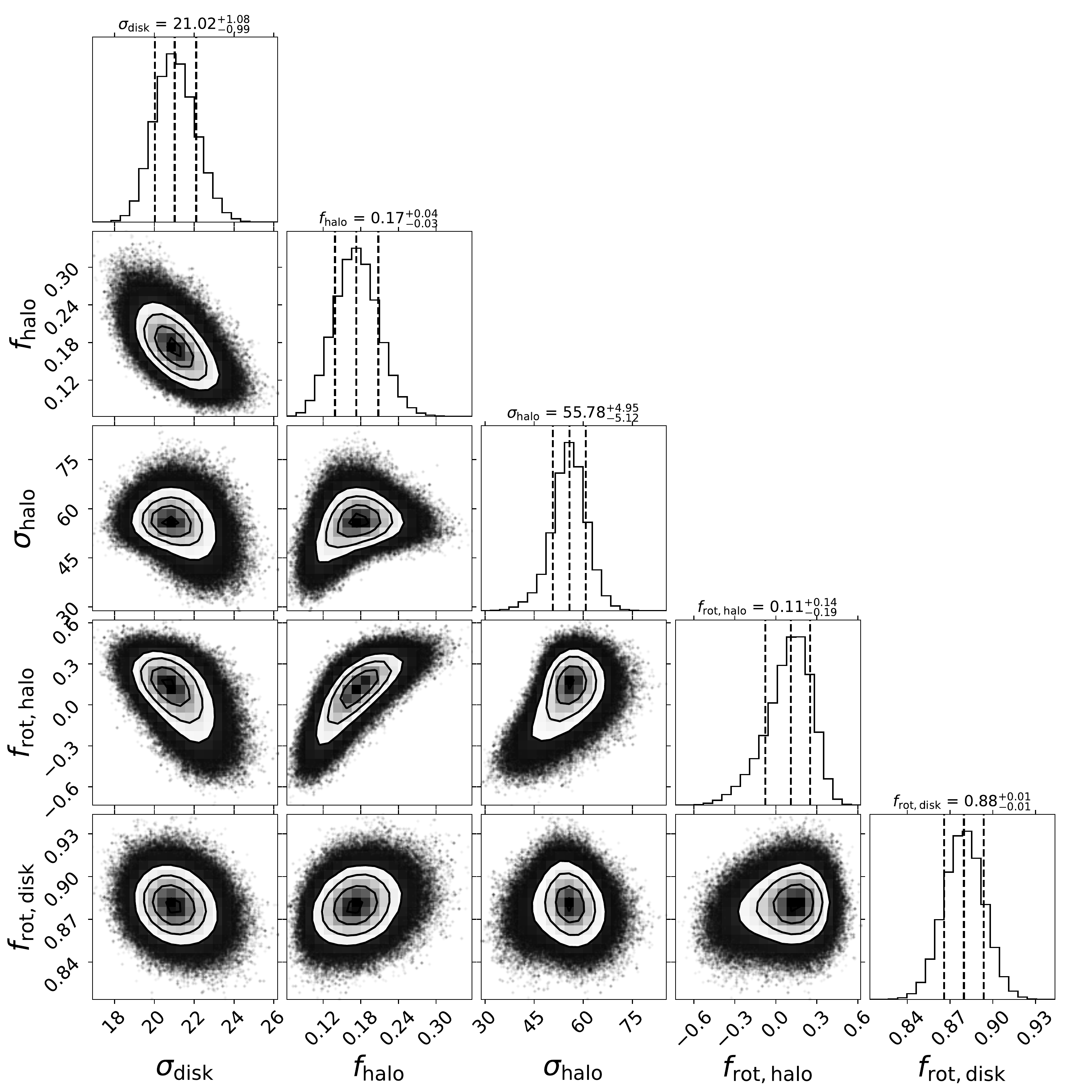}
\caption{
Same as Figure~\ref{fig:app_corner_radial_bins} for the independent two-component fit to the middle ($15\leq$\rdisk$<30$\arcmin) of the three radial bins shown in Figure~\ref{fig:fits_by_spatialposition} and discussed in Section~\ref{sec:models_bins}. 
}
\label{fig:ref_middlebin}
\end{figure*}

\begin{figure*}[tb!]
\plotone{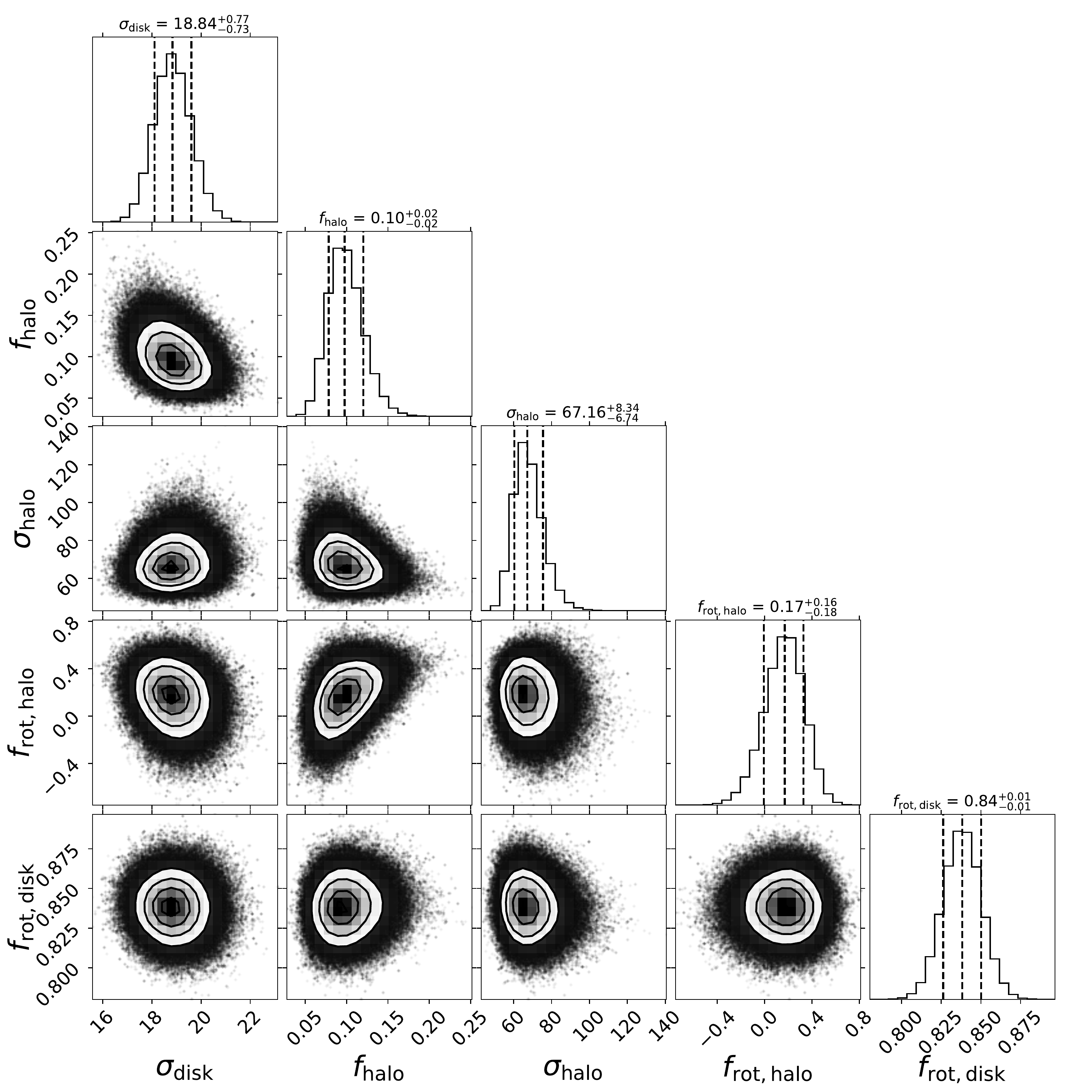}
\caption{
Same as Figure~\ref{fig:app_corner_radial_bins} for the independent two-component fit to the outer (\rdisk$\geq 30$\arcmin) of the three radial bins shown in Figure~\ref{fig:fits_by_spatialposition} and discussed in Section~\ref{sec:models_bins}. 
}
\label{fig:ref_outerbin}
\end{figure*}

\begin{figure*}[tb!]
\plottwo{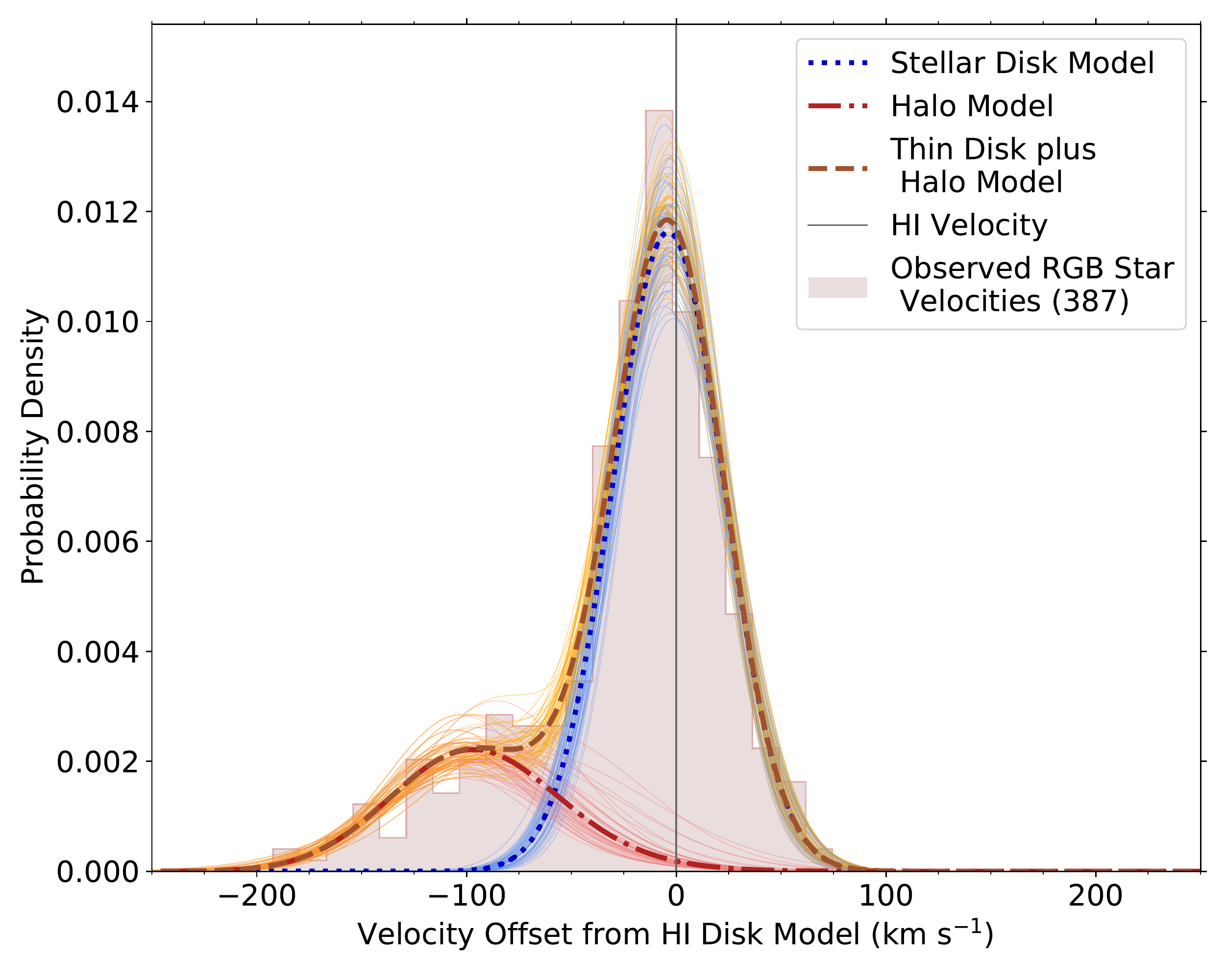}{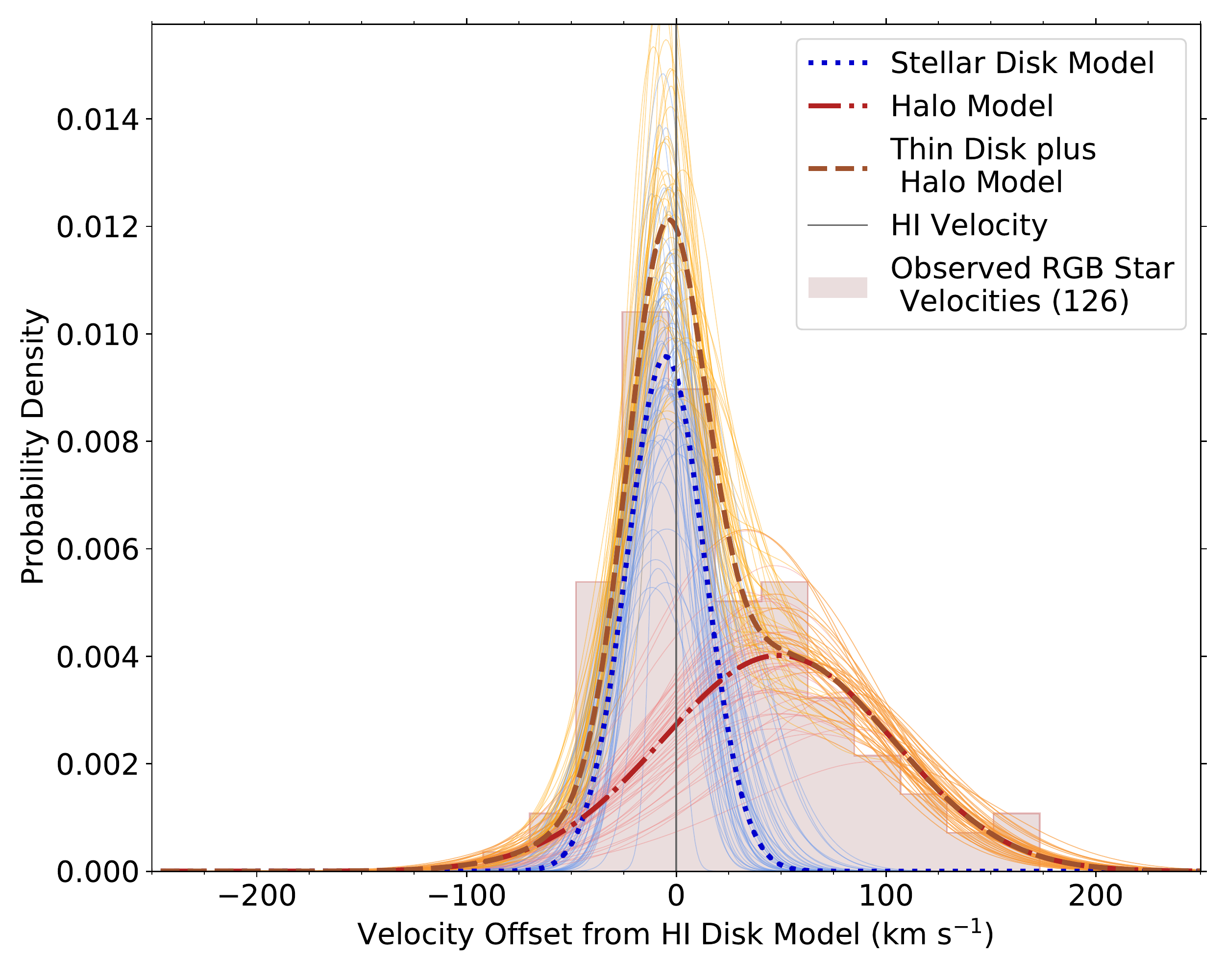}
\plottwo{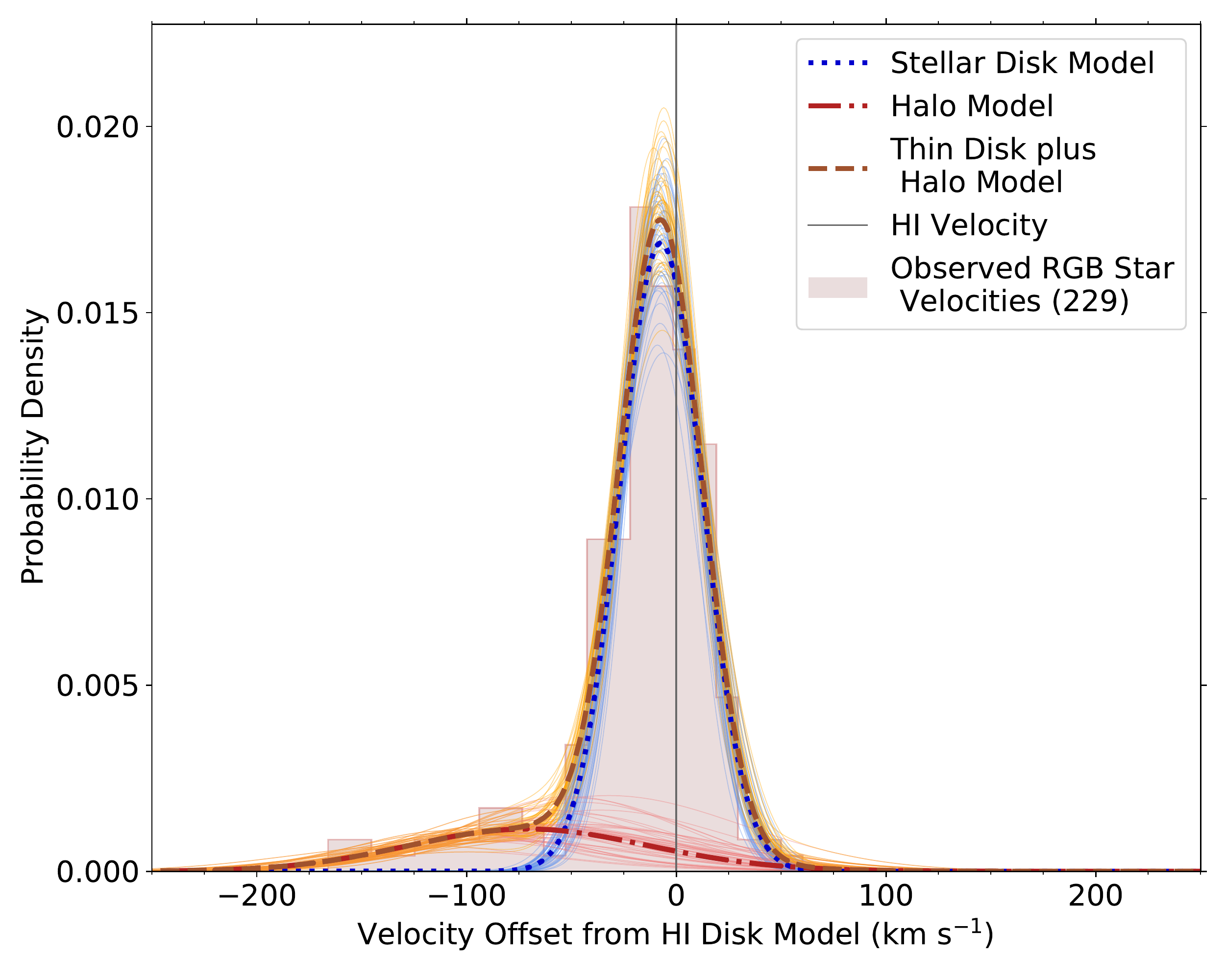}{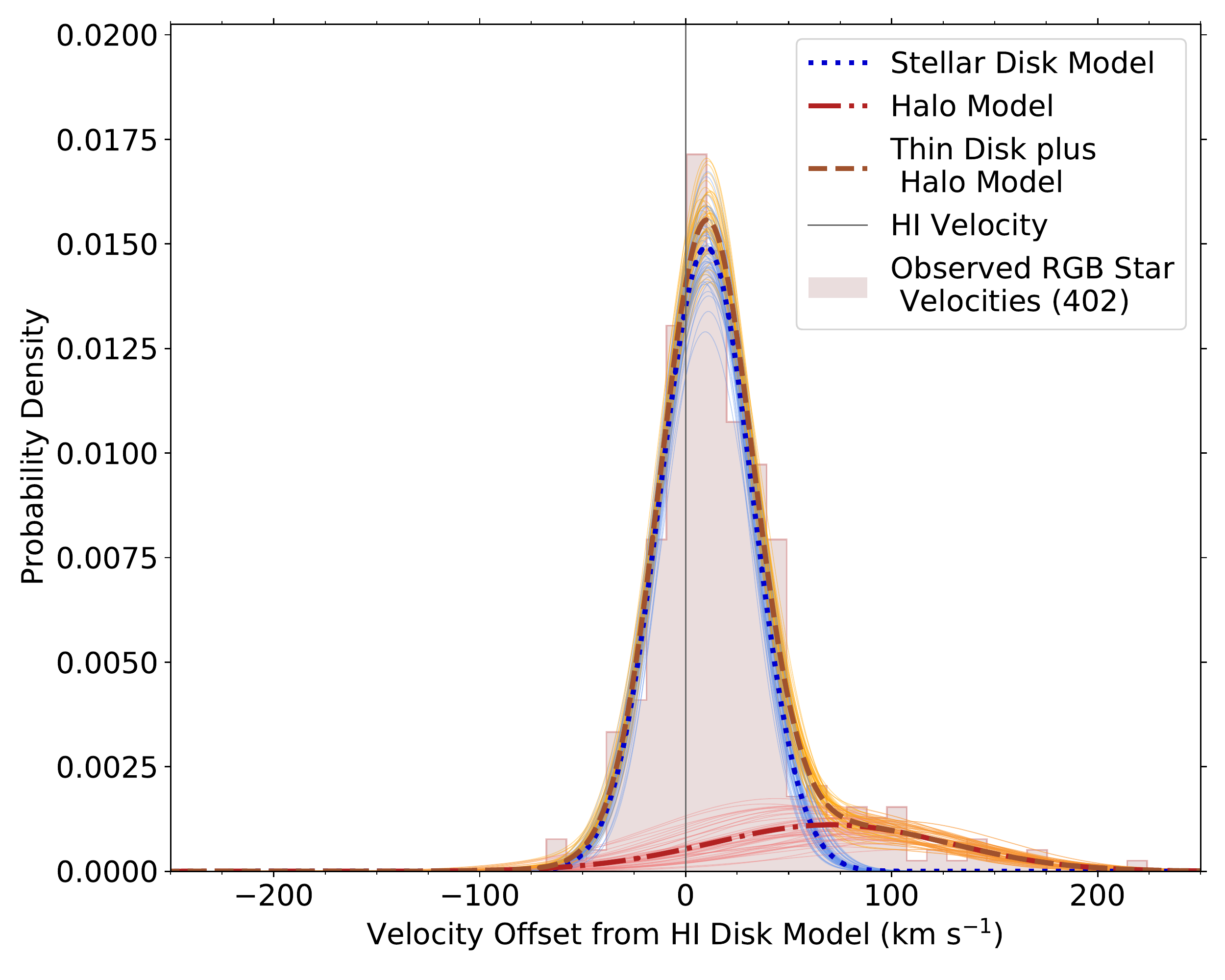}
\plottwo{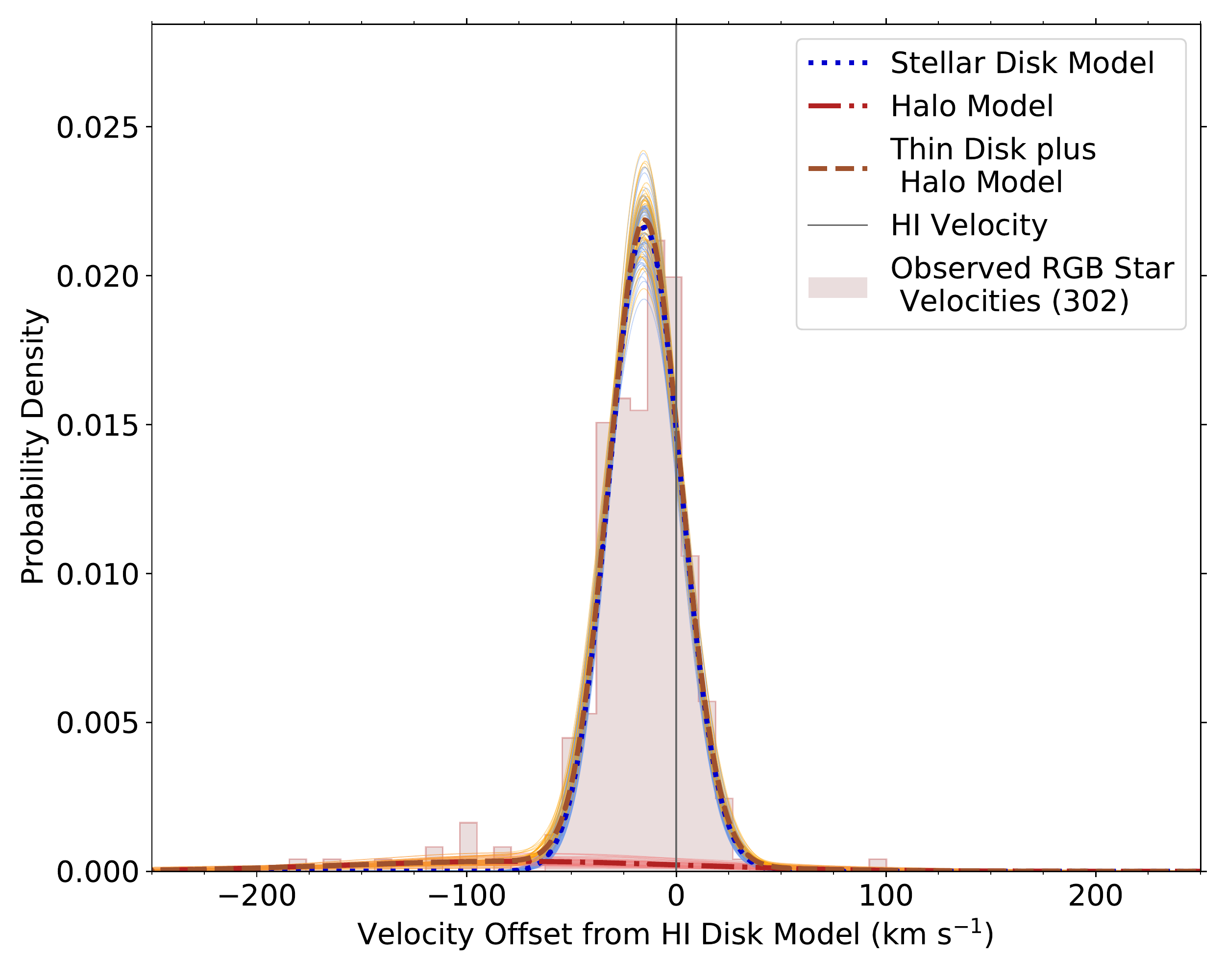}{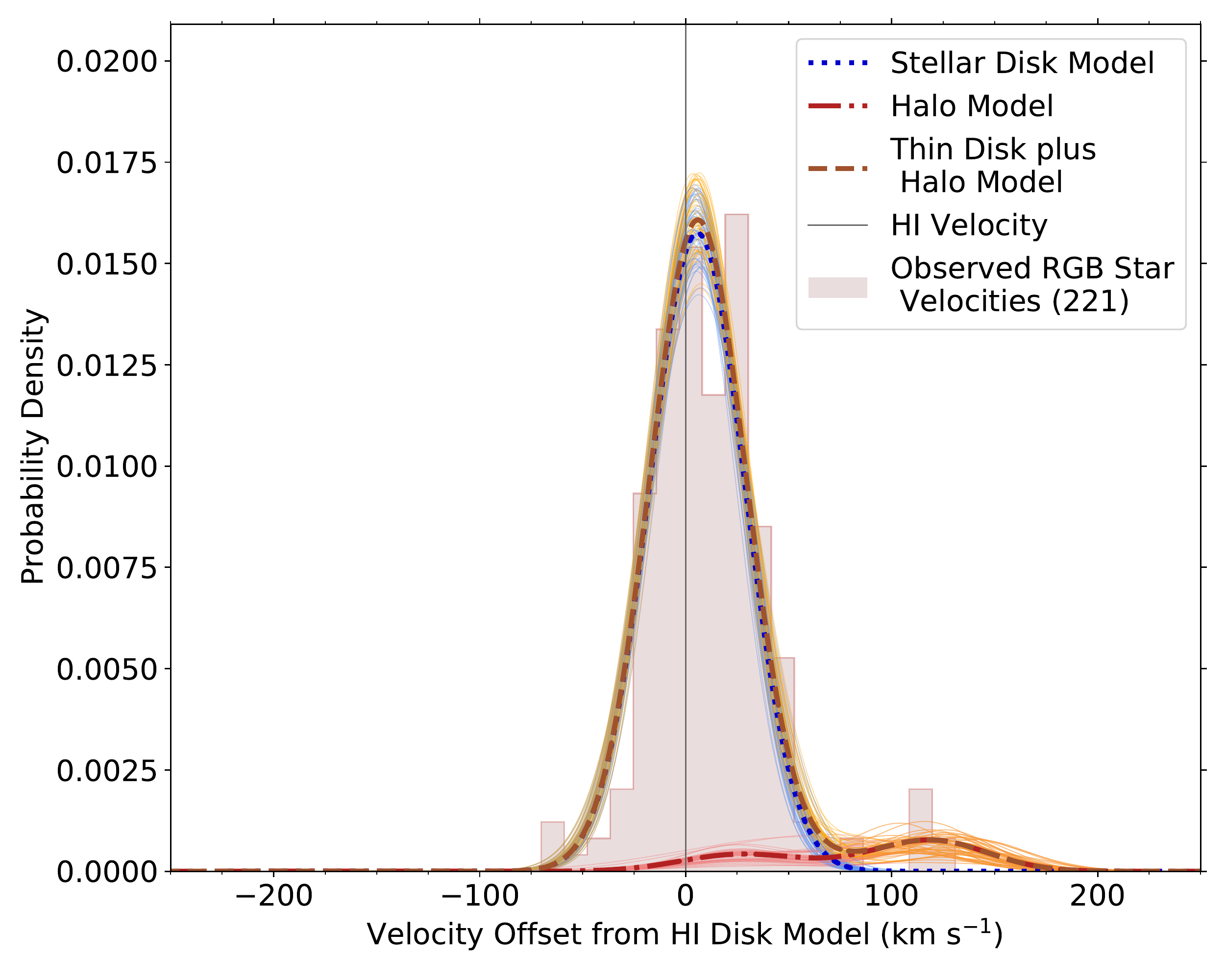}
\caption{
Distributions in \voff\ space of the data in the sub-samples shown in Figure~\ref{fig:fits_by_spatialposition} (Section~\ref{sec:models_bins}), along with the results of the model fits to each sub-sample (shown as described in Figure~\ref{fig:velhist_wmodel_halo}).  The left panels show the fits to the RGB sample in the approaching (northeast) half of M33's disk; the right panels show the receding (southwest) half.  The top row shows fits to the innermost radial bin (\rdisk$<$\rdiskcut\arcmin); the middle row to the middle radial bin (\rdiskcut\arcmin$\leq$\rdisk$<$\rdiskcuttwo\arcmin); and the bottom row to the outer radial bin (\rdisk$\geq$\rdiskcuttwo\arcmin).
}
\label{fig:app_fits_by_spatialposition}
\end{figure*}

\begin{figure*}[tb!]
\plotone{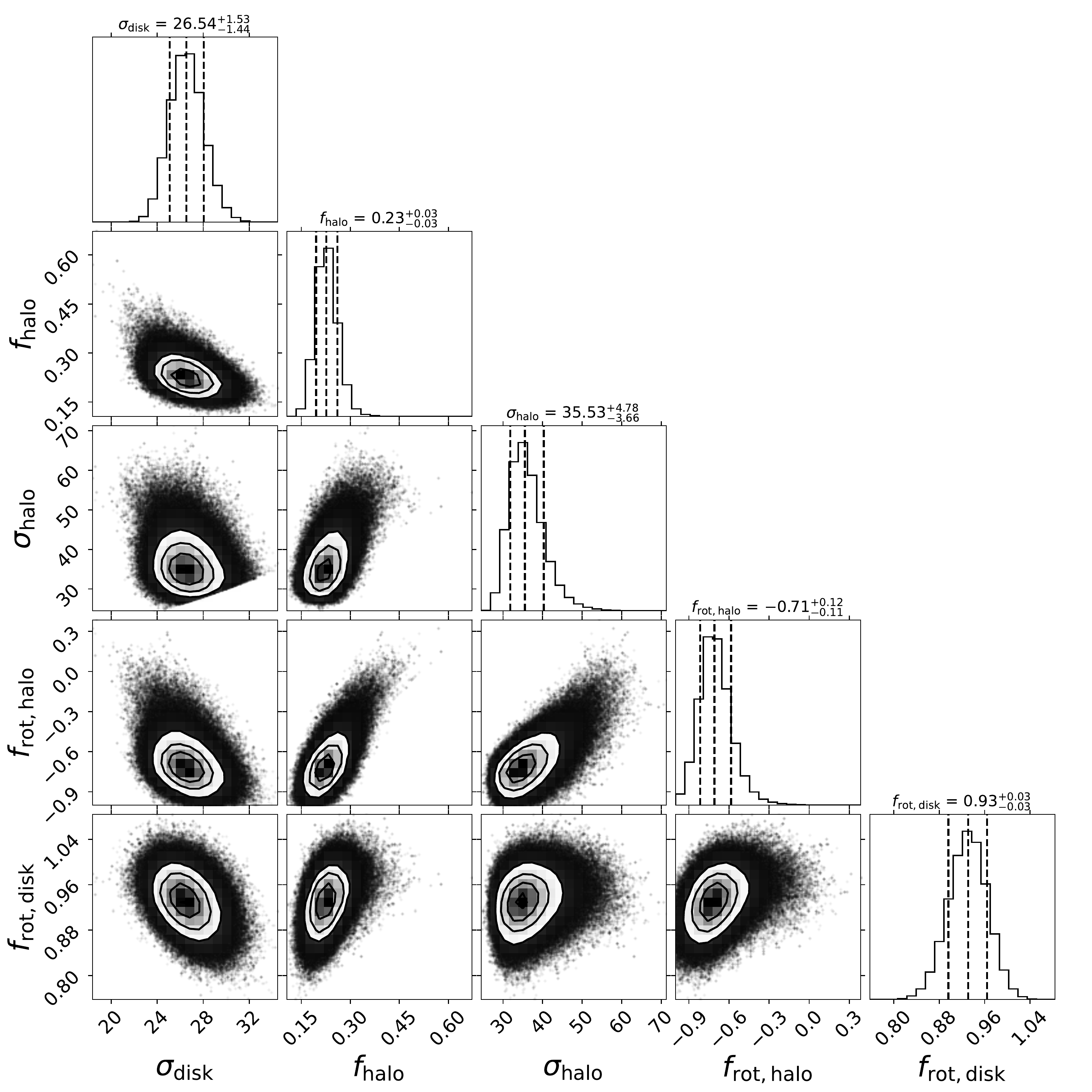}
\caption{
Same as Figure~\ref{fig:app_corner_all} for the independent two-component fit to the northeast half of the innermost radial bin (\rdisk$<15$\arcmin) 
shown in Figure~\ref{fig:fits_by_spatialposition} and discussed in Section~\ref{sec:models_bins}.  
}
\label{fig:app_corner_spatialbins}
\end{figure*}

\begin{figure*}[tb!]
\plotone{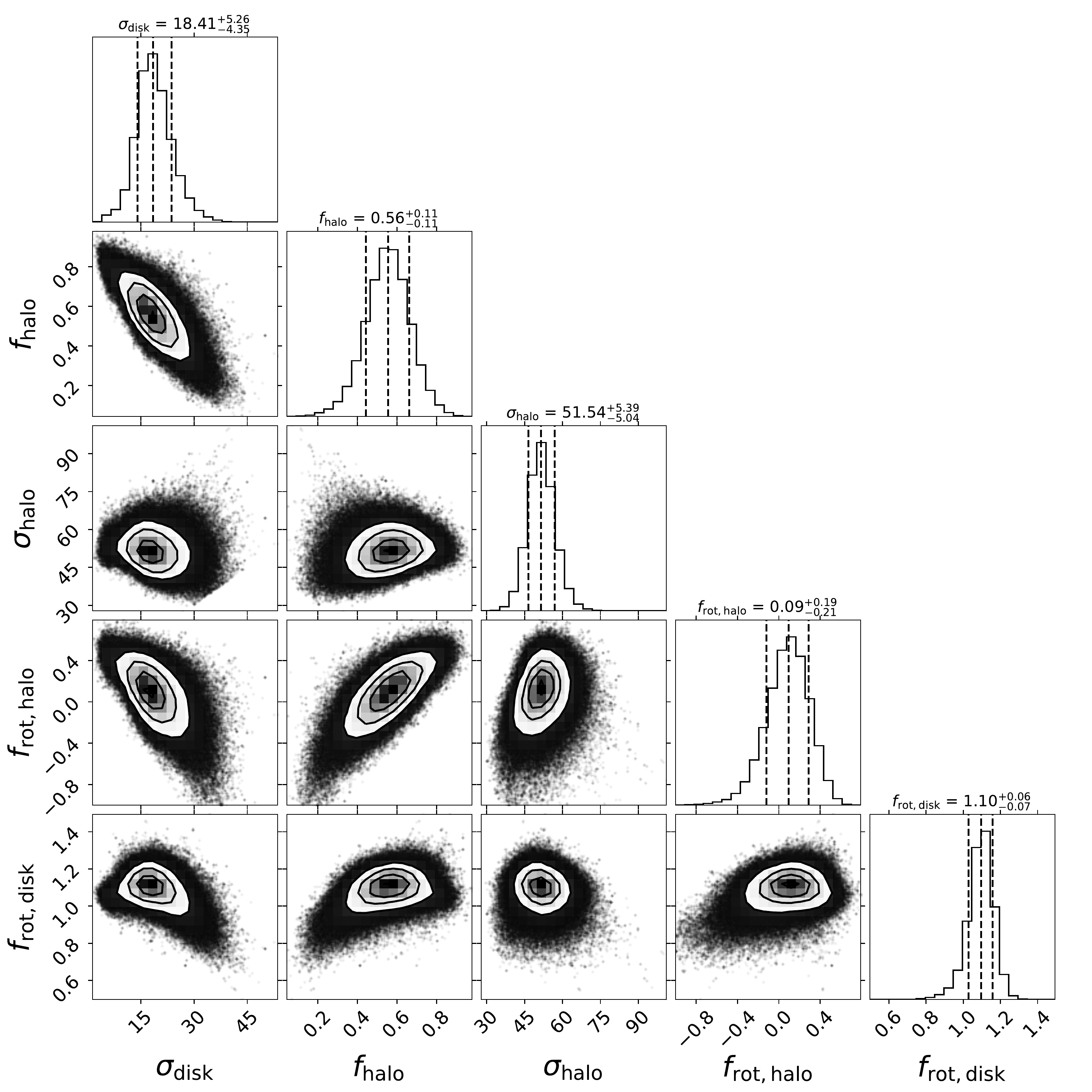}
\caption{
Same as Figure~\ref{fig:app_corner_all} for the independent two-component fit to the southwest half of the innermost radial bin (\rdisk$<15$\arcmin). 
}
\label{fig:ref_spatial2}
\end{figure*}

\begin{figure*}[tb!]
\plotone{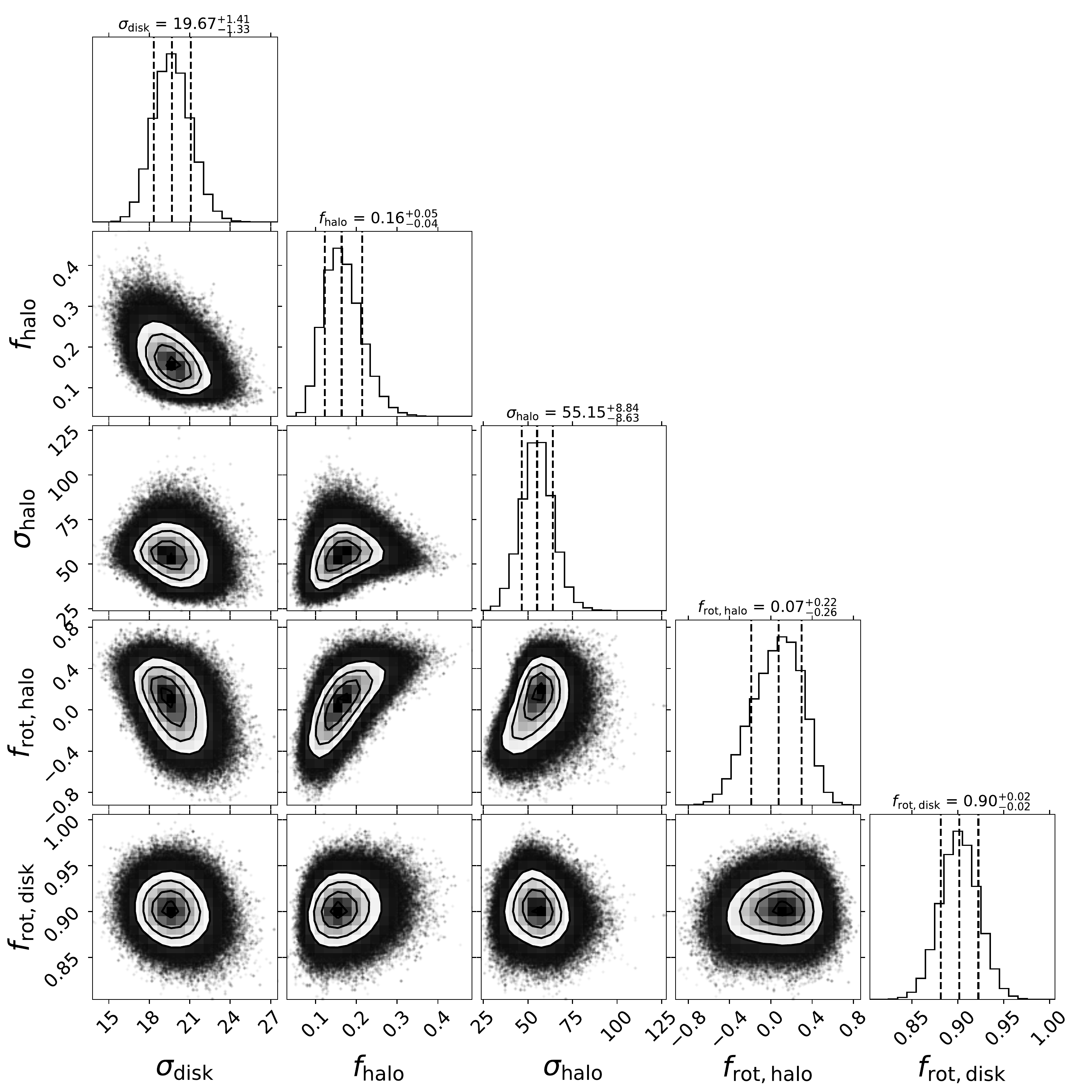}
\caption{
Same as Figure~\ref{fig:app_corner_all} for the fit to the northeast half of the middle radial bin ($15\leq$\rdisk$<30$\arcmin).   
}
\label{fig:ref_spatial3}
\end{figure*}

\begin{figure*}[tb!]
\plotone{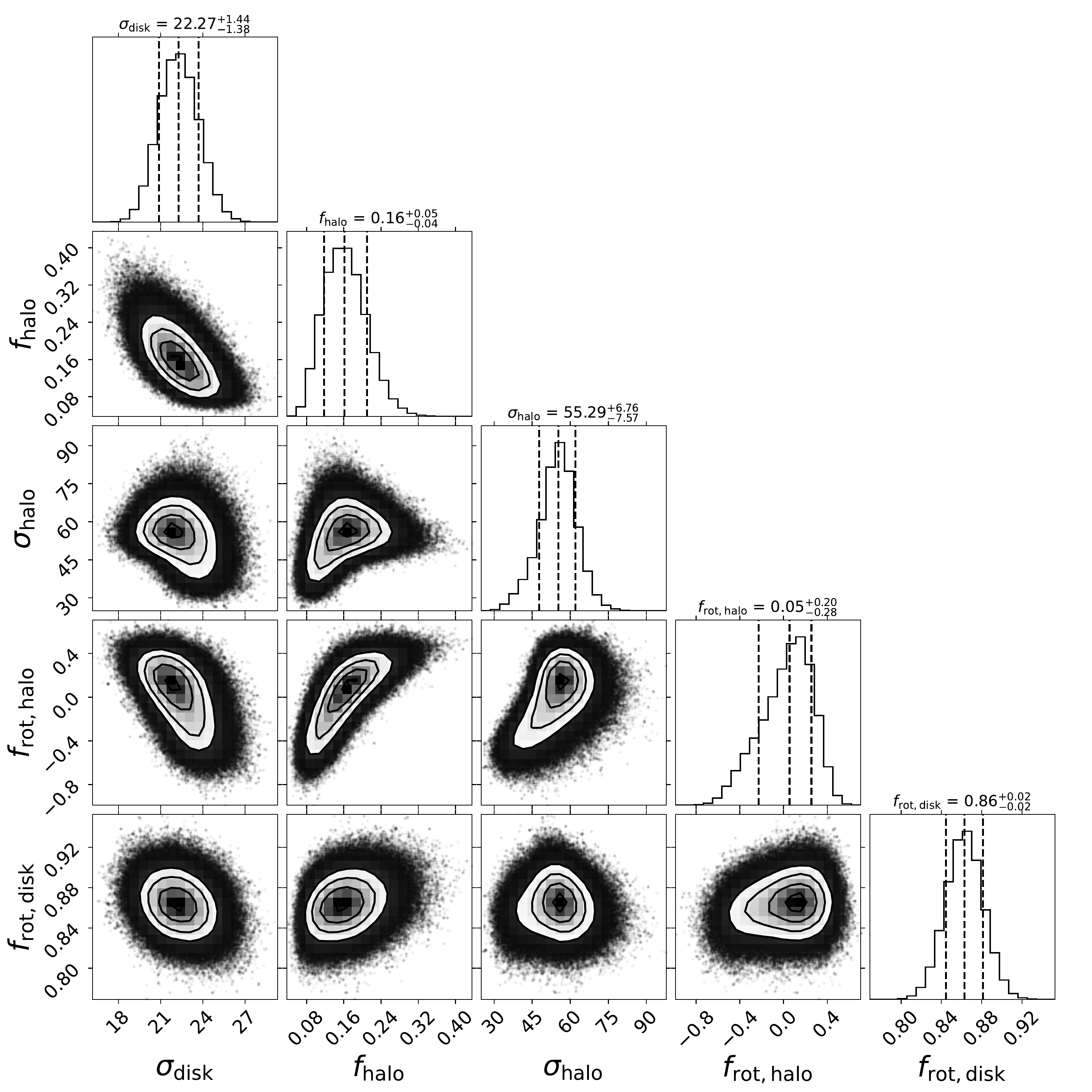}
\caption{
Same as Figure~\ref{fig:app_corner_all} for the fit to the southwest half of the middle radial bin ($15\leq$\rdisk$<30$\arcmin).   
}
\label{fig:ref_spatial4}
\end{figure*}

\begin{figure*}[tb!]
\plotone{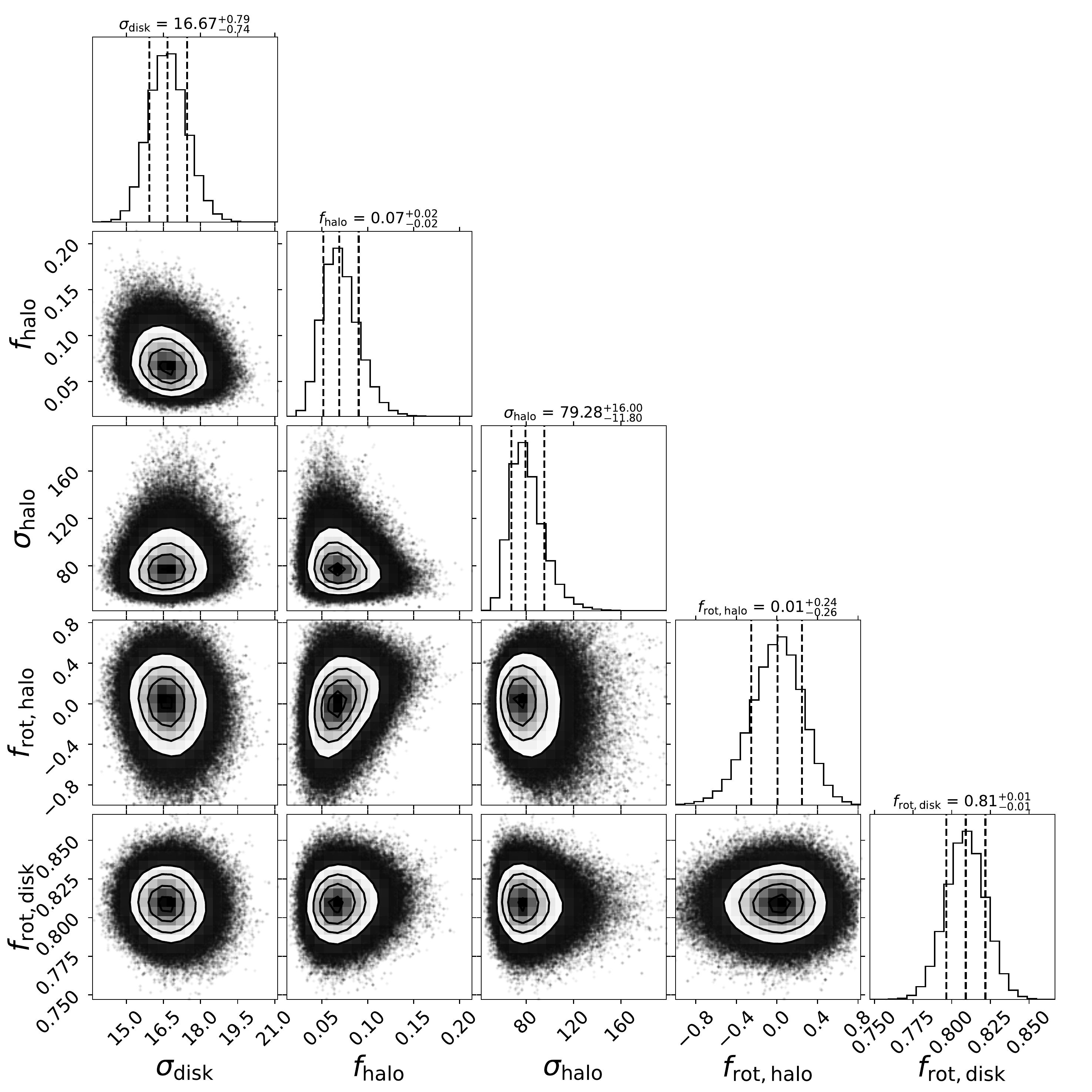}
\caption{
Same as Figure~\ref{fig:app_corner_all} for the fit to the northeast half of the outer radial bin (\rdisk$\geq 30$\arcmin).   
}
\label{fig:ref_spatial5}
\end{figure*}

\begin{figure*}[tb!]
\plotone{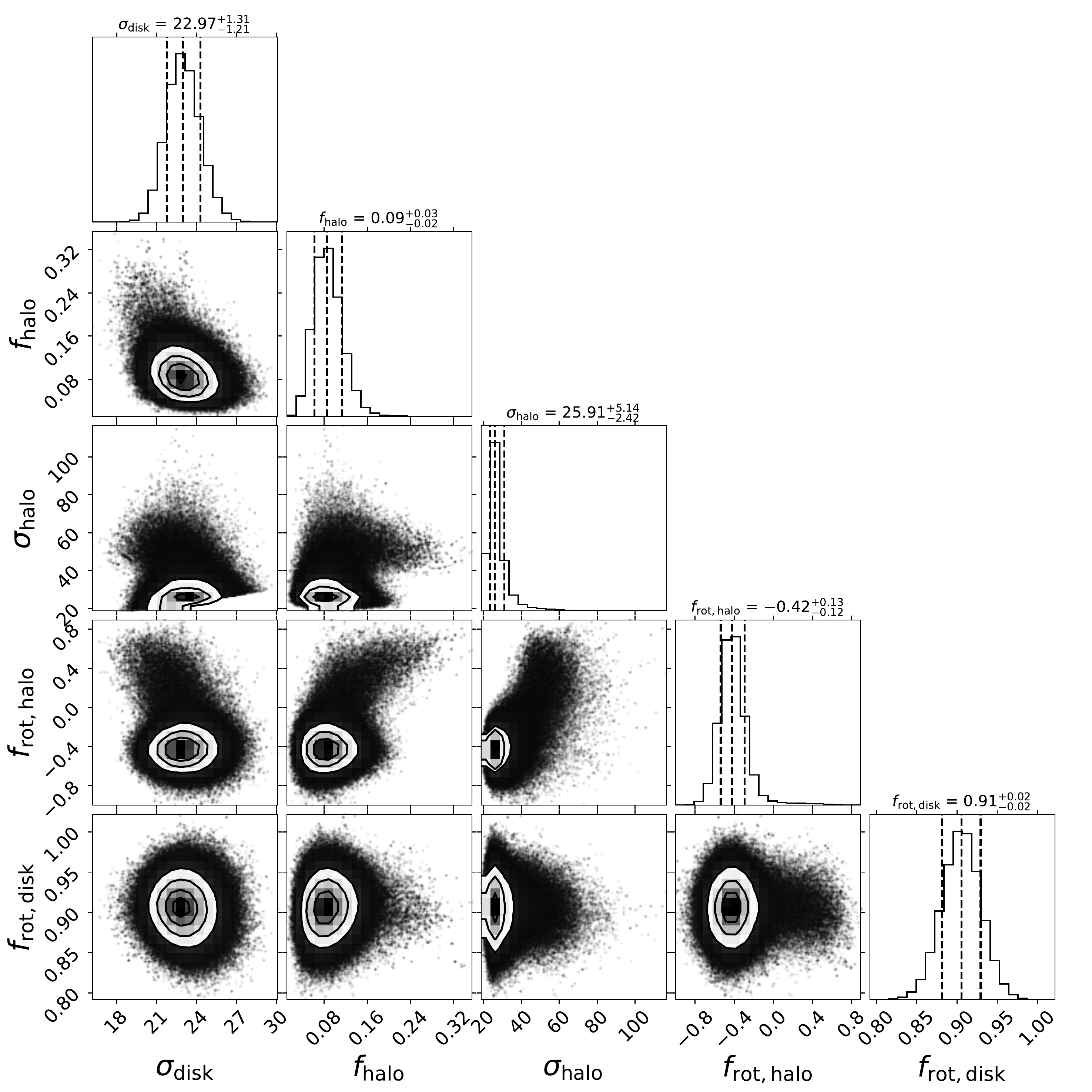}
\caption{
Same as Figure~\ref{fig:app_corner_all} for the fit to the southwest half of the outer radial bin (\rdisk$\geq 30$\arcmin).   
}
\label{fig:ref_spatial6}
\end{figure*}

Figure~\ref{fig:app_fits_full_radial_bins} shows the velocity distributions in \vlos\ and \voff, along with the model fits, for all stars in the inner, middle, and outer radial bins discussed in Section~\ref{sec:models_bins}.  Figures~\ref{fig:app_corner_radial_bins} through \ref{fig:ref_outerbin} show the marginalized 1- and 2-dimensional posterior probability distributions for each of these fits.  Figure~\ref{fig:app_fits_by_spatialposition} shows the velocity distributions and model fits in \voff\ corresponding to the spatial bins with \vlos\ distributions shown in Figure~\ref{fig:fits_by_spatialposition}; the corresponding posterior probability distributions are shown in Figures~\ref{fig:app_corner_spatialbins} through \ref{fig:ref_spatial6}.

\subsection{Justification for the Number of Model Components}\label{app:models_number_comps}

In Section~\ref{sec:one_comp_fits}, we argued that a single component was a poor fit to the RGB line of sight velocity distribution, and presented the results of a physically motivated two-component model.  As discussed in Section~\ref{sec:model_limitations}, the true velocity distribution of M33's disk regions may be more complex than captured by our two-component model.  

To quantify this trade-off between model complexity and the richness of our RGB dataset, we perform a Bayesian model comparison analysis on the full RGB spectroscopic sample.  We compare four different models. The simplest is a single component model ($N_{\rm comp}=1$).  This is straightforward to compare to the nominal model presented and discussed above, a two component model ($N_{\rm comp}=2$), which  we have interpreted as a disk and halo component.  We also compare these to models which have three and four components, where the additional components could account for additional complexity in the stellar disk and/or halo kinematics. 
Since the observed line-of-sight velocities indicate that the RGB stars in the disk rotate more slowly than the \hi\ model rotation velocity, we compared a one-component model that includes an \frot\ parameter to the nominal two-component model described above (e.g., the $N_{\rm comp}=1$ model follows the same formalism as the $N_{\rm comp}=2$ model for this test). Similarly, the three and four component models also follow this formalism, with the same model definitions as the nominal two component model.   For each of these models 
we employ broad priors on all model parameters (including allowing the \frot\ of each component to vary between -1 and 1), and we estimate the marginal likelihood $\mathcal{Z}$, often called the evidence:

\begin{equation}
P(\mathbf{D} | M) \equiv \mathcal{Z} = \int
\mathcal{L}(\boldsymbol{\Theta}) \pi(\boldsymbol{\Theta}) \,
d \boldsymbol{\Theta}.
\end{equation}

We estimate the log-evidence from our data ($\log{\hat{\mathcal{Z}}}$) using nested sampling \citep{Skilling06} as implemented in the Dynesty Python package \citep{Speagle20}. We then compare this estimate of $\log{\hat{\mathcal{Z}}}$  for each of the four models. We find that the one component model is disfavored over any of the $N_{\rm comp}>1$  models by $\Delta \log{\hat{\mathcal{Z}}}\gg 5$, the conventional ``decisive'' threshold \citep{jeffreys}, clearly demonstrating the need for an additional component. Among the $N_{\rm comp}>1$  models, the $N_{\rm comp}=3$ model is most favored, although less strongly than the comparison of the $N_{\rm comp}>1$ models versus the $N_{\rm comp}=1$ model. The $N_{\rm comp}=2$ (disk plus halo) model and the $N_{\rm comp}=3$ model are only distinguished by $\Delta \log{\hat{\mathcal{Z}}} \sim 0.4$ (``barely worth mentioning'').  The $N_{\rm comp}=3$ model is slightly more favored over the $N_{\rm comp}=4$ model
at $\Delta \log{\hat{\mathcal{Z}}} \sim 1.3$ (``strong'').

All $N_{\rm comp}>1$ models favor at least one component with a significantly lower rotation and larger dispersion, fitting the generally understood concept of ``halo.'' In the $N_{\rm comp}>=3$ models, there is some degeneracy between the model components,
for example between the highest dispersion, lowest \frot, halo-like component and intermediate (in dispersion and \frot, e.g., potentially representing a thicker disk) components. Hence, we conclude that a halo is clearly detected, and four (or more) major kinematical components are disfavored.  However, our data, combined with the present modelling assumptions, do not clearly distinguish between an $N_{\rm comp}=2$ or $N_{\rm comp}=3$ model.

The goal of this first contribution is to demonstrate that the dynamics of M33's stellar disk are more complicated than a single kinematically cold disk component and to present evidence for, and an initial characterization of, a significant, kinematically hot component present throughout M33's inner stellar disk. Given these goals, combined with the potential limitations of the current model for fully describing M33's stellar kinematics (Section~\ref{sec:model_limitations}), we present the $N_{\rm comp}=2$ model here, which, based on the model comparison described above, fits the data about as well as an $N_{\rm comp}=3$ model. 
As discussed in Section~\ref{sec:model_limitations}, in future work we will relax some of the assumptions that have been made in this initial analysis and which may impact our ability to identify and measure more complex disk and/or halo kinematics in a physically meaningful way. As part of that analysis, we will also revisit the evidence for the number of disk-like and halo-like components in our spectroscopic sample.

\listofchanges

\end{document}